\PassOptionsToPackage{square,numbers,sort&compress}{natbib}

\RequirePackage{fix-cm}
\documentclass{svjour3}                     % onecolumn (standard format)

\usepackage[top=3cm, bottom=3cm, left=3cm, right=3cm]{geometry}

\smartqed  % flush right qed marks, e.g. at end of proof

\makeatletter
\def\cl@chapter{\@elt {theorem}}
\makeatother

\usepackage{graphicx}

\usepackage{amsmath}
\usepackage{csquotes}
\usepackage{physics}
\usepackage{epstopdf}
\usepackage{mathtools}

\usepackage{cases}

\newcommand\restr[2]{{% we make the whole thing an ordinary symbol
  \left.\kern-\nulldelimiterspace % automatically resize the bar with \right
  #1 % the function
  \right|_{#2} % this is the delimiter
  }}

\newcommand\restrict[1]{\raisebox{-.5ex}{$|$}_{#1}}

\usepackage{amssymb}
\usepackage{latexsym}

\def\lunit{~\rm{\mu}m}
\def\tunit{~{\rm{ms}}}

\def\dunit{~\rm{\mu}m^2/{\rm{ms}}}
\def\kunit{~\rm{\mu}m/{\rm{ms}}}

\usepackage{url}
\usepackage{xcolor}
\definecolor{newcolor}{rgb}{.8,.349,.1}

 \journalname{Journal of Scientific Computing}

\usepackage[square,numbers,sort&compress]{natbib}

\relax 
\begin{document}

\title{Diffusion across semi-permeable barriers: spectral properties, efficient computation, and applications}

\titlerunning{Diffusion across semi-permeable barriers}        % if too long for running head

\author{Nicolas Moutal         \and
        Denis Grebenkov %etc.
}


\institute{N. Moutal\at
              PMC, CNRS – Ecole Polytechnique, F-91128, Palaiseau, France \\
			  Tel: +33 1 69 33 46 96 \\
              \email{nicolas.moutal@polytechnique.edu}
           \and
           D. Grebenkov \at
              PMC, CNRS – Ecole Polytechnique, F-91128, Palaiseau, France
}


\maketitle

\begin{abstract}
%%%
We present an efficient method to compute the eigenvalues and eigenmodes of the diffusion operator $\nabla(D\nabla)$ on one-dimensional heterogeneous structures with multiple semi-permeable barriers.
This method allows us to calculate the diffusion propagator and related quantities such as diffusion MRI signal or first exit time distribution analytically for regular geometries and numerically for arbitrary ones.
The effect of the barriers and the transition from infinite permeability (no barriers) to zero permeability (impermeable barriers) are investigated.
%%%%
\end{abstract}

\section{Introduction}
\label{intro}

Diffusion is a very broad transport mechanism which may describe heat conduction in solids as well as molecular exchanges in biological systems, among many examples. One often characterizes diffusion processes by the \textquote{diffusion propagator} (or \textquote{heat kernel}) $G(x_0\to x,t)$ which is the probability density of reaching position $x$ after a time $t$ starting from $x_0$. When diffusion takes place in a homogeneous medium without boundaries, the  propagator is a Gaussian distribution centered on $x_0$ with variance $2Dt$, where $D$ is the diffusion coefficient in the medium. On the other hand diffusion in complex systems such as biological cells or composite materials may exhibit non-Gaussian behavior due to confinement, hindrance by semi-permeable barriers or heterogeneity of the diffusion coefficient.

Generally speaking, the diffusion propagator obeys the diffusion equation:
\begin{equation}
\frac{\partial G}{\partial t}=\nabla(D\nabla G)\;, \quad G(x_0\to x,t=0)=\delta(x-x_0)\;,
\label{eq:Laplace_equation}
\end{equation}
where $\delta$ is the Dirac distribution, $\nabla=\frac{\partial}{\partial x}$ in the one-dimensional case, and the diffusion coefficient $D$ can in general be space and time dependent to capture heterogeneities of the medium \cite{Carslaw59,Crank75}. Throughout this article, we refer to $\nabla(D\nabla)$ as the \textquote{diffusion operator}. Note that if the diffusion coefficient is uniform, then the diffusion operator is simply proportional to the Laplace operator $\nabla^2$. The complexity of the geometry is hidden in the boundary conditions imposed on $G$ at the outer boundaries and possible inner semi-permeable barriers.
Analytical solutions of Eq. \eqref{eq:Laplace_equation} mainly rely on spectral decomposition over the diffusion operator eigenmodes which are explicitly known only for few geometries: slab, disk, sphere (and some simple extensions) \cite{Grebenkov13}. The study of more complicated structures requires numerical simulations such as  stochastic Monte-Carlo simulations \cite{Lejay12,Lejay2015a} or PDE solving with finite element or finite difference methods \cite{Hickson11a}. On top of being time-consuming these techniques give little theoretical insight into the dependence of the propagator on the physical parameters of the simulated medium.
In this situation, one-dimensional models of heterogeneous systems partitioned by semi-permeable barriers can help to uncover this dependence and to understand the role of diffusive exchange across the barriers.
Note that three-dimensional diffusion in a stack of parallel planes with lateral invariance is naturally reduced to one-dimensional models.
As a consequence, these models have a wide variety of applications, for example multilayer electrodes \cite{Diard05,Freger05,Ngameni14}, coating of electronic components and improving the performance of semi-conductors \cite{Graff04,Gurevich09,Aguirre00}, geophysics and thermal analyses of buildings \cite{Grossel98,Lu05,Lu06,DeMonte00,Barbaro88}, industrial processes \cite{Yuen94,Hickson09-1,Hickson09-2}, waste disposal and gas permeation in soils \cite{Shackleford91,Liu09,Shackleford13,Yates00}, drug delivery \cite{Siegel86,Pontrelli07,Todo13} and modeling tumor growth \cite{Mantzavinos16}. \textcolor{black}{They can also be applied as approximation schemes for finding the spectrum of Sturm-Liouville problems where the coefficients of the differential operator are replaced by piecewise constant (or polynomial) functions (the so-called \textquote{Pruess method}) \cite{Canosa1970a,Pruess1973a,Pruess1975a,Marletta1992a,Pruess1993a}.}
Two applications of particular interest to us are diffusion magnetic resonance imaging (dMRI), a powerful experimental technique for probing diffusion inside complex media such as biological tissues (see Sec. \ref{section:explication_dmri}), and first-passage phenomena (Sec. \ref{section:explication_temps_sortie}).
%
% or medical imaging \cite{Tanner78,Grebenkov10,VanAs07}.
%

Because of this diversity of applications, many authors have more or less independently tackled such models of one-dimensional diffusion in heterogeneous structures, with various computational techniques: spectral decompositions, Green functions, Laplace transforms and others (see \cite{Ozisik12,Mikhailov84} for a review of the subject). In this article we consider finite geometries, which are best treated by spectral decompositions (or \textquote{separation of variables}). To our knowledge, the most recent and complete work on this topic is the one by Hickson \textit{et al} \cite{Hickson11a,Hickson09-1,Hickson09-2}. However it was mainly devoted to the case of heterogeneous structures with distinct diffusivities and without barriers. Moreover the spectrum was computed numerically and only few analytical results were obtained. \textcolor{black}{On the other hand, some very general mathematical results were obtained by Gaveau \textit{et al} for generic heterogeneous media without barriers \cite{Gaveau1987a}.}
Another technique was proposed in the recent work by Carr and Turner \cite{Carr16}, in which the solution of Eq. \eqref{eq:Laplace_equation} was decomposed on the Laplacian eigenmodes of each compartment separately, instead of the eigenmodes of the whole structure.
This technique presents numerical advantages without providing analytical insights onto the spectrum of the diffusion operator.

\vskip 5mm

In this article we present an efficient method to compute the eigenvalues and eigenfunctions of the diffusion operator in one-dimensional domains with multiple barriers. This method allows us to calculate the diffusion propagator and related quantities such as dMRI signal or first exit time distribution analytically for sufficiently regular geometries such as a finite periodic geometry or a micro-structure inside a larger scale structure, and numerically for arbitrary structures.

{\color{black}
The article is organized as follows. Section \ref{section:calcul_general} is entirely devoted to analytics. We start with standard computations using transition matrices (Sec. \ref{section:general_case}) and obtain the equation of the spectrum as a transcendental equation $F(\lambda)=0$ (Eq. \eqref{eq:coherence_condition3}). Three following subsections are more technical and may be omitted in a first reading. In particular, we express the normalization constant of the eigenmodes as a function of $F$ (Eq. \eqref{eq:normalisation1}), and we derive general consequences of the symmetry or the periodicity of the medium (Sec. \ref{section:symmetry} and \ref{section:periodicity}, respectively). In Sec. \ref{section:study_spectrum}, we study in more detail the function $F$ and obtain simple estimates of its roots with respect to the geometrical parameters of the medium, in particular the permeability of the barriers. This part is crucial for the numerical implementation of the method. This section is concluded with some extensions of our model. Section \ref{section:example_periodic} illustrates our general approach on the example of a (finite) periodic structure with multiple identical barriers and compartments.
The numerical implementation of the method is presented in Sec. \ref{section:numerical_implementation}. In particular, we discuss the major numerical challenges related to finding very close zeros of the eigenspectrum equation \eqref{eq:coherence_condition3} and the proposed shortcuts based on the analytics from Sec. \ref{section:calcul_general}.} The application of our technique to the computation of the dMRI signal and the first exit-time distribution is briefly discussed in Sec. \ref{section:explication_dmri} and \ref{section:explication_temps_sortie}.
Section \ref{section:conclusion} concludes the paper and presents further perspectives and open problems.
%
%
%
%We start by presenting the general computation for an arbitrary one-dimensional geometry. The eigenvalues are expressed as roots of a function whose properties allow us to obtain simple estimates with respect to the geometrical parameters of the medium, in particular the permeability of the barriers.

The electronic Supplementary Material (SM) contains additional developments.
Section \ref{section:application_dmri} is devoted to the application to dMRI. The dependence of the acquired signal on the geometrical parameters of the medium is thoroughly discussed.
In Sec. \ref{section:application_temps_sortie}, the effect of semi-permeable barriers on the diffusive motion is studied from another viewpoint, namely the first exit time distribution. 
Some technical results are moved to Sec. \ref{section:non-degeneracy}, which contains proofs of the existence of infinitely many eigenvalues, their non-degeneracy, their monotonic growth with respect to the barrier permeabilities, as well as a Courant nodal theorem for our particular model of diffusion with barriers.

%%%%%%%%%%%%%%%%%%%%%%%%%%%%%%%%%%%%%%%%%%%%

\section{Computation of the eigenmodes of the diffusion operator}
\label{section:calcul_general}
\subsection{General case}
\label{section:general_case}

\begin{figure*}[tbp]
\begin{center}
\includegraphics[width=0.9\linewidth]{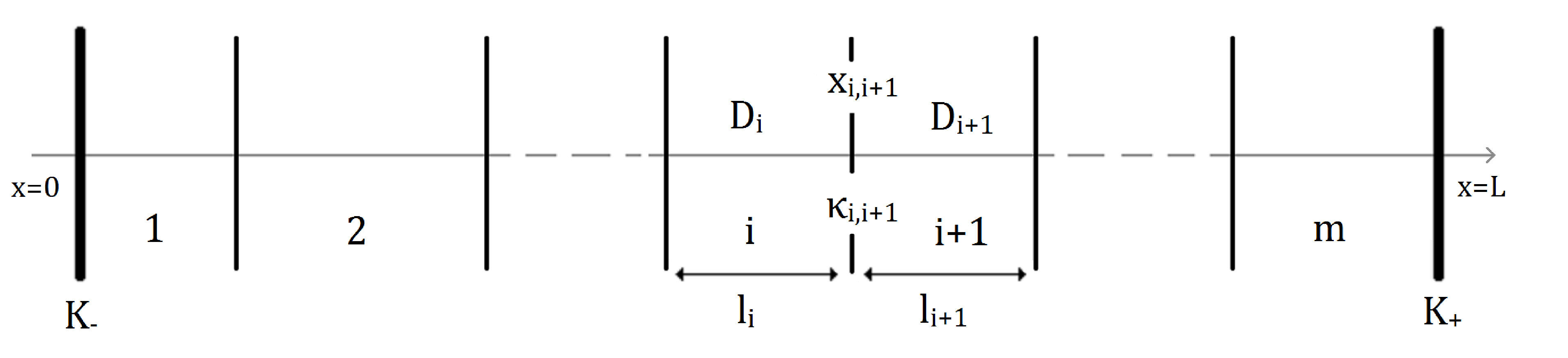}
\end{center}
\caption{Illustration of the geometry. Arbitrarily spaced barriers split the interval $[0,L]$ into $m$ compartments $\Omega_i$ of length $l_i$ and diffusion coefficient $D_i$. The positions of the barriers are denoted by $x_{i,i+1}$ and their permeabilities by $\kappa_{i,i+1}$. One can also take into account relaxation or leakage at the two outer barriers by permeabilities $K_-$, $K_+$.
}
\label{fig:couches}
\end{figure*}

In this section we study the  eigenmodes of the \textquote{diffusion operator} $\nabla(D\nabla)$ in a one-dimensional geometry (see Fig. \ref{fig:couches}). We reproduce the general computational scheme from Ref. \cite{Grebenkov10} and propose improvements specific to the one-dimensional geometry. An interval $\left[0,L\right]$ is divided by barriers into $m$ compartments (or \textquote{cells}) $\Omega_i=\left(x_{i-1,i},x_{i,i+1}\right)$, $i=1,\ldots,m$, where $x_{1,2},\ldots,x_{m-1,m}$ are the positions of $m-1$ inner barriers, and $x_{0,1}=0$ and $x_{m,m+1}=L$ correspond to the outer barriers. Each compartment is characterized by its length $l_i=x_{i,i+1}-x_{i-1,i}>0$ and diffusion coefficient $D_i> 0$ and each barrier by its permeability $\kappa_{i,i+1} \geq 0$ or equivalently by its \textquote{resistance} to diffusive exchange: $r_{i,i+1}=1/\kappa_{i,i+1}$. Finally one can take into account some relaxation or leakage at the  endpoints by non-negative permeabilities (or relaxaton coefficients) $K_-$ and $K_+$.

The diffusion coefficient $D$ is thus a piecewise constant function:
\begin{equation}
D(x)=\sum_{i=1}^m D_i I_{\Omega_i}(x),
\label{eq:def_D}
\end{equation}
where $I_{\Omega_i}$ denotes the indicator function of $\Omega_i$: $I_{\Omega_i}(x)=1$ if $x\in \Omega_i$ and $0$ otherwise. This implies that the diffusion operator can be split into two terms:
\begin{equation}
\nabla(D\nabla) = D\nabla^2 + (\nabla D)\nabla=D\nabla^2+\left(\sum_{i=1}^{m-1} (D_{i+1}-D_i)\delta(x-x_{i,i+1})\right)\nabla\;.
\label{eq:decomposition_operateur_diffusion}
\end{equation}
The second term vanishes at the interior points so that the diffusion operator is reduced to $D\nabla^2$. \textcolor{black}{The same is true for the general class of diffusion operators $\nabla(D^{\alpha}\nabla(D^{1-\alpha}\: \cdot))$, where $0\leq \alpha \leq 1$ is the It{\^o}-Stratonovitch  interpretation parameter (some authors use $1-\alpha$ instead of $\alpha$) \cite{Sokolov2010a,Haan2012a}. Here we consider heterogeneous diffusion coefficients with discontinuities at the barriers, hence these operators coincide inside the compartments but yield different boundary conditions at the barriers. Our choice $\nabla(D\nabla)$ corresponds to the H{\"a}nggi-Klimontovich interpretation \cite{Haenggi1978a,Haenggi1980a,Haenggi1982a,Klimontovich1990a,Klimontovich1994a} with $\alpha=1$, which is most often used in physical applications. The main reason is that it corresponds to the standard Fick law and that equilibrium solutions of the diffusion equation are constant, which is expected for, say, water diffusing in an isothermal medium. From a mathematical point of view, this choice ensures that the operator is self-adjoint, which allows us to use standard spectral methods.}

The $L^2$-normalized eigenmodes $u$ of the diffusion operator are then determined by the equation
\begin{equation}
D u''+\lambda u=0
\label{eq:equation_generale_modes}\;,
\end{equation}
with the boundary conditions
\begin{align}
&D_i u'\restrict{\Omega_i}=D_{i+1}u'\restrict{\Omega_{i+1}} & \text{at the barrier at $x_{i,i+1}$} \label{eq:equation_membranes_1}\\
&D_i u'\restrict{\Omega_i}=\kappa_{i,i+1}(u\restrict{\Omega_{i+1}}-u\restrict{\Omega_i}) & \text{at the barrier at $x_{i,i+1}$}\label{eq:equation_membranes_2}\\
&D_{1} u'(0)=K_{-}u(0) \label{eq:equation_bords_1}\\
&D_{m} u'(L)=-K_{+}u(L) \label{eq:equation_bords_2}\;,
\end{align}
and the normalization condition
\begin{equation}
\int_0^L u^2 =1\;,
\label{eq:equation_normalisation}
\end{equation}
where $u\restrict{\Omega_i}$ is the restriction of $u$ to the cell $\Omega_i$ ($i=1,\ldots,m$) and prime denotes the derivative with respect to $x$.
%\Cref{eq:equation_generale_modes}, combined with \cref{eq:equation_membranes_1,eq:equation_membranes_2,eq:equation_bords_1,eq:equation_bords_2}, and \cref{eq:equation_normalisation}, admits infinitely many solutions $u_n$, $n=1,2,\ldots$ associated with distinct non-negative eigenvalues $\lambda_n$ sorted by ascending order: $0\leq\lambda_1 < \lambda_2 < \ldots$. Moreover because of the symmetry of the diffusion operator these eigenmodes form a complete orthonormal basis (\textit{[Reference ?]}).
%A simple proof of the orthogonality of the eigenmodes is detailed in \labelcref{section:orthogonality}.
%

Eqs. \eqref{eq:equation_membranes_1} and \eqref{eq:equation_membranes_2} express the flux conservation across the barriers (no accumulation of diffusing particles) and the drop of particle density due to the non-zero resistance of the barriers, respectively. Note in particular that Eq. \eqref{eq:equation_membranes_1} ensures the continuity of $D\nabla u=Du'$.
%Another interpretation is that \cref{eq:equation_membranes_2} is also a flux conservation condition and that the particles at $x=x_{i,i+1}$ cross the barrier with a mean speed $\kappa_{i,i+1}$. 
The infinitely thin barriers that we consider can approximate barriers of thickness $h_{i,i+1}$ with the standard continuity conditions. When $h_{i,i+1}$ is much smaller than other length scales, one can interpret $\kappa_{i,i+1}h_{i,i+1}$ as the diffusion coefficient inside the barrier, whereas $(u\restrict{\Omega_{i+1}}-u\restrict{\Omega_i})/h_{i,i+1}$ approximates the derivative of $u$ across the barrier of thickness $h_{i,i+1}$.
If $\kappa_{i,i+1}=\infty$ there is no barrier and Eq. \eqref{eq:equation_membranes_2} becomes a continuity condition for $u$ at $x=x_{i,i+1}$. In the opposite limit $\kappa_{i,i+1}=0$ the compartments $\Omega_i$ and $\Omega_{i+1}$ do not communicate with each other: the flux $Du'$ is zero at the barrier and the discontinuity $(u\restrict{\Omega_{i+1}}-u\restrict{\Omega_i})(x_{i,i+1})$ is arbitrary. One can then study the two parts $[0,x_{i,i+1}]$ and $[x_{i,i+1},L]$ separately.

To avoid such trivial separations, we consider only non-zero permeabilities: $\kappa_{i,i+1}>0$ throughout this article. Under this assumption we prove in Sec. \ref{section:non-degeneracy} that there are infinitely many eigenvalues $\lambda_n, n=1,2,\ldots$, and all $\lambda_n$ are simple. One can also easily prove that they are non-negative, and we sort them by ascending order: $0\leq \lambda_1 < \lambda_2 < \ldots$. Moreover, thanks to the self-adjointness of the diffusion operator $\nabla(D\nabla)$ we know that the eigenmodes $u_n,n=1,2,\ldots$ form a complete orthonormal basis in the space $L^2(0,L)$ of square-integrable functions on $(0,L)$ \cite{Ozisik12,Mikhailov84}.

For simplicity we further assume that $K_- < \infty$, which allows us to write
\begin{equation}
u=\beta v\;,\quad v(0)=1\;,
\label{eq:u_beta_v}
\end{equation}
with $\beta$ being a normalization constant that ensures Eq. \eqref{eq:equation_normalisation}. The case of Dirichlet boundary conditions ($K_-=\infty$) requires another convention which is detailed in Sec. \ref{section:calculs_relax}. We study the (non-normalized) eigenmode $v$ first and then we compute the normalization constant $\beta$.

Throughout this section we assume $\lambda\neq0$. One can see that $\lambda=0$ is only possible if the relaxation coefficients $K_\pm$ are equal to zero and in this case one gets a constant eigenmode $v=1$ (and $\beta=1/\sqrt{L}$).

Equation \eqref{eq:equation_generale_modes} has a general solution
\begin{equation}
{v\restrict{\Omega_i}}(x) = a^l_i\cos(\sqrt{\lambda/D_i}(x-x_{i-1,i}))+b^l_i\sin(\sqrt{\lambda/D_i}(x-x_{i-1,i}))
\label{eq:formule_modes_gauche}\;,
\end{equation}
or equivalently
\begin{equation}
{v\restrict{\Omega_i}}(x) = a^r_i\cos(\sqrt{\lambda/D_i}(x-x_{i,i+1}))+b^r_i\sin(\sqrt{\lambda/D_i}(x-x_{i,i+1}))\;,
\label{eq:formule_modes_droite}
\end{equation}
where $a^l_i,b^l_i$ and $a^r_i,b^r_i$ are constants to be determined, related by
\begin{equation}
\begin{bmatrix}
a^r_i\\b^r_i
\end{bmatrix}
=
\mathcal{R}_i
\begin{bmatrix}
a^l_{i}\\b^l_{i}
\end{bmatrix}\;, \quad \text{ where } \quad 
\mathcal{R}_i=
\begin{bmatrix}
\cos(\sqrt{\lambda/D_i} l_i) & \sin(\sqrt{\lambda/D_i} l_i)\\
-\sin(\sqrt{\lambda/D_i} l_i) & \cos(\sqrt{\lambda/D_i} l_i)
\end{bmatrix}\;.
\label{eq:def_R}
\end{equation}
%where $\mathcal{R}(\theta)$ is the rotation matrix of angle $\theta$: $\mathcal{R}(\theta)=\begin{bmatrix}\cos\theta&-\sin\theta\\\sin\theta&\cos\theta\end{bmatrix}$.
%
%
%\Cref{eq:equation_generale_modes} has a general solution:
%\begin{equation}
%\label{eq:formule_modes_centre}
%{v\restrict{\Omega_i}}(x) = a_i\cos(\sqrt{\lambda/D_i}(x-x_i))+b_i\sin(\sqrt{\lambda/D_i}(x-x_i))\;,
%\end{equation}
%%with ${\beta_i}=\sqrt{\lambda_n/D_i}$.
%with two constants $a_i$ and $b_i$ to be determined.
%%
%We can choose $x_{i-1,i}$ as well as $x_i$ or $x_{i,i+1}$ as the origin point, so we can also write: 
%\begin{align}
%{v\restrict{\Omega_i}}(x) = a^r_i\cos(\sqrt{\lambda/D_i}(x-x_{i,i+1}))+b^r_i\sin(\sqrt{\lambda/D_i}(x-x_{i,i+1}))\;,\label{eq:formule_modes_droite}\\
%{v\restrict{\Omega_i}}(x) = a^l_i\cos(\sqrt{\lambda/D_i}(x-x_{i-1,i}))+b^l_i\sin(\sqrt{\lambda/D_i}(x-x_{i-1,i}))\label{eq:formule_modes_gauche}\;,
%\end{align}
%%
%with the new coefficients:
%\begin{equation*}
%\begin{bmatrix}
%a^r_i\\b^r_i
%\end{bmatrix}
%=
%\mathcal{R}(-\sqrt{\lambda/D_i} l_i/2)
%\begin{bmatrix}
%{a_i}\\{b_i}
%\end{bmatrix}\quad \text{and:} \quad
%\begin{bmatrix}
%a^l_{i}\\b^l_{i}
%\end{bmatrix}
%=
%\mathcal{R}(\sqrt{\lambda/D_{i}} l_{i}/2)
%\begin{bmatrix}
%{a_{i}}\\{b_{i}}
%\end{bmatrix}\;,
%\end{equation*}
%where $\mathcal{R}(\theta)$ is the rotation matrix of angle $\theta$: $\mathcal{R}(\theta)=\begin{bmatrix}\cos\theta&-\sin\theta\\\sin\theta&\cos\theta\end{bmatrix}$. 
%%
Note that
\begin{equation}
v\restrict{\Omega_i}(x_{i,i+1})=a^r_i\;,\quad D_i v'\restrict{\Omega_i}(x_{i,i+1})=\sqrt{\lambda D_i}b^r_i\;,
\label{eq:v_bords_matrice}
\end{equation}
with similar formulas for $a^l_i, b^l_i$, so that one can write the boundary equations \eqref{eq:equation_membranes_1} and  \eqref{eq:equation_membranes_2} as
%\begin{equation*}
%\sqrt{\lambda D_i}b^r_{i}=\sqrt{\lambda D_{i+1}} b^l_{i+1}= \kappa_{i,i+1}(a^l_{i+1}- a^r_{i}) \;,
%\end{equation*}
%which we can write in a matrix form:
\begin{equation}
\begin{bmatrix}
{a^l_{i+1}}\\{b^l_{i+1}}
\end{bmatrix}
=\mathcal{K}_{i,i+1}
\begin{bmatrix}
{a^r_i}\\{b^r_i}
\end{bmatrix}\;, \quad \text{ with } \quad
\mathcal{K}_{i,i+1}
=
\begin{bmatrix}
1&r_{i,i+1}\sqrt{\lambda D_i}\\
0&\sqrt{D_i/D_{i+1}}
\end{bmatrix}\;.
\label{eq:def_K}
\end{equation}
%
%
%
%By writing the formula for ${u_i}$ with matrices we can find the relationships between the coefficients $(a^r,b^r)$, $(a^l, b^l)$ and $(a,b)$. In fact,
%\begin{align*}
% {u_i}(x) &= 
%\begin{bmatrix}
%{a^r_i}&{b^r_i}
%\end{bmatrix}
%\begin{bmatrix}
%\cos{\beta_i}(x-x_{i,i+1})\\\sin{\beta_i}(x-x_{i,i+1})
%\end{bmatrix}\\
%&= 
%\begin{bmatrix}
%{a^r_i}&{b^r_i}
%\end{bmatrix}
%\mathcal{R}(-{\beta_i}{l_i}/2)
%\begin{bmatrix}
%\cos{\beta_i}(x-x_{i})\\\sin{\beta_i}(x-x_{i})
%\end{bmatrix}\;,
%\end{align*}
%where $\mathcal{R}(\theta)$ is the rotation matrix of angle $\theta$: $\begin{bmatrix}\cos\theta&-\sin\theta\\\sin\theta&\cos\theta\end{bmatrix}$. We will write in the following $\alpha_{n,i} = {\beta_i}l_{i}$, and $\tilde{\kappa}_{i,i+1}=\kappa_{i,i+1}{l_i}/D_i$.
%
%We obtain then:
%\begin{equation}
%\begin{bmatrix}
%a_{n,i}&b_{n,i}
%\end{bmatrix}
%=
%\begin{bmatrix}
%{a^r_i}&{b^r_i}
%\end{bmatrix}
%\mathcal{R}(-\alpha_{n,i}/2)\;.
%\end{equation}
%Finally,
%\begin{equation}
%\begin{bmatrix}
%a_{n,i}\\b_{n,i}
%\end{bmatrix}
%=
%\mathcal{R}(\alpha_{n,i}/2)
%\begin{bmatrix}
%{a^r_i}\\{b^r_i}
%\end{bmatrix}\;.
%\end{equation}
%Similarly we get:
%\begin{equation}
%\begin{bmatrix}
%a_{n,i+1}\\b_{n,i+1}
%\end{bmatrix}
%=
%\mathcal{R}(-\alpha_{n,i+1}/2)
%\begin{bmatrix}
%{a^l_{i+1}}\\{b^l_{i+1}}
%\end{bmatrix}\;.
%\end{equation}
%
%
%and we have introduced the notation: $\alpha_{n,i} = {\beta_i}l_{i}$. We will also write in the following $\tilde{\kappa}_{i,i+1}=\kappa_{i,i+1}{l_i}/D_i$ and $\tilde{r}_{i,i+1}=1/\tilde{\kappa}_{i,i+1}=D_i/l_i \kappa_{i,i+1}$.
%
The equations at the barriers can thus be restated in a matrix form:
%\begin{equation}
%\label{eq:membrane_equation_matrice}
%\begin{bmatrix}
%{a^l_{i+1}}\\{b^l_{i+1}}
%\end{bmatrix}
%=
%\begin{bmatrix}
%1&r_{i,i+1}\sqrt{\lambda D_i}\\
%0&\sqrt{D_i/D_{i+1}}
%\end{bmatrix}
%\mathcal{R}(\sqrt{\lambda/D_i}\cdot l_i)
%\begin{bmatrix}
%{a^l_i}\\{b^l_i}
%\end{bmatrix}\;.
%\end{equation}
%
%
\begin{equation}
\begin{bmatrix}
{a^l_{i+1}}\\{b^l_{i+1}}
\end{bmatrix}
=
\mathcal{M}_{i,i+1}
\begin{bmatrix}
{a^l_i}\\{b^l_i}
\end{bmatrix}\;,
\label{eq:membrane_equation_matrice}
\end{equation}
%with the notations:
%\begin{align}
%&\mathcal{M}_{i,i+1}=\mathcal{K}_{i,i+1}\mathcal{R}_i\;,\\
%&\mathcal{K}_{i,i+1}=
%\begin{bmatrix}
%1&r_{i,i+1}\sqrt{\lambda D_i}\\
%0&\sqrt{D_i/D_{i+1}}
%\end{bmatrix}\;,\\
%&\mathcal{R}_{i}=\mathcal{R}(-\sqrt{\lambda/D_i} l_i )
%%=\begin{bmatrix}
%%\cos(\sqrt{\lambda/D_i} l_i) & \sin(\sqrt{\lambda/D_i} l_i)\\
%%-\sin(\sqrt{\lambda/D_i} l_i) & \cos(\sqrt{\lambda/D_i} l_i)
%%\end{bmatrix}
%\;.
%\label{eq:abreviations}
%\end{align}
with the notation for the \textquote{transition matrix}:
\begin{equation}
\mathcal{M}_{i,i+1}=\mathcal{K}_{i,i+1}\mathcal{R}_i\;,
\label{eq:abreviations}
\end{equation}
with $\mathcal{R}_i$ and $\mathcal{K}_{i,i+1}$ defined by Eqs. \eqref{eq:def_R}, \eqref{eq:def_K}. In the same way, one can rewrite the endpoint conditions \eqref{eq:equation_bords_1}, \eqref{eq:equation_bords_2}:
\begin{equation*}
\begin{bmatrix}
-K_{-}&\sqrt{\lambda D_{1}}
\end{bmatrix}
\begin{bmatrix}
a^l_{1}\\b^l_{1}
\end{bmatrix}=0\quad \text{ and }\quad
\begin{bmatrix}
K_{+}&\sqrt{\lambda D_m}
\end{bmatrix}
\begin{bmatrix}
a^r_{m}\\b^r_{m}
\end{bmatrix}=0 \;.
\end{equation*}
We have the additional condition $a^l_1=v(0)=1$, therefore
\begin{equation}
\begin{bmatrix}
a^l_{1}\\b^l_{1}
\end{bmatrix}
=
\begin{bmatrix}
1\\ {K_{-}}/{\sqrt{\lambda D_{1}}}
\end{bmatrix} \quad\text{ and }\quad
\begin{bmatrix}
a^r_{m}\\b^r_{m}
\end{bmatrix}
=\epsilon
\begin{bmatrix}
1\\{-K_{+}}/{\sqrt{\lambda D_{m}}}
\end{bmatrix}\;,
\label{eq:conditions_limites_matrice}
\end{equation}
where $\epsilon$ is an unknown proportionality coefficient.
%and finally:
%\begin{equation}
%\begin{bmatrix}
%a_{n,1}\\b_{n,1}
%\end{bmatrix}
%\propto
%\mathcal{R}(-\alpha_{n,1}/2)
%\begin{bmatrix}
%\alpha_{n,1}\\\tilde{K}_{-}
%\end{bmatrix}
%\text{ and }
%\begin{bmatrix}
%a_{n,m}\\b_{n,m}
%\end{bmatrix}
%\propto
%\mathcal{R}(\alpha_{n,m}/2)
%\begin{bmatrix}
%\alpha_{n,m} \\ -\tilde{K}_{+}
%\end{bmatrix}\;,
%\end{equation}
%with the notation $\tilde{K}_-=\frac{K_- l_1}{D_1}$ and $\tilde{K}_+ = \frac{K_+l_m}{D_m}$.

Equation \eqref{eq:membrane_equation_matrice}, which relates the coefficients of one cell to those of the next cell, is compatible with Eq. \eqref{eq:conditions_limites_matrice}, which prescribes the first and last cell coefficients (up to a proportionality factor), only if $\lambda$ is an actual eigenvalue of the diffusion operator $\nabla(D \nabla)$. That is, by writing explicitly the condition that the product of all the transition matrices $\mathcal{M}_{i,i+1}$ should send the previously determined $(a^l_{1},b^l_{1})$ onto the $(a^l_{m},b^l_{m})$, we get the equation on the spectrum of the diffusion operator:
\begin{equation}
\mathcal{T}
\begin{bmatrix}
1\\ {K_{-}}/{\sqrt{\lambda D_{1}}}
\end{bmatrix}
=\epsilon
\begin{bmatrix}
1\\{-K_{+}}/{\sqrt{\lambda D_{m}}}
\end{bmatrix}\;,
\label{eq:coherence_condition}
\end{equation}
with 
\begin{equation}
\mathcal{T}=\mathcal{R}_m\mathcal{M}_{m-1,m}\dots \mathcal{M}_{1,2}\;.
\label{eq:formule_generale_T}
\end{equation} 
Note that this condition is equivalent to
\begin{equation}
\begin{bmatrix}
{K_{+}}/{\sqrt{\lambda D_{m}}}&~1
\end{bmatrix}
\mathcal{T}
=\eta
\begin{bmatrix}
{-K_{-}}/{\sqrt{\lambda D_{1}}}&~1
\end{bmatrix}\;,
\label{eq:coherence_condition2}
\end{equation}
and to
\begin{equation}
F(\lambda):=
\begin{bmatrix}
{K_{+}}/{\sqrt{\lambda D_{m}}}&~1
\end{bmatrix}\mathcal{T}(\lambda)
\begin{bmatrix}
1\\ {K_{-}}/{\sqrt{\lambda D_{1}}}
\end{bmatrix}=0\;.
\label{eq:coherence_condition3}
\end{equation}
The proportionality coefficients $\epsilon$ and $\eta$ are constrained by the relation: $\epsilon\eta=\det \mathcal{T}=\sqrt{\frac{D_1}{D_m}}$. 

%%%%%%%%%%%%%%%%%%%%%%%%%%%%%%%%%%%%%%%%

\subsection{Computation of the norm}
\label{section:computation_norm}

Now we compute the normalization constant $\beta$. Since the eigenmode $v$ is a piecewise combination of sine and cosine functions, the constant $\beta$ can be obtained by a direct integration (see Ref. \cite{Grebenkov10}). This approach is convenient for numerical computations. Here we present another approach which is more suitable for analytical derivations.
The starting point of the method is the spectral decomposition of the diffusion propagator:
\begin{equation}
G(t,x_0\to x) = \sum_{n=1}^{\infty} u_n(x_0) u_n(x) e^{-\lambda_n t} = \sum_{n=1}^{\infty} {\beta_n}^2 v_n(x_0)v_n(x) e^{-\lambda_n t}\;,
\label{eq:decomposition_G}
\end{equation}
where $n=1,2,\ldots$ spans the infinitely many eigenmodes of the diffusion operator.
%We now compute this propagator in a different way by solving explicitly the differential equation $D\frac{\partial^2G}{\partial x^2}=\frac{\partial G}{\partial t}$ with the initial condition $G(t=0,x_0\to x)=\delta(x-x_0)$. 
We now compute this propagator in a different way by solving explicitly Eq. \eqref{eq:Laplace_equation}.
Again, we use Eq. \eqref{eq:decomposition_operateur_diffusion} to transform $\nabla(D\nabla)$ into $D\nabla^2$ at the interior points. 
Let $\tilde{G}(s,x_0\to x)$ denote the Laplace transform of the propagator: $\tilde{G}(s,x_0\to x)=\int_0^{\infty} e^{-st} G(t,x_0\to x)\,\mathrm{d}t$.
Then $\tilde{G}$ obeys the equation
\begin{equation*}
D(x)\tilde{G}''(s,x_0\to x) = s\tilde{G}(s,x_0\to x) - \delta(x-x_0)\;,
\end{equation*}
with the same boundary conditions \eqref{eq:equation_membranes_1}-\eqref{eq:equation_bords_2} as for the propagator $G$ in time domain. As in the previous section, prime denotes derivative with respect to $x$. We use the method from Sec. \ref{section:general_case} to solve the homogeneous equation with the inner boundary conditions \eqref{eq:equation_membranes_1}, \eqref{eq:equation_membranes_2} imposed at the barriers: if $s\neq 0$ we can build two solutions $\phi(s,x)$ and $\psi(s,x)$ such that:
\begin{itemize}
\item $\phi(s,x)$ is built from $\begin{bmatrix} a^l_1\\b^l_1 \end{bmatrix}=\begin{bmatrix} 1\\0 \end{bmatrix}$: at the left endpoint its derivative with respect to $x$ is zero and its value is one.
\item $\psi(s,x)$ is built from  $\begin{bmatrix} a^l_1\\b^l_1 \end{bmatrix}=\begin{bmatrix} 0\\1 \end{bmatrix}$: at the left endpoint its derivative with respect to $x$ is $\sqrt{s/D_1}$ and its value is zero.
\end{itemize}
It is then easy to obtain the complete solution because the Wronskian matrix $\mathcal{W} = \begin{bmatrix} \phi(s,x) & \psi(s,x) \\\phi'(s,x) & \psi'(s,x)\end{bmatrix}$ is quite simple. Indeed over any layer $\Omega_i$ the determinant of $\mathcal{W}$ is constant and equal to $\sqrt{sD_1}/D_i$. This is obtained from the differential equation obeyed by $\phi(s,x)$ and $\psi(s,x)$ and the boundary conditions at each barrier.
The standard method for solving the second order differential equations then yields
\begin{equation*}
%\tilde{G}(s,x) = \mu(s,x) \phi(s,x) + \nu(s,x) \psi(s,x) \;,
\tilde{G}=\mu \phi + \nu \psi\;,
\end{equation*}
with the equation on $\mu$, $\nu$:
\begin{equation*}
D(x)\begin{bmatrix}\mu'(s,x) \\ \nu'(s,x)\end{bmatrix} = \mathcal{W}^{-1}\begin{bmatrix}0\\ -\delta(x-x_0)\end{bmatrix} = -\frac{D(x)}{\sqrt{D_1s}}\delta(x-x_0) \begin{bmatrix}-\psi(s,x)\\\phi(s,x)\end{bmatrix} \;.
\end{equation*}
After a straightforward integration, we obtain
\begin{align*}
\tilde{G}(x_0\to x,s)&=\left(A +\frac{1}{\sqrt{D_1s}}\psi(s,x_0) H(x-x_0)\right)\phi(s,x) \\&+ \left(B - \frac{1}{\sqrt{D_1s}}\phi(s,x_0)H(x-x_0)\right)\psi(s,x) \;,
\end{align*}
which is valid for any $x_0, x\in [0,L]$, and $s\neq 0$, where $H$ is the Heaviside function and the constants $A$ and $B$ remain to be determined. We consider general relaxing conditions at the endpoints:
\begin{equation*}
\begin{cases}
D_1\frac{\partial \tilde{G}}{\partial x}(x=0)=K_- \tilde{G}(x=0) \\
D_m \frac{\partial \tilde{G}}{\partial x}(x=L)=-K_+ \tilde{G}(x=L)
\end{cases}\;,
\end{equation*}
from which
\begin{align*}
A&=\frac{\phi(s,x_0)(D_m\psi'(s,L)+K_+\psi(s,L))-\psi(s,x_0)(D_m\phi'(s,L)+K_+\phi(s,L))}{D_mK_-\psi'(s,L)+K_+K_-\psi(s,L)+D_m\sqrt{D_1s}\phi'(s,L)+K_+\sqrt{D_1s}\phi(s,L)}\;,\\
B&=\frac{K_- A}{\sqrt{D_1 s}}\;.
\end{align*}
Now we simplify the above expressions. We anticipate that the non-normalized eigenmodes are $v_n(x)=v(\lambda_n,x)$, with
\begin{equation*}
v(s,x)=\phi(s,x)+\frac{K_-}{\sqrt{D_1s}}\psi(s,x)\;,
\end{equation*}
and we use Eq. \eqref{eq:v_bords_matrice} to get
%
%Let us denote $\mathcal{T}(s)$ the transition matrix from $\begin{bmatrix}a^l_{1}\\b^l_{1}\end{bmatrix}$ to $\begin{bmatrix} a^r_{m}\\b^r_{m}\end{bmatrix}$. We have then:
%\begin{align*}
%&\phi_s(L)=\begin{bmatrix}1&0\end{bmatrix}\mathcal{T}(s)\begin{bmatrix}1\\0\end{bmatrix}\; , \quad   \psi_s(L)=\begin{bmatrix}1&0\end{bmatrix}\mathcal{T}(s)\begin{bmatrix}0\\1\end{bmatrix},\\
%&D_m\phi_s'(L)=\begin{bmatrix}0&\sqrt{D_ms}\end{bmatrix}\mathcal{T}(s)\begin{bmatrix}1\\0\end{bmatrix}\; ,\quad   D_m\psi_s'(L)=\begin{bmatrix}0&\sqrt{D_ms}\end{bmatrix}\mathcal{T}(s)\begin{bmatrix}0\\1\end{bmatrix}.
%\end{align*}
%As a consequence, the denominator in the expression of $A$ is equal to
%%$F(s)=\begin{bmatrix}K_+ & \sqrt{D_m s}\end{bmatrix}\mathcal{T}(s)\begin{bmatrix}\sqrt{D_1 s}\\K_-\end{bmatrix}$. 
%$F(s)$ (see \cref{eq:coherence_condition3}).
%Moreover, one has:
%\begin{equation*}
%A\phi(s,x)+B\psi(s,x)=\frac{\phi(s,x)+\frac{K_-}{\sqrt{D_1 s}}\psi(s,x)}{F(s)} \left( \begin{bmatrix}K_+&\sqrt{D_m s}\end{bmatrix}\mathcal{T}(s)\begin{bmatrix}-\psi(s,x_0)\\\phi(s,x_0)\end{bmatrix}\right)\;.
%\end{equation*}
%
%The equation of the spectrum is obtained by setting the denominator $F(s)$ of the above expression to zero. The (non-normalized) eigenmodes of the diffusion operator are:
%\begin{equation*}
%v_s=\phi_s+\frac{K_-}{\sqrt{D_1s}}\psi_s, \quad s=\lambda_n\;,
%\end{equation*}
%%If we write 
%%\begin{equation*}
%%-\psi(s,x_0)=\frac{\sqrt{D_1s}}{K_-}\phi(s,x_0)-\frac{\sqrt{D_1s}}{K_-}v(s,x_0)\;,
%%\end{equation*}
%%then we get:
%and:
\begin{align*}
A\phi(s,x)+B\psi(s,x)&=\frac{v(s,x)\phi(s,x_0)}{K_-}-\frac{\sqrt{D_1 s}}{K_-}v(s,x)v(s,x_0)\frac{\begin{bmatrix}K_+&\sqrt{D_m s}\end{bmatrix}\mathcal{T}(s)\begin{bmatrix}1\\0\end{bmatrix}}{F(s)}\;,
\end{align*}
with $\mathcal{T}$ and $F$ defined in Eqs. \eqref{eq:formule_generale_T}, \eqref{eq:coherence_condition3}, respectively, in which $\lambda$ is replaced by $s$.
To obtain the propagator in time domain, one needs to perform an inverse Laplace transform. This is done by looking for the poles $s=\lambda_n$ of $\tilde{G}$ and the above formula shows that they are given by the zeros of $F(s)$, as expected. We prove in Sec. \ref{section:simplicity} that these zeros are simple.
%We recall that at $s=\lambda_n$, \cref{eq:coherence_condition,eq:coherence_condition2} yield:
%\begin{equation}
%\begin{bmatrix}K_+ & \sqrt{\lambda_n D_m}\end{bmatrix}\mathcal{T}(\lambda_n)
%=\eta_n \begin{bmatrix}-K_- & \sqrt{\lambda_n D_m}\end{bmatrix}\;,\quad
%\mathcal{T}(\lambda_n)\begin{bmatrix}\sqrt{\lambda_n D_1}\\ K_-\end{bmatrix}
%=\epsilon_n
%\begin{bmatrix}
%\sqrt{\lambda_n D_m} \\ -K_+
%\end{bmatrix}\;,
%\label{eq:simplif}
%\end{equation}
%from which we get simply: 
At $s=\lambda_n$, one can use Eqs. \eqref{eq:coherence_condition} and \eqref{eq:coherence_condition2} to compute the residue of $\tilde{G}$, which yields simply
\begin{equation*}
\mathrm{Res}_{s=\lambda_n}(\tilde{G})=\left.\frac{-\eta_n \sqrt{D_1 s}\: v(s,x)v(s,x_0)}{\frac{\mathrm{d}F}{\mathrm{d}s}}\right|_{s=\lambda_n}\;.
\end{equation*}
By comparison with Eq. \eqref{eq:decomposition_G}, this allows us to conclude:
\begin{equation}
{\beta_n}^{-2}=-\frac{1}{\eta_n\sqrt{D_1\lambda_n}}\frac{\mathrm{d}F}{\mathrm{d}\lambda}(\lambda_n)\;.
\label{eq:normalisation1}
\end{equation}
%which implies that the derivative and $\eta_n$ have opposite signs.
%One can develop further the derivative. By using the relations \labelcref{eq:simplif}, one gets:
%\begin{equation}
%{\beta_n}^{-2}=\frac{1}{2s\sqrt{D_1}}\left[-\frac{\epsilon_n}{\det\mathcal{T}(s)}\left(\begin{bmatrix}K_+ & \sqrt{D_m s}\end{bmatrix}\frac{\mathrm{d}\mathcal{T}}{\mathrm{d}\sqrt{s}}\begin{bmatrix}\sqrt{D_1 s}\\K_-\end{bmatrix}\right)+\left(\frac{{\epsilon_n}^2}{\det\mathcal{T}(s)}\sqrt{D_m}K_++\sqrt{D_1}K_-\right)\right]_{s=\lambda_n}
%\label{eq:normalisation}
%\end{equation}
%
In general, one obtains $\eta_n$  by computing the matrix product in Eq. \eqref{eq:coherence_condition2}. A great simplification occurs in the case of symmetric geometries, which is the topic of the next section.

%%%%%%%%%%%%%%%%%%%%%%%%%%%%%%%%%

\subsection{Symmetry properties}
\label{section:symmetry}
For a geometry which is symmetric with respect to the middle of the interval $[0,L]$, some simplifications occur. In fact the symmetry of the geometry implies that the eigenmodes are either symmetric or anti-symmetric with respect to the middle of the interval, and as a consequence $\epsilon=\eta=+1$ or $\epsilon=\eta=-1$, respectively. 
%($+1$ corresponds to the symmetric eigenmodes and $-1$ to the anti-symmetric ones).
These statements can be easily proved with the above matrix formalism. In fact, the symmetry of the geometry is equivalent to the two properties:
\begin{enumerate}
\item The endpoints vectors $\mathcal{V}_+=\begin{bmatrix}1 \\ {- K_{+}}/{\sqrt{\lambda D_{1}}}\end{bmatrix}$ and $\mathcal{V}_-=\begin{bmatrix}1 \\ {K_{-}}/{\sqrt{\lambda D_{m}}}\end{bmatrix}$ have equal first components and opposite second components, which follows from the symmetry $K_-=K_+$, $D_1=D_m$. With the notation $\mathcal{S}=\begin{bmatrix}1&0\\0&-1\end{bmatrix}$, this can be restated as $\mathcal{V}_\pm=\mathcal{S}\mathcal{V}_\mp$.
\item The inverse of the transition matrix $\mathcal{T}$ is obtained by replacing the off-diagonal terms by their opposite in its expression (note that this corresponds to the transformation $\sqrt{\lambda}\to-\sqrt{\lambda}$). In fact, this property is clearly true for the \textquote{elementary blocks} $\mathcal{K}$ and $\mathcal{R}$ and thus it is also the case for $\mathcal{R}_m\mathcal{K}_{m-1,m}\mathcal{R}_{m-1}\dots\mathcal{K}_{1,2}\mathcal{R}_1$ because $\mathcal{R}_i=\mathcal{R}_{m+1-i}$ and $\mathcal{K}_{i,i+1}=\mathcal{K}_{m-i,m+1-i}$. In other words, $\mathcal{T}^{-1}=\mathcal{S}\mathcal{T}\mathcal{S}$.
\end{enumerate}

The consequence of these two properties is that Eq. \eqref{eq:coherence_condition} can be restated as: \textquote{$\mathcal{V}_-$ is an eigenvector of $\mathcal{S}\mathcal{T}$} and that this matrix is equal to its inverse:
\begin{equation*}
\left(\mathcal{S}\mathcal{T}\right)^{-1}=\mathcal{T}^{-1}\mathcal{S}^{-1}=\mathcal{S}\mathcal{T}\;.
\end{equation*}
This implies that the eigenvalues of this matrix, hence the proportionality coefficients $\epsilon,\eta$ in Eqs. \eqref{eq:coherence_condition} and \eqref{eq:coherence_condition2}, are equal to $\pm 1$.
%\begin{equation}
%\epsilon=\eta=\pm1\;.
%\end{equation}
%
We can also easily prove the symmetry or anti-symmetry of the eigenmodes. In fact, one has
\begin{align*}
\begin{bmatrix}{a^l_i} \\{b^l_i}\end{bmatrix}&=\mathcal{K}_{i-1,i}\mathcal{R}_{i-1}\ldots\mathcal{R}_1  \mathcal{V}_- \\
\begin{bmatrix}a^r_{m+1-i} \\b^r_{m+1-i}\end{bmatrix}&=\mathcal{K}_{m+1-i,m+2-k}^{-1}\mathcal{R}_{m+2-k}^{-1}\ldots\mathcal{R}_m^{-1}  \epsilon \mathcal{V}_+
\end{align*}
Hence
\begin{equation}
\begin{bmatrix}a^r_{m+1-i} \\b^r_{m+1-i}\end{bmatrix}=\mathcal{S}\mathcal{K}_{i-1,i}\mathcal{S}\mathcal{S}\mathcal{R}_{i-1}\mathcal{S}\ldots\mathcal{S}\mathcal{R}_1\mathcal{S}  \epsilon \mathcal{V}_+ = \epsilon\mathcal{S}\begin{bmatrix}{a^l_i} \\{b^l_i}\end{bmatrix}\;.
\label{eq:symetrie_coefficients}
\end{equation}
Let $x\in \Omega_i$, we write $x=x_{i-1,i}+\xi$, with $0<\xi<l_i$, which implies by symmetry that $L-x=x_{m+1-i,m+2-i}-\xi$. According to Eqs. \eqref{eq:formule_modes_gauche}, \eqref{eq:formule_modes_droite}, and \eqref{eq:symetrie_coefficients}, we have then
\begin{align*}
v(x)&=
\begin{bmatrix}{a^l_i} &{b^l_i}\end{bmatrix}
\begin{bmatrix}\cos(\xi\sqrt{\lambda/D_i})\\\sin(\xi\sqrt{\lambda/D_i})\end{bmatrix}
\\&=\epsilon
\begin{bmatrix}a^r_{m+1-i} &b^r_{m+1-i}\end{bmatrix}
\begin{bmatrix}\cos(-\xi\sqrt{\lambda/D_{m+1-i}})\\\sin(-\xi\sqrt{\lambda/D_{m+1-i}})\end{bmatrix}
=\epsilon v(L-x)\;,
\end{align*}
%which is equivalent to:
%\begin{equation}
%{u_i}(x_{i-1,i}+x)=\epsilon u_{m+1-i}(x_{m+1-i,m+2-k}-x)\;,
%\end{equation}
since $D_i=D_{m+1-i}$.
Therefore the eigenmode is symmetric if $\epsilon=+1$ and anti-symmetric if $\epsilon=-1$.
Moreover from Eq. \eqref{eq:normalisation1} we deduce that the derivative $\frac{\mathrm{d}F}{\mathrm{d}\lambda}(\lambda_n)$ and $\eta_n$ have opposite signs. Because the eigenvalues $\lambda_n$ are the zeros of $F$, the derivative alternates between positive and negative sign, and so do $\eta_n$ and $\epsilon_n$. In particular, in the case of a symmetric geometry, the modes $u_n$ are alternately symmetric and anti-symmetric. One can show that the first mode $u_1$ is always symmetric ($\epsilon_1=\eta_1=1$), hence
\begin{equation}
\epsilon_n=\eta_n=(-1)^{n-1}\;.
\end{equation}

%%%%%%%%%%%%%%%%%%%%%%%%%%%%%%%%%%%%%%%%%%%

\subsection{Periodicity properties}
\label{section:periodicity}
A finite periodic geometry is an $M$-times repetition of an elementary block composed of $N$ compartments: $(D_1;l_1),$ $(D_2;l_2),$ $\ldots ,$ $(D_N;l_N)$. The transition matrix of the block is
\begin{equation}
\mathcal{M}=\mathcal{K}_{inter} \mathcal{R}_N\mathcal{K}_{N-1,N}\dots\mathcal{R}_{1}\;,
\label{eq:matrice_bloc_periodique}
\end{equation}
where $\mathcal{K}_{inter}$ is the matrix corresponding to the inter-block barriers. Then the complete transition matrix $\mathcal{T}$ is equal to
\begin{equation}
\mathcal{T}=\mathcal{K}_{inter}^{-1}\mathcal{M}^M\;.
\label{eq:matrice_complete_periodique}
\end{equation}
Because of the periodicity,
\begin{equation*}
\det\mathcal{M}=\underbrace{\sqrt{\frac{D_N}{D_1}}}_{\det\mathcal{K}_{inter}}\sqrt{\frac{D_{N-1}}{D_N}}\dots \sqrt{\frac{D_1}{D_2}}=1\;.
\end{equation*}
This property makes the computation of $\mathcal{M}^M$ easier, thanks to the formula
\begin{equation}
\mathcal{M}^M=\frac{\sin M\psi}{\sin\psi}\mathcal{M}-\frac{\sin(M-1)\psi}{\sin\psi}\mathcal{I}_2\;,
\label{eq:M_puissance}
\end{equation}
where $\mathcal{I}_2$ is the $2\times 2$ identity matrix and $\psi$ is implicitly defined by
\begin{equation}\cos\psi=\frac{1}{2}\Tr\mathcal{M}\;.
\label{eq:definition_cospsi}
\end{equation}
Formula \eqref{eq:M_puissance} implies that the inter-block variation of the coefficients $a$, $b$ has the form:
\begin{equation}
a_{i_0+N(j-1)}=A\cos(j\psi)+B\sin(j\psi)\;, \qquad j=1,\ldots,M\;,
\label{eq:variation_inter_bloc}
\end{equation}
with a similar formula for $b$, where $A$ and $B$ are coefficients which depend on the choice of the origin $i_0 \in \{1,\dots,N-1\}$.
%
%
%where $j=1,\ldots,M$ is the index of the block.
Thus $\psi$ governs the global behavior of the mode (when the number $M$ of repeated blocks is sufficiently large).

 %Note that as $\mathcal{M}$ is a real matrix, $\psi$ is real or imaginary, and in the latter case one should have in mind the following formula:
%\begin{align}
%\cos(ix)&=\cosh(x),\\
%\frac{\sin Mix}{\sin ix}&=\frac{\sinh Mx}{\sinh x},
%\end{align}
%if $x$ is real.
%Note also that a periodic geometry is symmetric if and only if the endpoint boundary conditions are symmetric, that is if $K_+=K_-$.

%\vskip 5mm
%
%Let us summarize the computational steps: (i) to compute the transition matrix $\mathcal{M}_{i,i+1}$ in Eq. \eqref{eq:abreviations} for each compartment; (ii) to apply Eq. \eqref{eq:formule_generale_T} to get the complete transition matrix; in case of a periodic geometry, the computation is simplified by the use of Eq. \eqref{eq:M_puissance}; (iii) to solve Eq. \eqref{eq:coherence_condition3} to get the spectrum  of the diffusion operator; each solution of Eq. \eqref{eq:coherence_condition3} determines one eigenvalue whereas Eqs. \eqref{eq:conditions_limites_matrice}, \eqref{eq:membrane_equation_matrice}, and \eqref{eq:normalisation1} yield the coefficients $a^l_i$, $b^l_i$, $k=1,\ldots,m$ for each (non-normalized) mode and the corresponding normalization constant; combined with Eq. \eqref{eq:formule_modes_gauche} it allows one to compute the eigenmode at any point of the interval.

%%%%%%%%%%%%%%%%%%%%%%%%%%%%%%%%%%%%%%%%%%%%%

%%%%%%%%%%%%%%%%%%%%%%%%%%%%%%%%%
\subsection{Study of the spectrum}
\label{section:study_spectrum}

\begin{figure*}[ptb]
\begin{center}
\includegraphics[width=0.9\linewidth]{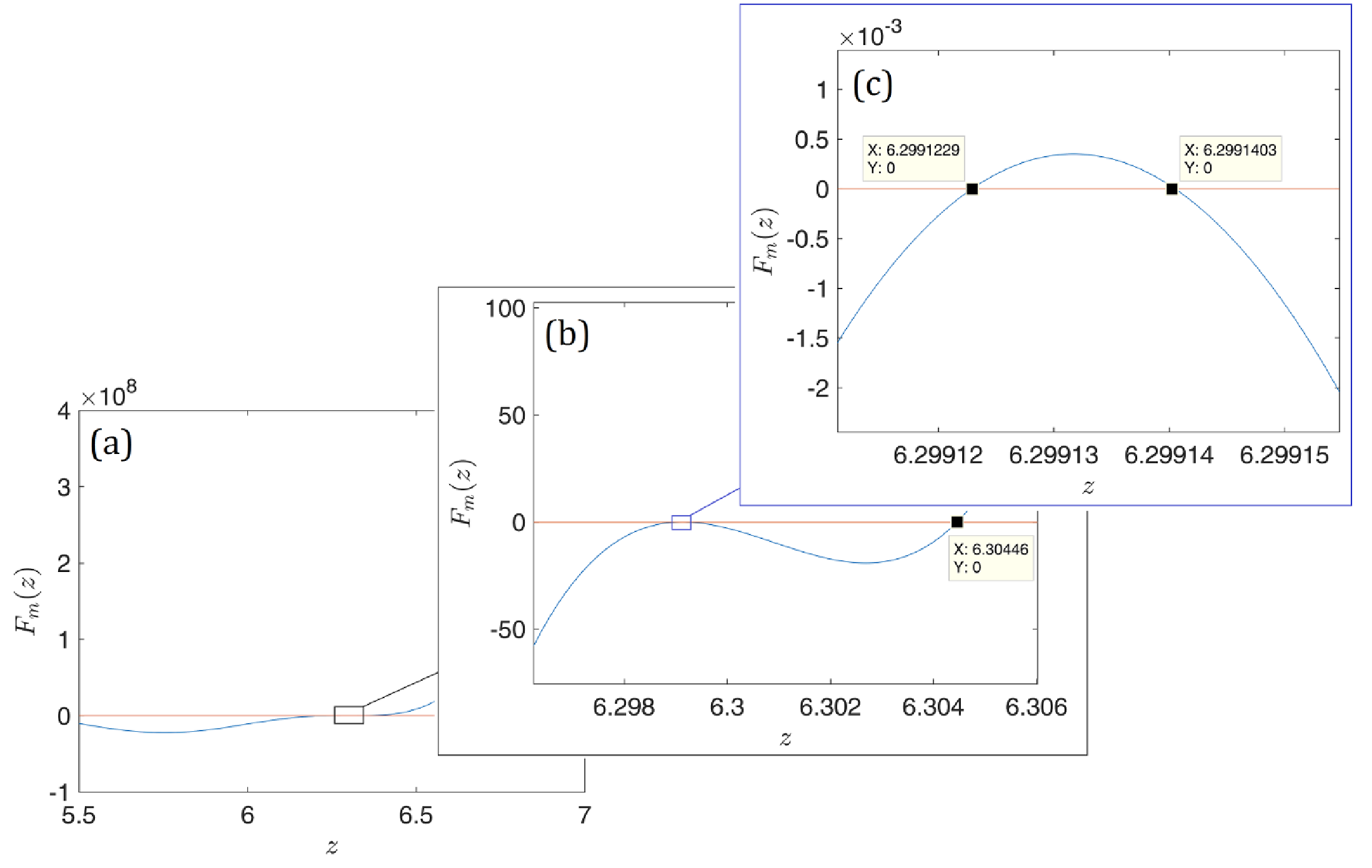}
\end{center}
\caption{Example of roots which may prove challenging to find numerically with standard methods. We consider five compartments and $D_1=\ldots=D_5=1$, $r_{1,2}=\ldots=r_{4,5}=10$ and the lengths $l_i$ of the five compartments are: $1; 1.2; 1.5; 1.2; 1$, with reflecting boundary conditions at the endpoints: $K_{\pm}=0$. The root $z=6.30446$ (b) corresponds to $l_3=1.5$, with $n=3$, $\zeta=2$, whereas the two roots $z_\pm=6.2991316\pm 8.7 \cdot 10^{-6}$ (c) correspond to $l_1=l_5=1$, with $n=2$, $\zeta=1$ (see explanations in the text). Notice the scale changes, horizontally and vertically, between (a), (b) and (c).
}
\label{fig:graphes-zoom}
\end{figure*}

The main numerical difficulty of the above method is to solve Eq. \eqref{eq:coherence_condition3} on the spectrum, that is to find the zeros of $F(\lambda)$. 
In fact, a standard method to find \emph{all} the zeros of a function in a given interval is to compute the function on a fine array $(0,\epsilon,2\epsilon,\ldots)$ and to look for the sign changes, that indicate the presence of at least one zero. By decreasing $\epsilon$, one is assured at some point to find all the zeros of the function. However, in general one knows neither the number of zeros of the function in a given interval nor the minimal spacing between the zeros. In turn, missing some zeros would result in missed eigenmodes, and thus in inaccurate computation of the propagator and the related diffusion quantities. An example of $F(\lambda)$ shown in Fig. \ref{fig:graphes-zoom} illustrates that some roots may be very close to each other.
We provide here a rough analysis of Eq. \eqref{eq:coherence_condition3} in order to study this phenomenon.
%
%%%%%%%%%%%%%%%%%%

%%%%%%%%%%%%%%%%%

We discard the elementary case of a single interval ($m=1$) where the roots of $F$ are explicitly known \cite{Carslaw59,Crank75}. Let us assume for simplicity that all the diffusion coefficients $D_i$ and the barrier resistances $r_{i,i+1}$ are identical (denoted $D$ and $r$, respectively). Furthermore we set the relaxation coefficients $K_\pm$ to zero. We change the variable $\lambda$ by $z=\sqrt{\lambda/D}$ and reveal an explicit dependence of $F$ on the geometry (omitting $D$ and $r$ for the sake of clarity):
\begin{equation*}
F(\lambda)=F_m(z;l_1,\ldots,l_m)\;.
\end{equation*}
\subsubsection{Regime $r\to 0$}
First we consider the regime of quasi-permeable barriers, that is $r\to 0$.
One has
\begin{equation*}
\mathcal{K}=\mathcal{I}_2+rDz\mathcal{N}\;, \quad \mathcal{N}=\begin{bmatrix}0&1\\0&0\end{bmatrix}\;,
\end{equation*}
from which we deduce the first-order expansion
\begin{equation*}
F_m(z; l_1,\ldots,l_m)\approx \!-\!\sin(zL)+rDz\sum_{i=1}^{m}\sin(z(l_1\!+\!\dots\!+\!l_i))\sin(z(l_{i+1}\!+\!\dots\!+\!l_m))\;.
\end{equation*}
This formula implies that the roots are approximately equal to $z_0={n\pi}/{L}$, with an integer $n$. In fact, one can compute the first order correction to this formula, which yields
\begin{equation}
z\approx\frac{n\pi}{L} \left(1-\frac{rD}{l}\left[\frac{1}{m}\sum_{i=1}^m \sin^2\left(n\pi\frac{l_1+\dots+l_i}{L}\right)\right]\right),
\end{equation}
where $l=L/m$ is the arithmetic mean of the $l_i$. The factor inside the brackets is always less than $1$, hence the (first order) relative perturbation of the roots is at most $rD/l$. 
%For identical compartments ($l_1=\dots=l_m=l$) the bracket is equal to $1/2$ if $n$ is not a multiple of $m$ and $0$ otherwise.
%
Therefore in the regime of quasi-permeable inner barriers ($rD/l \ll 1$) the roots are easy to find numerically because we have a good estimate of their position and a good lower bound of the distance between them. 

\subsubsection{Regime $r\to \infty$}
Now we turn to the opposite regime of almost impermeable barriers: $r\to\infty$. In this case one writes
\begin{equation*}
\mathcal{K}=rDz\left(\mathcal{N}+\frac{1}{rDz} \mathcal{I}_2\right)\;.
\end{equation*}
For $z$ large enough such that $rDz \gg 1$, this yields
\begin{align}
F_m(z; l_1,\ldots,l_m)&\approx (-rDz)^{m-1} \sin(zl_1)\cdots\sin(zl_m)
\left(1-\frac{1}{rDz}\sum_{i=1}^{m-1} \frac{\sin(z(l_i+l_{i+1}))}{\sin(zl_i)\sin(zl_{i+1})}+\dots\right)\;.
\label{eq:formule_generale_approchee}
\end{align}
%+\frac{1}{(rDz)^2}\sum_{i=2}^{m-1}\frac{\sin(z(l_{i-1}+l_i+l_{i+1}))}{\sin(zl_{i-1})\sin(zl_i)\sin(zl_{i+1})}+\frac{1}{{rDz)^2}\sum_{i,k'>k+1}\frac{\sin(z(l_i+l_{i+1}))\sin(z(l_{i'}+l_{i'+1}))}{\sin(zl_i)\sin(zl_{i+1})\sin(zl_{i'})\sin(zl_{i'+1})}
From this expression one gets the approximate roots $z_0={n\pi}/{l_i}$ with an integer $n$, as expected. The non-zero permeability of the barriers increases the values of the roots by coupling the compartments to their nearest neighbors. The higher-order terms of the expansion \eqref{eq:formule_generale_approchee} involve coupling between next-nearest neighbors, etc. From the above formula we expect the increase to be of order ${(rDz_0)}^{-1}$. The case $n=0$ (that is, $z_0=0$) is special and we treat it later.
Note that the above expansion is valid around $z_0={n\pi}/{l_i}$ (with $n>0$) if $rDz_0 \gg 1$, that is $rD/l_i \gg 1$.

If we consider an \emph{isolated} root  $z_0={n\pi}/{l_i}$ (which means that all the other ${n'\pi}/{l_{i'}}$ are located at a relative distance much greater that ${(rDz_0)}^{-1}$), then we get
%\begin{equation*}
%\begin{cases}
%z\approx \frac{n\pi}{l_i}+\frac{2}{n\pi rD} & \text{if $1 < k < m$;} \\
%z\approx \frac{n\pi}{l_i}+\frac{1}{n\pi rD} & \text{if $k=1$ or $k=m$.}
%\end{cases}
%\end{equation*}
\begin{equation}
z\approx \frac{n\pi}{l_i} + \frac{\zeta_i}{n\pi r D},
\label{eq:formule_approchee_racine_isolee}
\end{equation}
where $\zeta_i$ is the number of neighbors of the cell $i$ ($\zeta_i=2$ if $1<i<m$, otherwise $\zeta_i=1$).

The case of \emph{non-isolated} roots is more complicated but also more interesting. In fact all the numerical difficulties come from this case. From the equation
\begin{equation*}
\mathcal{R}_i \begin{bmatrix}1\\0\end{bmatrix}\begin{bmatrix}0&1\end{bmatrix}\mathcal{R}_i - \begin{bmatrix}1\\0\end{bmatrix}\begin{bmatrix}0&1\end{bmatrix}=-\sin(zl_i) \mathcal{R}_i\;,
\end{equation*}
we deduce the following general relation which is valid for any $i$ from $1$ to $m-1$:
\begin{equation}
F_m(z;l_1,\ldots,l_m)=\frac{\left[{F_i(z;l_1,\ldots,l_i)F_{m+1-i}(z;l_i,\ldots,l_m)}{-F_{i-1}(z;l_1,\ldots,l_{i-1})F_{m-i}(z;l_{i+1},\ldots,l_m)}\right]}{F_1(z;l_i)}.
\label{eq:formule_decoupage}
\end{equation}
% Note that this relation holds only for reflecting conditions at the outer boundaries ($K_\pm=0$) but may be easily generalized to arbitrary boundary conditions.
Now we assume that there exist $i_1 < i_2$ such that
\begin{equation*}
z_0=\frac{n_1\pi}{l_{i_1}}=\frac{n_2\pi}{l_{i_2}}\;,
\end{equation*}
with $n_1$, $n_2$ integers. Note that $n_1/n_2=l_{i_1}/l_{i_2}$. We look for an approximate root of the form $z=z_0(1+\eta)$, with $\eta\sim (rDz_0)^{-1}$ (where $\sim$ means \textquote{is of the same order of magnitude as}).

First let us consider the case where two compartments $i_1$ and $i_2$ are not neighbors, that is $i_1 +1 <i_2$. From Eq. \eqref{eq:formule_generale_approchee} we infer
\begin{align*}
%&F_m(z;l_1,\ldots,l_m)\sim (rDz)^{m-3}\;;\\
&F_{i_1+1}(z;l_1,\ldots,l_{i_1+1})\sim (rDz)^{i_1}\eta \sim (rDz)^{i_1-1}\;,\\
&F_{m-i_1}(z;l_{i_1+1},\ldots,l_m)\sim (rDz)^{m-1-i_1}\eta \sim (rDz)^{m-2-i_1}\;,\\
&F_{i_1}(z;l_1,\ldots,l_{i_1})\sim (rDz)^{i_1-1}\eta \sim (rDz)^{i_1-2}\;,\\
&F_{m-1-i_1}(z;l_{i_1+2},\ldots,l_m)\sim (rDz)^{m-2-i_1}\eta \sim (rDz)^{m-3-i_1}\;,
\end{align*}
hence Eq. \eqref{eq:formule_decoupage} becomes
\begin{equation*}
F_m(z;l_1,\ldots,l_m)= \frac{F_{i_1}(z;l_1,\ldots,l_{i_1})F_{m+1-{i_1}}(z;l_{i_1},\ldots,l_m)}{F_1(z;l_{i_1})}\left(1+O((rDz)^{-2})\right)\;.
\end{equation*}
We deduce that the roots of $F_m(z;l_1,\ldots,l_m)$ are given by the roots of the functions $F_{i_1}(z;l_1,\ldots,l_{i_1})$ and $F_{m+1-{i_1}}(z;l_{i_1},\ldots,l_m)$, which are not coupled to the first order in ${(rDz)}^{-1}$:
%This implies that the previous analysis \eqref{eq:formule_approchee_racine_isolee} of an isolated root is still valid to the first order in $\frac{1}{rDz}$, thus we will have two roots:
\begin{equation}
z\approx z_0+\frac{\zeta_{i_1}}{n_1\pi rD} \quad \text{ and } \quad z\approx z_0+\frac{\zeta_{{i_2}}}{n_2\pi rD}\;.
\end{equation}
Note that the same is true for any number of \textquote{coinciding} roots as long as they correspond to non-adjacent compartments. The roots are at a relative distance of order $(rDz_0)^{-1}$ if $n_1/\zeta_{i_1}\neq n_2/\zeta_{i_2}$. If $n_1/\zeta_{i_1}= n_2/\zeta_{i_2}$ one has to compute the next-order corrections which involve the length of the other compartments, as explained previously. One can show that the term of order $(rDz_0)^{i_1-i_2}$ is always non-zero; for symmetric geometries $(rDz_0)^{i_1-i_2}$ may be the first non-zero term of the expansion \textcolor{black}{of the relative difference of the roots}.

Now we consider the case $i_2=i_1+1$. We use Eq. \eqref{eq:formule_generale_approchee} to get
\begin{align*}
F_m(z;l_1,\ldots,l_m)&\approx(-rDz)^{m-3}\left(\prod_{i\neq i_1,i_1+1}\sin(zl_i)\right)\left(n_1 n_2 X^2-(\zeta_{i_1} n_1+\zeta_{i_2} n_2)X+(\zeta_{i_1}\zeta_{i_2}-1) \right)\;,
\end{align*}
where $X=rD\pi\eta$. Thus we obtain two roots:
\begin{equation}
z_\pm = z_0+\frac{X_\pm}{rD\pi}\;, \quad \text{with} \quad X_\pm=\frac{\zeta_{i_1} n_1+\zeta_{i_2}n_2 \pm\sqrt{(\zeta_{i_1} n_1-\zeta_{i_2}n_2)^2+4n_1n_2}}{2n_1n_2}\;.
\end{equation}
Note that $z_+-z_-\geq\frac{2}{\pi\sqrt{n_1n_2}rD}$.
One can perform the same computations for a larger number of adjacent cells with \textquote{coinciding} roots: at the end one has to solve a polynomial equation in the variable $X$. The roots are always distinct and separated by a relative distance of order $(rDz_0)^{-1}$. Section \ref{section:modes_simple_periodic} is devoted to the exact computation of the roots for an array of identical cells, which is a good example of such a situation.

In all the above computations we assumed $z_0 =n\pi/l_i$ with positive $n$. However there are also $m$ roots located near zero. To find them we expand the sine and cosine functions in Eq. \eqref{eq:formule_generale_approchee} and get to the first order in $zl$ a polynomial equation of degree $m$ in the variable $Z=rDlz^2$, where $l$ is the harmonic mean of the $l_i$. Hence we obtain $m$ roots of the form:
\begin{equation}
z_n=\sqrt{\frac{Z_n}{rDl}}\;, \quad n=1,\ldots,m
\label{eq:formule_approchee_petit_z}
\end{equation}
with $Z_n$ spanning the solutions of the polynomial equation. Note that we assumed $rD/l_i \gg 1$ hence one has $zl \ll 1$, which legitimates \textit{a posteriori} the polynomial expansion. Furthermore, the first coefficients of the polynomial expansion are readily available from Eq. \eqref{eq:formule_generale_approchee} and we get from them that:
\begin{equation}
\sum_{n=1}^m Z_n \approx 2m\;.
\end{equation}
This formula is valid in the regime $rD/l\gg 1$ and its simplicity comes from the particular choice of $l$ we made (harmonic mean of the $l_i$). If one assumes that the roots $Z_n$ are approximately equispaced at small $n$, then one obtains immediately that the first roots $Z_n$, and hence $\lambda_n$, follow a $1/m^2$ dependence on $m$.

From this analysis of the low permeability regime ($rD/l_i \gg 1$ for all $i$) we can draw several conclusions, partly illustrated in Fig. \ref{fig:graphes-zoom}.
\begin{itemize}
\item the $m$ first roots ($zl \ll 1$) behave differently than the other ones. They typically spread over a distance $(rDl)^{-1/2}$.
\end{itemize}
The following points only apply to the other roots ($zl \gtrsim 1$).
\begin{itemize}
\item all the roots increase from the limits $z_0={n\pi}/{l_i}$ with the permeability of the inner barriers (a general mathematical proof of this statement is given in Sec. \ref{section:monotonicity}). The relative increase is of the first order in $(rDz_0)^{-1}$;
\item very close roots associated to adjacent cells are coupled by the permeability of their barrier and separate from each other by a relative distance of order $(rDz_0)^{-1}$;
\item very close roots associated to non-adjacent cells are not coupled to the first order in $(rDz)^{-1}$. The difficult case is when the two cells have the same length: then $n_1=n_2$ and the relative distance between the two roots is in the best case of order $(z_0 rD)^{-2}$. In fact, it depends on the length of all other cells. For example, symmetric geometries typically lead to a relative distance between roots of order $(z_0 rD)^{-|i_2-i_1|}$.
\end{itemize}
All the previous computations are somewhat schematic because we made a particular choice of geometry (same diffusion coefficients, same permeability and no relaxation at the outer boundaries) from the beginning. However, the above conclusions are globally still valid in the general case, with appropriate modifications. For example if one considers perfectly relaxing condition at the endpoints ($K_\pm=\infty$), then in the low-permeability limit the roots corresponding to the outer compartments are $z_0={(n+1/2)\pi}/{l_i}$ ($i=1$ or $m$), whereas the roots corresponding to the other compartments are $z_0={n\pi}/{l_i}$, $1<i<m$ (with an integer $n$). Thus one has to consider separately the case of the outer compartments depending on the conditions at the outer boundaries. We come back to the relaxing case in Sec. \ref{section:application_temps_sortie} and Sec. \ref{section:calculs_relax}. Moreover, the case of heterogeneous diffusion coefficients is treated analytically in the simplest case of a bi-periodic structure in Sec. \ref{section:bi-periodique}.

%%%%%%%%%%%%%%%%%%%%%%%%%%%%%%%%%%%%%%%%%%%%%%%%%%%%

%%%%%%%%%%%%%%%%%%%%%%%%%%%%%%%%%%%%%%%%%%%%%%%%%%%

\subsection{Extensions}
\label{section:problemes_generaux}

The above analysis may be extended in many ways. First, one can consider more general boundary conditions. In particular, many experiments in heat conduction are done with one end of the system in contact with a heat source (acting as a constant heat flux or as a thermostat with a constant temperature). One should then replace our homogeneous outer boundary conditions \eqref{eq:equation_bords_1}, \eqref{eq:equation_bords_2} by inhomogeneous boundary conditions. The only difference is in the steady-state solution ($\lambda=0$) which is easy to obtain, whereas the transient solution remains the same (see \cite{Ozisik12,Hickson09-1}). One is then often interested in the \textquote{critical time}, i.e. the typical time required to reach the steady-state solution. More precisely, one definition of the critical time is the time at which the average temperature over the sample is equal to some fraction $\alpha <1$ of the average steady-state temperature over the sample. Other definitions and a thorough comparison of these definitions are detailed in \cite{Hickson11b,Hickson11c}. This time is essentially given by the study of the first non-zero eigenvalue of the diffusion operator, for which we are able to obtain  estimates with respect to the geometrical parameters of the medium (such as Eq. \eqref{eq:formule_approchee_petit_z}, which yields $\lambda\sim (rlm^2)^{-1}$, in the low-permeability regime).
 The situation is different when the boundaries are subject to modulated heating, which is the case in geophysics and building design \cite{Grossel98,Lu05,Lu06,DeMonte00,Barbaro88}, and in photothermal measurements \cite{Gurevich09,Aguirre00}. One can still transform the problem into an homogeneous boundary problem but it requires adding a suitable source term to the diffusion equation \cite{Ozisik12}. In some cases the main mechanism of heat relaxation at the outer boundaries is not conduction-convection but radiation, with a non-linear $T^4$ heat flux \cite{Miller03}. Finally, when considering diffusion of ions in multilayer chemical system such as electrodes, one writes chemical equilibrium condition at the interfaces: the ratio of concentrations on both sides of the interface is equal to the partition coefficient \cite{Diard05,Freger05,Ngameni14,Fukuda94,Fukuda95}. This is another type of inner boundary condition, which leads to different $\mathcal{K}$ matrices, quite similar to the case of heterogeneous diffusion coefficients and no barriers.

Another possible generalization is the inclusion of bulk reaction rates inside the compartments. That is, to change Eq. \eqref{eq:Laplace_equation} to a reaction-diffusion equation:
\begin{equation}
\frac{\partial G}{\partial t}=D\nabla^2 G + \mu G\;,
\label{eq:reaction_diffusion}
\end{equation}
where $\mu$ may depend on space and $G$ \cite{Hickson11b}. If $\mu$ is constant, then one gets the solution of Eq. \eqref{eq:reaction_diffusion} by multiplying the solution of Eq. \eqref{eq:Laplace_equation} by $\exp(\mu t)$. The case of piecewise constant $\mu$ ($\mu=\mu_i$ on $\Omega_i$) is slightly more complicated but may be easily incorporated into our computations. Such reaction-diffusion models may describe diffusion of molecules that can be trapped, killed, destroyed, or loose their activity \cite{Grebenkov17b,Meerson15,Yuste13,Biess11,Carranza10} or, on the opposite, self-heating by temperature-induced oxidation \cite{Gray90} ($\mu > 0$). Other applications include ecology dynamics \cite{Okubo01} and fabrication of multilayer foil materials \cite{Mann97,Gachon05}. 

Last, one can consider other equations than the diffusion equation \eqref{eq:Laplace_equation}, for example:
\begin{itemize}
\item inhomogeneous Laplace (Poisson) equation: $\nabla(D\nabla \Psi) = F$,
\item inhomogeneous Helmholtz ($s>0$) or modified Helmholtz ($s < 0$) equations: $(s + \nabla D\nabla)\Psi=F$,
\item inhomogeneous diffusion equation: $\frac{\partial \Psi}{\partial t} - \nabla(D\nabla \Psi)=F$, $\Psi(x,t=0)=U(x)$,
\item inhomogeneous wave equation: $\frac{\partial^2 \Psi}{\partial t^2}- \nabla(D\nabla \Psi)=F$, $\Psi(x,t=0)=U(x)$, $\frac{\partial \Psi}{\partial t}(x,t=0)=V(x)$,
\end{itemize}
where $F,U,V$ are given functions, and with the boundary conditions
\eqref{eq:equation_membranes_1}, \eqref{eq:equation_membranes_2}, \eqref{eq:equation_bords_1}, and \eqref{eq:equation_bords_2}.
Thanks to the knowledge of the eigenmodes basis of the diffusion operator $\nabla(D \nabla)$, the above equations may be solved by decomposing $u$ and $F$ over this basis \cite{Carslaw59,Crank75}.

The computational method that we presented is therefore relevant to many models and applications. In the Supplementary Material we discuss two particular examples: diffusion MRI (Sec. \ref{section:application_dmri}) and first exit time distribution (Sec. \ref{section:application_temps_sortie}).

%%%%%%%%%%%%%%%%%%%%%%%%%%%%%%%%%%%%%%%%%%%%%

\section{Example: simple periodic geometry}
\label{section:example_periodic}

In this section, we illustrate the application of our general method to the case of a (finite) periodic structure which is relevant for various applications. Throughout this section, we assume that all $l_i$, $D_i$, $\kappa_{i,i+1}$ are the same (denoted $l$, $D$, $\kappa$ in the following). We apply the results of Sec. \ref{section:calcul_general} and obtain the eigenmodes and eigenvalues $u_n$, $\lambda_n$. Similar computations for more complicated structures are presented in Sec. \ref{section:bi-periodique} (bi-periodic geometry) and Sec. \ref{section:autres_structures_imbrique} (two-scale geometry).

\subsection{Eigenmodes}
\label{section:modes_simple_periodic}

We assume reflecting boundary conditions at the endpoints ($K_\pm=0$) and introduce the dimensionless parameters
\begin{equation}
\alpha=\sqrt{\lambda/D} l\; \quad \text{ and } \quad \tilde{r}=1/\tilde{\kappa}=rD/l\;.
\label{eq:notations_simple}
\end{equation}
%
%Moreover the relaxation coefficients $\tilde{K_\pm}$ are also identical (noted $\tilde{K}$ in the following).
Then the transition matrix of the elementary block is simply
\begin{equation}
\mathcal{M}=\mathcal{K}\mathcal{R}=\begin{bmatrix}\cos\alpha-\tilde{r}\alpha\sin\alpha&\sin\alpha+\tilde{r}\alpha\cos\alpha\\
-\sin\alpha&\cos\alpha\end{bmatrix}\;,
\label{eq:M_simple}
\end{equation} and  Eq. \eqref{eq:coherence_condition} on the spectrum becomes
%\begin{equation}
%\label{eq:coherence_periodic}
%\mathcal{K}^{-1}
%\mathcal{M}^{m}
%\begin{bmatrix}
%{\alpha_{n}}\\\tilde{K}
%\end{bmatrix}
%=\epsilon_n
%\begin{bmatrix}
%{\alpha_{n}} \\ -\tilde{K}
%\end{bmatrix}\;.
%\end{equation}
\begin{equation}
\mathcal{K}^{-1}
\mathcal{M}^{m}
\begin{bmatrix}
\alpha\\ 0
\end{bmatrix}
=\epsilon
\begin{bmatrix}
\alpha \\ 0
\end{bmatrix}\;.
\label{eq:coherence_periodic}
\end{equation}
Since the geometry is  symmetric, we already know that $\epsilon=\pm 1$.
Furthermore we use the results of Sec. \ref{section:periodicity} to compute $\mathcal{M}^m$: first we apply Eq. \eqref{eq:definition_cospsi} to define $\psi$:
\begin{equation}
\cos\psi=\cos\alpha-\frac{\tilde{r}}{2}\alpha\sin\alpha\;,
\label{eq:equation_alpha_psi}
\end{equation}
then from Eq. \eqref{eq:M_puissance}, we get
\begin{equation}
\mathcal{M}^m=\begin{bmatrix}\left(\cos\alpha-\tilde{r}\alpha\sin\alpha\right)\frac{\sin m\psi}{\sin\psi}-\frac{\sin (m-1)\psi}{\sin\psi}&\left(\sin\alpha+\tilde{r}\alpha\cos\alpha\right)\frac{\sin m\psi}{\sin\psi}\\
-\sin\alpha\frac{\sin m\psi}{\sin\psi}&\cos\alpha\frac{\sin m\psi}{\sin\psi}-\frac{\sin (m-1)\psi}{\sin\psi}\end{bmatrix}\;.
\label{eq:M_puissance_simple}
\end{equation}
%
%In order to compute $\mathcal{M}^m$, we perform the diagonalization of the matrix $\mathcal{M}$, which yields:
%\begin{equation}
%\frac{i}{2\sin\alpha_n\sin\psi_n}
%\begin{bmatrix}\cos\alpha_n-e^{i\psi_n}&\cos\alpha_n-e^{-i\psi_n}\\\sin\alpha_n&\sin\alpha_n\end{bmatrix}\begin{bmatrix}e^{i\psi_n}&0\\0&e^{-i\psi_n}\end{bmatrix}\begin{bmatrix}\sin\alpha_n&e^{-i\psi_n}-\cos\alpha_n\\-\sin\alpha_n&\cos\alpha_n-e^{i\psi_n}
%\end{bmatrix}\
%\end{equation}
%where the eigenvalues are determined by the equation:
%\begin{equation}
%\label{eq:equation_alpha_psi}
%\cos\psi_n = \cos\alpha_n - \frac{\tilde{r}}{2}\alpha_n\sin\alpha_n\;.
%\end{equation}
%
%The matrix $\mathcal{M}$ is not diagonalizable if $\cos\psi_n=\pm1$.
%
Equation \eqref{eq:coherence_periodic} can be further simplified by using the fact that $\mathcal{K}\begin{bmatrix}1\\0\end{bmatrix}=\begin{bmatrix}1\\0\end{bmatrix}$. We thus have the simple condition
\begin{equation}
\mathcal{M}^m
\begin{bmatrix}
1\\0
\end{bmatrix}
=\epsilon
\begin{bmatrix}
1\\0
\end{bmatrix}\;,
\label{eq:coherence_periodic_simple}
\end{equation}
which gives the equation on $\alpha$ (and thus on eigenvalues $\lambda$)
\begin{equation}
\sin\alpha\frac{\sin m\psi}{\sin\psi}=0\;.
\label{eq:coherence_periodic_explicite}
\end{equation}
 This corresponds to two cases:
\begin{itemize}
\item  $\sin\alpha=0$, that is $\alpha = j\pi$, with $j=0,1,2,\ldots$. We denote these solutions by $\alpha_{j,0}$ if $j$ is even and $\alpha_{j,m}$ if $j$ is odd. The vector $\begin{bmatrix}1\\0\end{bmatrix}$ is an eigenvector of the matrix $\mathcal{M}$ with the eigenvalue $(-1)^j$, thus $\epsilon=(-1)^{jm}$.
%This is exactly the solution for one isolated interval of length $l$ and it corresponds to the modes where the compartments do not communicate with each other.
%In this case we have $\epsilon=(-1)^{jm}$.
\item $\frac{\sin m\psi}{\sin\psi}=0$, which gives $m\psi = p\pi$, where $p \in \{1,\ldots,m-1\}$, and can be restated according to Eq. \eqref{eq:equation_alpha_psi} as:
\begin{equation}
\cos\alpha - \frac{\tilde{r}}{2}\alpha\sin\alpha = \cos p\pi/m\;, \quad p\in \{1,\ldots,m-1\}\;.
\label{eq:spectre_periodic}
\end{equation}
For each value of $p$ this yields an infinite array of solutions that we will denote as $\alpha_{j,p}$, where the $j$ index means $j\pi \leq \alpha_{j,p} < (j+1)\pi$ ($j=0,1,\ldots$). We have $\mathcal{M}^m=(-1)^p\mathcal{I}_2$, therefore $\epsilon=(-1)^p$.
\end{itemize}

\begin{figure*}[tb]
\begin{center}
\includegraphics[width=0.9\linewidth]{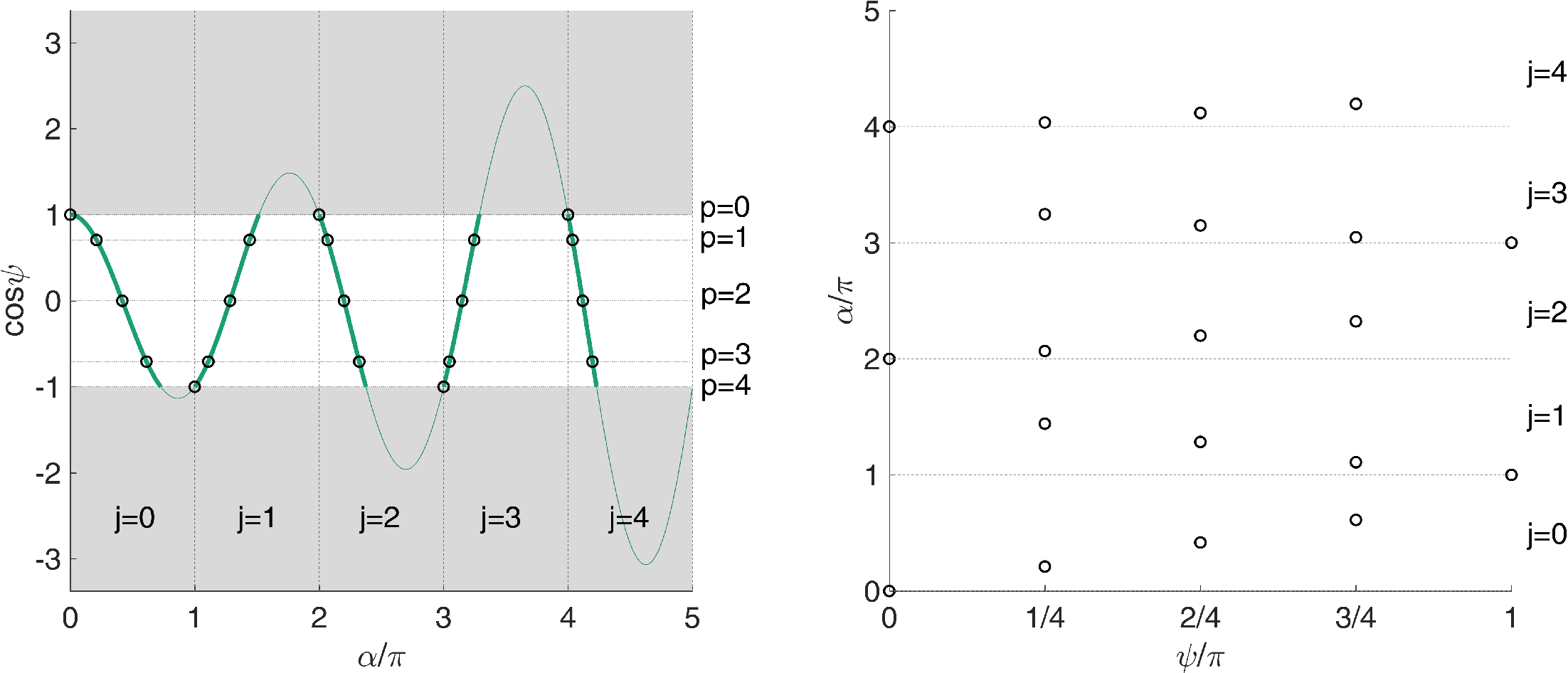}
\end{center}
%\caption{Solutions $\alpha_{j,p}$ plotted versus $\psi_p=p\pi/m$, with $m=10$ and $\tilde{r}=0.4$ (left) and the corresponding graph of $\cos\psi=\cos\alpha - \frac{\tilde{r}}{2}\alpha\sin\alpha$ (right). One can see branches beginning at $j\pi$ and ending below $(j+1)\pi$. As $\alpha$ increases, the graph of $\cos\psi$ crosses the $[-1;1]$ interval with a steeper slope, which results in solutions closer to $j\pi$ as $j$ increases.
\caption{(left) Plot of $\cos\psi=\cos\alpha-\frac{\tilde{r}}{2}\alpha\sin\alpha$ with $\tilde{r}=0.4$. Horizontal dotted lines indicate $\cos\psi=\cos p\pi/m$, $p=0,\ldots,m$, with $m=4$ and the circles represent the solutions $\alpha_{j,p}$. (right) An equivalent representation is the plot of $\alpha_{j,p}$ versus $\psi_p = p\pi/m$. One can see branches beginning at $j\pi$ and ending below $(j+1)\pi$. As $\alpha$ increases, the graph of $\cos\psi$ crosses the $[-1;1]$ interval with a steeper slope, which results in solutions closer to $j\pi$ as $j$ increases.
}
\label{fig:spectre_ref02}
\end{figure*}

Figure \ref{fig:spectre_ref02} illustrates the solutions $\alpha_{j,p}$ in the case $m=4$ and $\tilde{r}=0.4$. One can see that the solutions are grouped in branches of $m$ values. Each branch begins at a multiple of $\pi$ and ends below the next one. The branches of even $j$ begin with $\psi=0$ ($p=0$) and increase with increasing $p$, whereas the odd $j$ branches begin with $\psi=\pi$ ($p=m$) and increase with decreasing $p$. Note that we discard the branches with negative $j$ because $\alpha \geq 0$ according to Eq. \eqref{eq:notations_simple}.

\begin{figure*}[tbp]
\begin{center}
\includegraphics[width=0.9\linewidth]{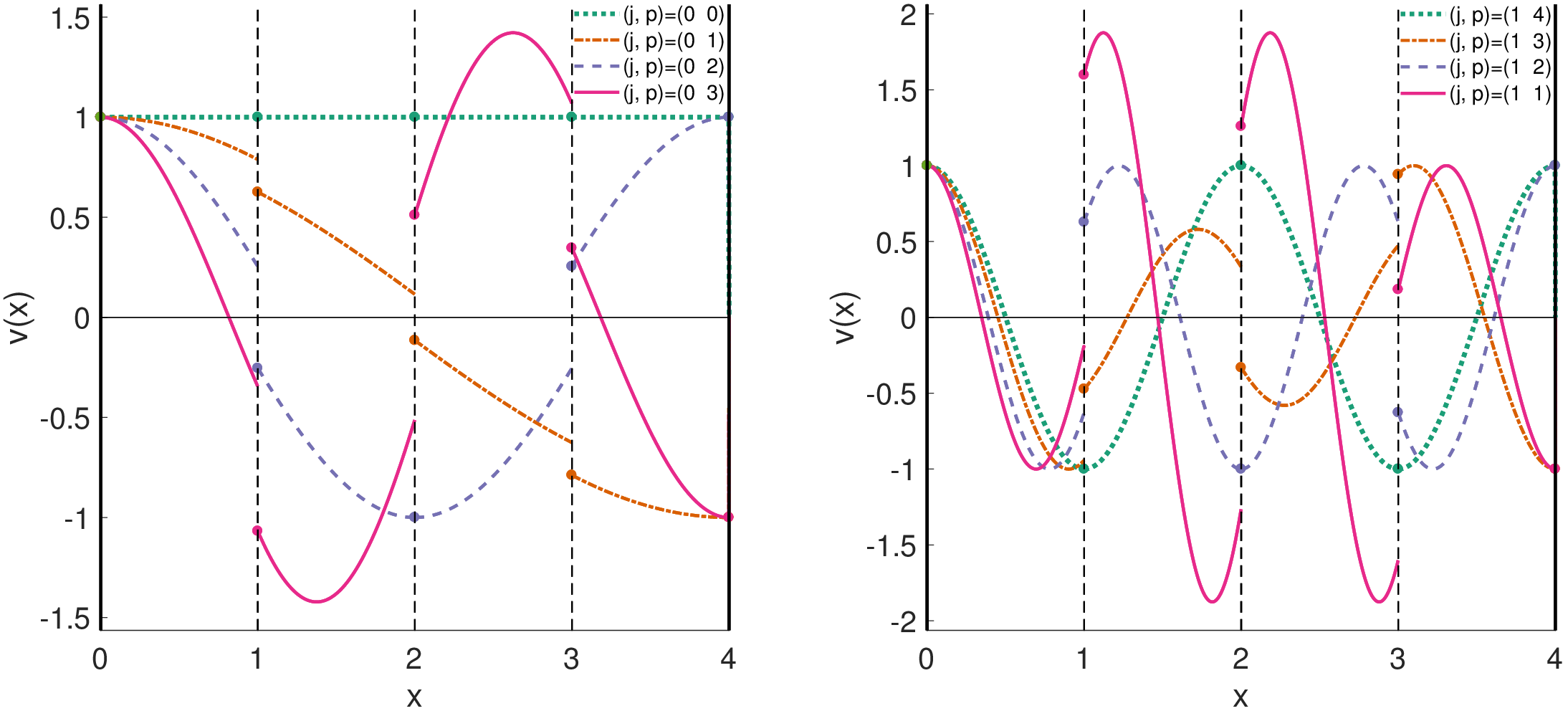}
\end{center}
\caption{Plot of the diffusion operator eigenmodes for the array of $m=4$ identical cells of length $1$ with impermeable outer boundaries and $\tilde{r}=0.4$. (left) $j=0$, $p=0,\ldots,m-1$; (right) $j=1$, $p=m,\ldots,1$. Note the discontinuities at the barriers which increase when $\alpha_{j,p}$ increases.
}

\label{fig:modes_j}
\end{figure*}

Note that $\alpha$ (or $j$) dictates the intra-compartment variation of the mode, whereas $\psi$ (or $p$) is related to its inter-compartment variation (as we explained in Sec. \ref{section:periodicity}). In fact, the index $j$ is equal to the number of extrema of the mode in the first compartment (not counting the one at $x=0$). If one is interested in the inter-compartment variation only, for example by looking at the value of the mode at the beginning of each compartment, then $p$ represents the number of extrema of this variation over the whole interval.
\textcolor{black}{Moreover, the Courant nodal theorem (proved for our particular model in Sec. \ref{section:courant_nodal_theorem}) states that each eigenmode changes sign $p+jm$ times.}
%(one should count the left endpoint in case it is an extremum but not the right endpoint).
Figure \ref{fig:modes_j} shows the first modes of an array of $m=4$ identical cells with impermeable outer barriers.
The first two branches are represented. We have additionally plotted dots at the beginning of each compartment to make the inter-compartment variation more visible.

One can compare the results of this section with Bloch waves in solid state physics. Indeed the branches of solutions $\alpha_{j,p}$ are similar to energy bands, where $j$ and $p$ are analogous to the band index $n$ and the wavenumber $k$, respectively. This is no surprise because we are dealing with a (finite) periodic geometry. Although the periodicity is not expressed through an energy potential but boundary conditions, the mathematical framework is the same. This explains the striking similarity between Fig. \ref{fig:spectre_ref02} and energy band diagrams (where only the $k\geq 0$ half would be represented).

%This is of course easier to understand when the number of cells is large. On Fig. \ref{fig:modes_m100} we show a plot of some modes for $m=100$. We have chosen $p=7$ and $j=0,1,2$. One can see that although $\alpha$ is very different between the different modes, the overall behaviour of the modes is the same: the dots form a sine function with $p=7$ extrema.

%\begin{figure*}[tbp]
%\begin{center}
%\includegraphics[width=120mm]{Modes_m100_p7_j_bis.eps}
%
%\end{center}
%\caption{Plot of the modes for the simple periodic geometry with impermeable boundaries. The interval is composed of $m=100$ cells and $1/2\tilde{\kappa}=0.2$. The modes are taken from different branches ($j=0,1,2$) but have all $p=7$.
%}
%\label{fig:modes_m100}
%\end{figure*}

\subsection{Computation of the norm}
Because the geometry is symmetric and the relaxation coefficients $K_\pm$ are equal to zero, one can transform the formula \eqref{eq:normalisation1} of the normalization constant  into
\begin{align}
\beta_{j,p}^{-2}&=\frac{l}{2}\left|\begin{bmatrix}0 & 1\end{bmatrix}\frac{\mathrm{d}\mathcal{T}}{\mathrm{d}\alpha}\begin{bmatrix}1\\0\end{bmatrix}\right|_{\alpha=\alpha_{j,p}}=\frac{l}{2}\left|\begin{bmatrix}0 & 1\end{bmatrix}\frac{\mathrm{d}\mathcal{M}^m}{\mathrm{d}\alpha}\begin{bmatrix}1\\0\end{bmatrix}\right|_{\alpha=\alpha_{j,p}}=\frac{l}{2}\left|\frac{\mathrm{d}}{\mathrm{d}\alpha}\left(\sin\alpha\frac{\sin(m\psi)}{\sin{\psi}}\right)\right|_{\alpha=\alpha_{j,p}}\;.
\label{eq:normalisation_simple_periodic}
\end{align}
%Now we use the fact that $\begin{bmatrix}0 & 1\end{bmatrix}\mathcal{M}^m\begin{bmatrix}1\\0\end{bmatrix}=\frac{\sin m\psi}{\sin \psi} \begin{bmatrix}0&1\end{bmatrix}\mathcal{M}\begin{bmatrix}1\\0\end{bmatrix}=-\frac{\sin m\psi}{\sin \psi} \sin\alpha$. We recall that $\cos\psi=\frac{1}{2}\Tr\mathcal{M}=\cos\alpha-\frac{\alpha\sin\alpha}{2\tilde{\kappa}}$. 
Now we use Eq. \eqref{eq:coherence_periodic_explicite}, which leads us to distinguish the two cases as above:
\begin{itemize}
\item $\sin\alpha = 0$: it corresponds to $\alpha=j\pi$, with a positive integer $j$ (recall that we discard $\alpha=0$). Then $\cos\psi=(-1)^j$ and $\frac{\sin m\psi}{\sin \psi}=m(-1)^{j(m-1)}$. We conclude that the norm of the mode is:
\begin{equation}
\beta_{j,p}^2=\frac{2}{ml}\;.
\end{equation}
\item $\frac{\sin m\psi}{\sin \psi}=0$: it corresponds to $\alpha_{j,p}$ ($\psi=p\pi/m$), $p=1,\ldots,m-1$ and $j=0,1,\ldots$. In this case, the derivative in Eq. \eqref{eq:normalisation_simple_periodic} is easily computed by the chain rule:
\begin{align*}
&\frac{\mathrm{d}}{\mathrm{d}\alpha}\left(\frac{\sin m\psi}{\sin \psi}\right)=\frac{\mathrm{d}\cos\psi}{\mathrm{d}\alpha}\frac{\mathrm{d}\psi}{\mathrm{d}\cos\psi}\frac{\mathrm{d}}{\mathrm{d}\psi}\left(\frac{\sin m\psi}{\sin \psi}\right)\nonumber\\
&=-\!\left(\sin\alpha\left(1+\frac{\tilde{r}}{2}\right)+\frac{\tilde{r}}{2}\alpha\cos\alpha\right)\!\left(\frac{-1}{\sin\psi}\right)\frac{m\cos m\psi \sin\psi-\sin m \psi\cos\psi}{\sin^2 \psi}\;,
\end{align*}
which by evaluation at $\alpha_{j,p}$ yields:
\begin{equation}
\beta_{j,p}^2=\frac{2}{ml}\frac{\sin^2 p\pi/m}{\sin\alpha_{j,p}\left(\sin\alpha_{j,p}\left(1+\frac{\tilde{r}}{2}\right)+\frac{\tilde{r}}{2}\alpha_{j,p}\cos\alpha_{j,p}\right)}\;.
\label{eq:beta_j_p}
\end{equation}
\end{itemize}

%%%%%%%%%%%%%%%%%%%%%%%%%%%%%%%%%%%%%%%%%%

\color{black}
\section{Implementation and Applications}

\subsection{Numerical Implementation}

\label{section:numerical_implementation}

From a numerical point of view, the computational steps are the following: (i) to compute the transition matrix $\mathcal{M}_{i,i+1}$ in Eq. \eqref{eq:abreviations} for each compartment; (ii) to apply Eq. \eqref{eq:formule_generale_T} to get the complete transition matrix; (iii) to solve Eq. \eqref{eq:coherence_condition3} to get the spectrum  of the diffusion operator; each solution of Eq. \eqref{eq:coherence_condition3} determines one eigenvalue whereas Eqs. \eqref{eq:membrane_equation_matrice} and \eqref{eq:conditions_limites_matrice}  yield the coefficients $a^l_i$, $b^l_i$, $k=1,\ldots,m$ for each (non-normalized) mode; (iv) to compute the normalization constant; combined with Eq. \eqref{eq:formule_modes_gauche} it allows one to compute the eigenmode at any point of the interval.

Steps (i) and (ii) are easy and fast since we are dealing with $2\times 2$ matrices. Step (iv) can be done either with Eq. \eqref{eq:normalisation1}, which involves a numerical derivative, or by a direct computation, using:
%\begin{equation}
%\int_{0}^{l_i} \left(a^l_i \cos(x\sqrt{\lambda/D_i}) + b^l_i \sin(x\sqrt{\lambda/D_i})\right)^2 \,\mathrm{d}x=\frac{1}{2}\left[({a^l_i}^2+{b^l_i}^2)l_i+\frac{({a^l_i}^2-{b^l_i}^2)\sqrt{D_i}}{2\sqrt{\lambda}}\sin(2l_i\sqrt{\lambda/D_i})+\frac{a^l_i b^l_i \sqrt{D_i}}{2\sqrt{\lambda}}(1-\cos(2l_i\sqrt{\lambda/D_i})\right]
%\end{equation}
\begin{align}
\int_{0}^{l} \left(a \cos(kx) + b \sin(kx)\right)^2 \,\mathrm{d}x&=\frac{(a^2+b^2)l}{2}+\frac{(a^2-b^2)}{4k}\sin(2kl)+\frac{a b}{4k}(1-\cos(2kl))\;.
\end{align}

The most complicated and time-consuming step is (iii). As we explained in Sec. \ref{section:study_spectrum}, two or more solutions of Eq. \eqref{eq:coherence_condition3} may be very close to each other in the case of low-permeability barriers (typically $\kappa \ll D/l$). The estimates we derived allow us to localize the roots that speeds up the computation.
This is the crucial point and one of the major practical achievements of the paper. This numerical improvement allows us to detect very close zeros (as those shown in Fig. \ref{fig:graphes-zoom}) and to compute the eigenmodes of the diffusion operator in heterogeneous structures with hundreds of barriers.
Moreover, Fig. \ref{fig:graphes-zoom} illustrates an interesting property of $F_m(z;l_1,\ldots,l_m)$ as a function of $z$: two local extrema are apparently always separated by a zero. Although we have no mathematical proof for this observation, it is very helpful because it allows us to detect pairs of close zeros by the change of sign of the derivative of the function, which may take place on a much larger scale than the change of sign of the function itself. One can also take advantage of the Courant nodal theorem (which is proven for our particular model in Sec. \ref{section:courant_nodal_theorem}): the $n$-th eigenmode has $n$ nodal domains (connected components on which the eigenmode has a constant sign), or equivalently, the $n$-th eigenmode changes sign $n-1$ times (possibly at the barriers). This can be used as an efficient test to check \textit{a posteriori} that no eigenvalue is missed.

In practice, the standard floating-point precision limits the relative accuracy of a numerical computation to about $10^{-15}$. Let us assume that we are dealing with a geometry such that two eigenvalues $\lambda_1$ and $\lambda_2$ are much closer than this limit; for example they coincide up to $10^{-20}$. With the above tricks we are still able to detect those roots and even to compute accurately their position and spacing. However, the subsequent computations performed on $\lambda_1$ and $\lambda_2$ (for example, the computation of the eigenmodes or their norm) treat $\lambda_1$ and $\lambda_2$ as equal numbers. Even worse: the closeness of $\lambda_1$ and $\lambda_2$ is related to the very fast local variations of $F(\lambda)$ with $\lambda$, and as a consequence of the coefficients $(a^l_i,b^l_i)$ and of the norm of the eigenmode. Therefore it is very difficult to compute accurately these quantities for two eigenmodes corresponding to very close eigenvalues. The estimates derived in Sec. \ref{section:study_spectrum} can be used to detect \textit{a priori} such situations in which the spectral decomposition can numerically fail.

If one is interested in the diffusion propagator \eqref{eq:decomposition_G} or related quantities, the infinite collection of eigenmodes has to be truncated. This is done by sorting the eigenvalues $\lambda_n$ in ascending order and then cutting off the ones such that $\lambda_n t \gg 1$, where $t$ is the smallest diffusion time for which the computation is needed. The precise choice of the truncation threshold is a compromise between precision and speed of computation. Practically, one can check the validity of the truncation by re-doing the computation with a higher threshold and then comparing the two results.

We have implemented the proposed method for an arbitrary configuration of barriers and diffusion coefficients as a Matlab code. The numerical results presented in the Supplementary Materials were obtained on a basic laptop computer by using this code. The code can be sent upon request.

%%%%%%%%%%%%%%%%%%%%%%%%%%%%%%%%%%%%%

%%%%%%%%%%%%%%%%%%%%%%%%%%%%%%%%%

\subsection{Application to diffusion MRI}
\label{section:explication_dmri}

Diffusion of spin-bearing particles (such as nuclei of hydrogen atoms in water molecules) may be  surveyed by diffusion magnetic resonance imaging (dMRI), which is a powerful imaging technique with many biomedical applications \cite{Callaghan91,Price09,Grebenkov07,Kiselev17}. From the knowledge of the diffusion propagator one can access the dMRI signal under the so-called Narrow-Pulse Approximation (NPA), thus motivating numerous theoretical and experimental works on diffusion in complex geometries.
As explained previously, restricted diffusion in simple domains such as slab, cylinder, sphere, can be treated analytically  \cite{Tanner68,Callaghan92,Coy94,Callaghan95}. In contrast, most works devoted to multi-layered systems with semi-permeable barriers are numerical.
Tanner took advantage of the simple expression of the Laplace eigenmodes in a slab geometry to study a finite periodic repetition of semi-permeable barriers \cite{Tanner78}.
%
%
%Tanner studied diffusion in one-dimensional array of identical cells separated by semi-permeable membranes numerically by relying on the simple expression of the Laplace eigenmodes in a slab geometry\cite{Tanner78}. 
The same method was applied later by Kuchel and Durrant to unevenly spaced membranes \cite{Kuchel99}. These approaches were generalized by Grebenkov with a matrix formalism allowing efficient computation of the signal in general multi-layered planar, cylindrical or spherical structures, without the NPA restriction \cite{Grebenkov10}. 
Powles and co-workers proposed in \cite{Powles92} an opposite approach based on the (one-dimensional) analytical solution of $G$ for one semi-permeable barrier extended to several barriers by multiple reflections. Other numerical techniques such as a finite differences method were reported \cite{Novikov98}.
The first analytical expression of the dMRI signal in a one-dimensional geometry with periodic permeable barriers was provided by Sukstanskii \textit{et al.} \cite{Sukstanskii04}. Relying on the periodicity of the system they computed directly the signal in Laplace domain without having to derive the diffusion propagator.
Unevenly spaced membranes were treated in \cite{Grebenkov14-1,Grebenkov14-2} from the analytical solution for one membrane and under the assumption that the diffusing time is sufficiently short so that the layers are independent.
Note that in contrast to almost all previously cited works the analysis performed in \cite{Grebenkov14-2} does not confine to infinitely narrow pulses.
Finally, Novikov \textit{et al.} studied the effect of randomly placed semi-permeable barriers on the diffusive motion \cite{Novikov11,Novikov14}.  Using a renormalization group technique, they obtained structural universality classes characterized by the disorder introduced by the barriers, which in turn govern the long-time asymptotic behavior of the mean square displacement.

\begin{figure*}[hpb]
\begin{center}
\includegraphics[width=0.9\linewidth]{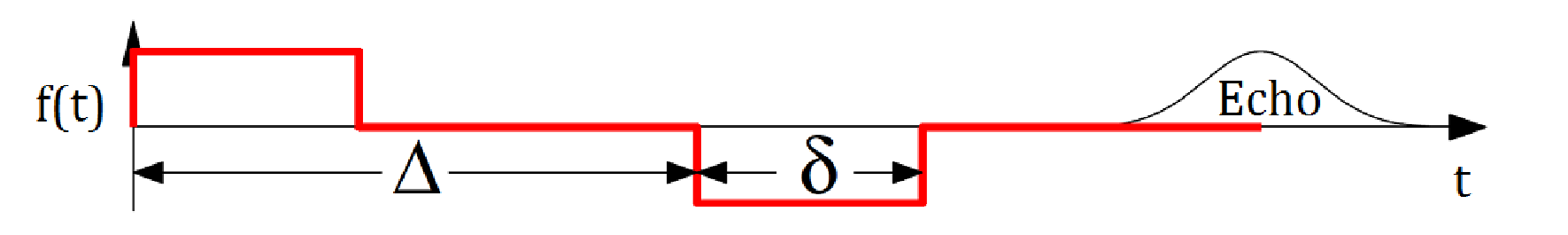}
\end{center}
\caption{Temporal profile $f(t)$ for Pulsed-Gradient-Spin-Echo (PGSE) experiments. The Narrow-Pulse Approximation (NPA) is the limit case $\delta\to 0$ while $\gamma g\delta$ remains constant.
}
\label{fig:PGSE}
\end{figure*}

In the general case, the signal is obtained by solving the Bloch-Torrey equation for the local magnetization $m(x,t)$: 
\begin{equation}
\frac{\partial m}{\partial t}=D\nabla^2 m + i\gamma gx f(t)m \;,
\label{eq:Bloch-Torrey}
\end{equation}
where $D$ is the diffusion coefficient, $\gamma$ the gyromagnetic ratio of the nuclei, $g$ the magnetic field gradient and $f(t)$ a customizable temporal profile \cite{Callaghan91,Price09,Grebenkov07}. In our one-dimensional geometry, the signal is then given by
\begin{equation}
S(t)=\frac{1}{L}\int _0^L m(x,t)\,\mathrm{d}x\;.
\label{eq:equation_signal_m}
\end{equation}
The method developed in Sec. \ref{section:calcul_general} for computing the diffusion operator eigenmodes allows us to calculate the signal analytically for infinitely narrow gradient pulses, or numerically for arbitrary pulse sequences (such as the one in Fig. \ref{fig:PGSE}). In particular, this method generalizes earlier approaches \cite{Tanner78,Novikov98,Sukstanskii04,Grebenkov14-1} and opens unprecedented opportunities for studying more sophisticated configurations of barriers such as microstructures inside larger scale structures.

The computations are detailed in Sec. \ref{section:application_dmri}. We explain how one can obtain the dMRI signal from the Fourier transform of the eigenmodes $u_n$ in the so-called narrow pulse regime, then we derive the expression of the signal for the periodic geometry presented in Sec. \ref{section:example_periodic}. We discuss the effect of the permeability of the barriers on the dMRI signal in the regimes of short and long diffusion time. In particular, we obtain a scaling law of the form $\tilde{\kappa}t/(\tilde{\kappa}+1)$ involving $t$ and $\tilde{\kappa}=\kappa l/D$, which is valid in the long time regime ($t\gg l^2/D$). Computations for more sophisticated geometries are presented in Sec. \ref{section:autres_structures_relax} (relaxation at the outer boundaries), \ref{section:bi-periodique} (bi-periodic geometry), and \ref{section:autres_structures_imbrique} (two-scale geometry).

% In the case of regular geometries such as a finite periodic geometry, one can perform all the computations explicitly. This allowed us to derive the dMRI signal under the narrow pulse approximation, which generalizes previous results.
 % In particular, we discussed in detail the influence of the inner barrier permeability $\tilde{\kappa}$ on the signal for a finite periodic geometry with reflecting conditions at the outer boundaries. Switching to perfectly relaxing conditions we studied the distribution of the first exit time out of this structure, which gives another insight into the effect of semi-permeable barriers on the diffusive motion. We extended some of our results to an irregular structure by relying on physical arguments and numerical computations. An interesting perspective would be to investigate the dMRI signal for irregular geometries such as the ones considered in \cite{Novikov11,Novikov14}. In these articles, the authors were concerned with the mean square displacement only, whereas our method enables us to compute the dMRI signal for any gradient sequence.

\subsection{First exit time distribution}
\label{section:explication_temps_sortie}

Another application of the diffusion operator eigenmodes is the computation of the first exit time distribution. First exit times are a particular case of first passage phenomena, which find many applications in physics, chemistry, biology, or economy. In particular, one-dimensional models are relevant to a wide variety of phenomena in which an event is triggered when a fluctuating variable reaches a given threshold (examples include avalanches, neuron firing, or sell/buy orders) as well as diffusion controlled reactions such as fluorescence quenching or predation \cite{Redner01,Metzler14}. In general planar domains, exit times were thoroughly investigated in the so-called \textquote{narrow-escape limit} \cite{Holcman14} and few results are available for arbitrary escape areas \cite{Grebenkov16,Rupprecht15}.

For this purpose, let us consider \emph{perfectly relaxing} conditions at the outer boundaries of the interval $[0,L]$: $K_{\pm}=\infty$. Then the quantity
\begin{equation*}
\int_0^L G(x\to x',\tau)\,\mathrm{d}x'
\end{equation*}
represents the probability of not reaching the outer boundaries for a particle starting at $x$, up to the time $\tau$. In other words, if one denotes by $T_x$ the random variable equal to the first exit time of a particle starting at $x$, then the tail distribution and the probability density of $T_x$ are respectively given by:
\begin{align}
&\mathbb{P}(T_x > \tau)=\int_0^L G(x\to x',\tau)\,\mathrm{d}x'=\sum_{n=1}^{\infty}   e^{-\lambda_n \tau}u_n(x)\left(\int_0^L u_n(x')\,\mathrm{d}x'\right)\;,\label{eq:formule_temps_sortie_P}\\
&\rho_{T_x}(\tau)=\frac{\mathbb{P}(\tau<T_x<\tau+\mathrm{d}\tau)}{\mathrm{d}\tau}=\sum_{n=1}^{\infty}\lambda_n  e^{-\lambda_n \tau}u_n(x)\left(\int_0^L u_n(x')\,\mathrm{d}x'\right)\;.\label{eq:formule_temps_sortie_f}
\end{align}

The computations are detailed in Sec. \ref{section:application_temps_sortie}. We rely on the computation of the eigenmodes for a periodic geometry with perfectly relaxing outer boundaries performed in Sec. \ref{section:autres_structures_relax} and obtain the first exit time distribution for this structure. We study the limit of a large number of barriers (where the size $L$ of the large interval remains constant). Similarly to the computation of the dMRI signal, we obtain a scaling law of the form $\tilde{\kappa}t/(\tilde{\kappa}+1)$. Then we turn to irregular geometries where $l_i$ and $\kappa_{i,i+1}$ are randomly distributed and we observe the same scaling law, with a new definition for $\tilde{\kappa}$ which depends on permeabilities and positions of the barriers. Numerical computations show a very good agreement even for a moderate number of barriers ($m\approx 10$). Moreover, we analyze the regime of very low permeability, where the diffusive motion can be replaced by a discrete hopping model, and exhibit a perfect agreement with previously obtained results.
 
%%%%%%%%%%%%%%%%%%%%%%%%%%%%%%%%%%%%%%%%%%%%%%%%%%%%%%%%%%%%%%%%%%%%%%%%%

\section{Conclusion}
\label{section:conclusion}

We presented an efficient method to compute the eigenmodes of the diffusion operator on a one-dimensional interval segmented by semi-permeable barriers, which in turn give access to the diffusion propagator. One can then compute several diffusion-related quantities such as the dMRI signal for any pulse sequence or the first exit time distribution.

Although the general matrix formalism is applicable to other multi-layered structures such as concentric cylindrical or spherical shells \cite{Grebenkov10}, the main analytical simplifications follow from the translation invariance of the Laplacian eigenmodes which is specific to one-dimensional models. In particular we derived some estimates that help us to accurately compute the eigenvalues, even when they are extremely close to each other. This is the crucial numerical step that allowed us to deal with heterogeneous structures with hundreds of semi-permeable barriers. This efficient method opens unprecedented opportunities to investigate the impact of microstructure onto diffusive motion.

%%%%%%%%%%%%%%%%%%%%%%%%%%%%%%%%%%%%%%%%%%%%%%%%%%%%%%%%%%%%%%%%%%%%%% 

\begin{acknowledgements}

We acknowledge the support under Grant No. ANR-13-JSV5-0006-01 of the
French National Research Agency.

\end{acknowledgements}

\clearpage
\newpage

\setcounter{equation}{0}
\setcounter{section}{0}
\setcounter{figure}{0}
\setcounter{table}{0}
\setcounter{page}{1}
\makeatletter
\renewcommand{\theequation}{S\arabic{equation}}
\renewcommand{\thefigure}{S\arabic{figure}}
\renewcommand{\thesection}{SM. \Roman{section}}

%\onecolumn
%\onecolumngrid

\noindent
{\textbf{\large Supplementary Material for the article ``Diffusion across semi-permeable barriers: spectral properties, efficient computation, and applications''}}

\section{Computation of the dMRI Signal}
\label{section:application_dmri}
\subsection{General case}

For a general geometry and an arbitrary pulse sequence one may solve numerically the Bloch-Torrey equation \eqref{eq:Bloch-Torrey} by decomposing $m(x,t)$ over the diffusion operator eigenmodes basis $(u_n)_{n\in\mathbb{N}}$:
\begin{equation}
m(x,t)=\sum_{n=1}^{\infty} m_n(t) u_n(x) \;, \quad \text{with} \quad m_n(t)=\int_0^L  u_n^*(x) m(x,t)\,\mathrm{d}x\;,
\label{eq:m_decomposition}
\end{equation}
where the asterisk denotes complex conjugation \cite{Grebenkov07,Grebenkov08}.
%wth complex coefficients $m_n$. 
Truncating the decomposition \eqref{eq:m_decomposition} to a finite number of terms $n_{max}$, one can represent the solution of the Bloch-Torrey equation as a vector:
\begin{equation*}
\mathbf{m}(t)=\begin{pmatrix}
m_1(t)\\m_2(t)\\\vdots\\m_{n_{max}}(t)\end{pmatrix} \;.
\end{equation*}
The Bloch-Torrey equation can then be rewritten as
\begin{equation*}
\frac{\partial \mathbf{m}}{\partial t} = -{\Lambda} \mathbf{m} + i\gamma g f(t) {B}\mathbf{m}\;,
\label{eq:Bloch-Torrey_matrices}
\end{equation*}
with the following matrices:
\begin{align*}
&{\Lambda}_{n,n'}=\int_0^L u_n^*(x)\, (-D\nabla^2 u_{n'}(x))\,\mathrm{d}x =\lambda_n \delta_{n,n'}\;, \\
&{B}_{n,n'}=\int_0^L u_n^*(x)  x u_{n'}(x)\, \mathrm{d}x\;, 
\label{eq:matrices}
\end{align*}
where $\delta_{n,n'}$ is the Kronecker symbol: $\delta_{n,n'}=1$ if $n=n'$ and $\delta_{n,n'}=0$ otherwise. Moreover, one can write
\begin{align*}
S(t)&=\frac{1}{L}\int_0^L m(x,t)\,\mathrm{d}x = \left(\mathbf{a}\cdot \mathbf{m}(t)\right)=\sum_{n=0}^{n_{max}} a_n m_n(t)\;, \quad \text{with} \qquad
{a}_n=\frac{1}{L}\int_0^L u_n(x)\,\mathrm{d}x\;.
\end{align*}
When the gradient profile $f(t)$ is made of two rectangular pulses of duration $\delta$, separated by time $\Delta$ (Fig. \ref{fig:PGSE}), the corresponding PGSE signal is obtained by computing the following scalar product:
\begin{equation}
S=\mathbf{a}\cdot\left[e^{-(\Delta-\delta)\Lambda}e^{-\delta(\Lambda+i\gamma g B)}e^{-(\Delta-\delta)\Lambda}e^{-\delta(\Lambda-i\gamma g B)}\right]\mathbf{m}(t=0)\;,
\end{equation}
with matrix exponentials in square brackets.
More generally, approximating the profile $f(t)$ by a piecewise constant function, one can accurately compute the dMRI signal for an arbitrary profile $f(t)$  \cite{Grebenkov10,Grebenkov07,Grebenkov08}. The initial condition for the magnetization is often uniform, $m(t=0)=1/L$, in which case $\mathbf{m}(t=0)=\mathbf{a}$.

%%%%%%%%%%%%%%%%%%%%%%%%%%%%%%%%

\subsection{NPA approximation for an array of identical cells and reflecting conditions at the outer boundaries}

The Narrow-Pulse Approximation (NPA) is the limit $\delta\to 0$ while $\gamma g \delta$ remains constant. In this regime the signal is directly linked to the diffusion propagator $G$ by
\begin{equation}
S(\Delta) = \int_0^L\!\!\int_0^L \rho(x_0)G(x_0\to x,\Delta)\cos(\gamma g \delta (x-x_0))\,\mathrm{d}x\,\mathrm{d}x_0 \;,
\end{equation}
where $\rho(x_0)$ is the initial spin density \cite{Callaghan91,Price09,Grebenkov07}. The spectral decomposition \eqref{eq:decomposition_G} yields
%Let us decompose $G$ on the orthogonal basis of the eigenmodes $u_{n}$ of the Laplace operator: $G(x_0\to x,t) = \sum_{n} g_n(t)u_{n}(x)$, with $g_n(t)=\int G(x_0\to x,t)u_{n}(x)\,\mathrm{d}x / \int u_{n}(\xi)^2 \,\mathrm{d}\xi$.
%Because $u_n$ is an eigenmode with eigenvalue $\lambda_n$, we know that $g_n(t) = g_n(0) \exp(-\lambda_n t)$ so we just need to compute $g_n(0)$. From $G(x_0 \to x,0) =\delta(x-x_0)$ we obtain $g_n(0)=u_{n}(x_0)/\int u_{n}(\xi)^2\,\mathrm{d}\xi$. We can then write: 
\begin{equation*}
%\label{NPA_modes_formula}
S(\Delta)=\sum_{n=1}^{\infty} e^{-\lambda_n \Delta}\int_0^L\!\!\int_0^L \rho(x_0){u_{n}(x_0)u_{n}(x)}\cos(\gamma g \delta(x-x_0))\,\mathrm{d}x\,\mathrm{d}x_0 \;,
\end{equation*}
If the initial density is uniform $\rho(x_0) = 1/L$, the symmetry between $x$ and $x_0$ leads to the following simplification:
\begin{equation}
%S &= \frac{1}{L}\sum_{n=1}^\infty e^{-\lambda_n t} \iint \frac{u_n(x) u_n(x_0)}{\int {u_n}^2} e^{i\gamma g \delta(x-x_0)} \,\mathrm{d}x\,\mathrm{d}x_0\\&= 
S(\Delta)=\frac{1}{L}\sum_{n=1}^{\infty} e^{-\lambda_n \Delta}{\left|{\int_0^L u_n(x)e^{i\gamma g \delta x} \,\mathrm{d}x} \right |}^2 \;.
\label{eq:NPA_modes_formula}
\end{equation}

This formula is the basis of the NPA and was initially introduced in \cite{Tanner68} to study the signal coming from a single isolated interval. Later the effect of semi-permeable barriers was numerically studied in \cite{Tanner78} for the most simple one-dimensional geometry where all $l_i$, $D_i$, $\kappa_{i,i+1}$ are the same (denoted $l$, $D$, $\kappa$ in the following).

In this section we apply the results of Sec. \ref{section:example_periodic} and extend the results of Ref. \cite{Tanner78}. In addition to Sec. \ref{section:example_periodic} we compute the Fourier transform of the modes which gives us the signal $S$. In Sec. \ref{section:autres_structures_relax} we extend this computation to  relaxing conditions at the outer boundaries. A more complicated geometry consisting of a microstructure inside a larger scale structure is treated in Sec. \ref{section:autres_structures_imbrique}.

We temporarily use the subscript $k$ instead of $i$ for the compartments in order to avoid any confusion with the imaginary unit $i=\sqrt{-1}$. As previously we use the position of the barrier to the left as the origin in the formula \eqref{eq:formule_modes_gauche} of the eigenmodes. This means that we have to compute integrals of the form:
\begin{align*}
\int_0^{l_k}\!\!\! e^{i\gamma g \delta x}\cos(\! x\sqrt{\lambda/D_k}\!) \mathrm{d}x
&=\frac{l_k}{2}\!\!\left(\! \frac{e^{i(\gamma g \delta +\sqrt{\lambda/D_k})l_k}-1}{il_k(\gamma g \delta +\sqrt{\lambda/D_k})} \!+\! \frac{e^{i(\gamma g \delta -\sqrt{\lambda/D_k})l_k}-1}{il_k(\gamma g \delta -\sqrt{\lambda/D_k})} \!\!\right),\\
\int_0^{l_k}\!\!\! e^{i\gamma g \delta x}\sin(\! x\sqrt{\lambda/D_k}\!) \,\mathrm{d}x
&=\frac{l_k}{2i}\!\!\left(\! \frac{e^{i(\gamma g \delta +\sqrt{\lambda/D_k})l_k}-1}{il_k(\gamma g \delta +\sqrt{\lambda/D_k})} \!-\! \frac{e^{i(\gamma g \delta -\sqrt{\lambda/D_k})l_k}-1}{il_k(\gamma g \delta -\sqrt{\lambda/D_k})}\!\! \right).
\end{align*}
We denote by $\mathcal{L}_k$ %(where $q_k=\gamma g \delta l_k$)
the row vector whose components are the above integrals.
%\begin{equation}
%\mathcal{L}_i(q_i)=\frac{l_i}{2}\begin{bmatrix}
%-i\left( \frac{e^{i(q_i-\alpha_{n,k})}-1}{q_i-\alpha_{n,k}}+\frac{e^{i(q_i+\alpha_{n,k})}-1}{q_i+\alpha_{n,k}}  \right) &\left( \frac{e^{i(q_i-\alpha_{n,k})}-1}{q_i-\alpha_{n,k}}-\frac{e^{i(q_i+\alpha_{n,k})}-1}{q_i+\alpha_{n,k}}  \right)
%\end{bmatrix}\;.
%\end{equation}
%\begin{equation*}
%\mathcal{L}_k = \frac{l_k}{2}
%\begin{bmatrix}\displaystyle
%-i\left(\frac{e^{i(\gamma g \delta -\sqrt{\lambda/D_k})l_k}-1}{(\gamma g \delta -\sqrt{\lambda/D_k})l_k}+ \frac{e^{i(\gamma g \delta +\sqrt{\lambda/D_k})l_k}-1}{(\gamma g \delta +\sqrt{\lambda/D_k})l_k} \right)& \displaystyle
%\left(\frac{e^{i(\gamma g \delta -\sqrt{\lambda/D_k})l_k}-1}{(\gamma g \delta -\sqrt{\lambda/D_k})l_k}- \frac{e^{i(\gamma g \delta +\sqrt{\lambda/D_k})l_k}-1}{(\gamma g \delta +\sqrt{\lambda/D_k})l_k} \right)
%\end{bmatrix}\;.
%\end{equation*}
%(We used $k$ instead of $i$ to index the compartments in order to avoid any confusion with the imadinary unit $i=\sqrt{-1}$).
The Fourier transform of the eigenmode $v$ is then simply
\begin{equation}
\int_0^L v(x)e^{i\gamma g \delta x} \,\mathrm{d}x=
\sum_{k=1}^m e^{i\gamma g \delta x_{k-1,k}}\mathcal{L}_k \begin{bmatrix}a^l_{k}\\b^l_{k}  \end{bmatrix}\;.
\label{eq:Fourier_general}
\end{equation}

\label{section:Fourier_simple}
Now we apply this general formula to our finite periodic geometry. The sum can be simplified because all $\mathcal{L}_k$ are the same:
\begin{equation}
\mathcal{L}=\frac{l}{2}\begin{bmatrix}\displaystyle
-i\left( \frac{e^{i(q-\alpha)}-1}{q-\alpha}+\frac{e^{i(q+\alpha)}-1}{q+\alpha} \right) &\displaystyle
\left( \frac{e^{i(q-\alpha)}-1}{q-\alpha}-\frac{e^{i(q+\alpha)}-1}{q+\alpha}  \right)
\end{bmatrix}\;,
\end{equation}
where $q=\gamma g \delta l$.
Moreover $x_{k-1,k}=(k-1)l$ so we can rewrite the sum \eqref{eq:Fourier_general}:
\begin{align}
\int_0^L v(x)e^{i\gamma g \delta x} \,\mathrm{d}x&=\sum_{k=1}^m e^{i(k-1)q}\mathcal{L} \begin{bmatrix}a^l_{k}\\b^l_{k} \end{bmatrix} = \mathcal{L}  \sum_{k=0}^{m-1} e^{ikq}\mathcal{M}^k \begin{bmatrix}1\\0\end{bmatrix}\nonumber\\
&= \mathcal{L}  (\mathcal{I}_2 - e^{iq}\mathcal{M})^{-1}(\mathcal{I}_2 - e^{imq}\mathcal{M}^{m})\begin{bmatrix}1\\0\end{bmatrix}=(1-(-1)^p e^{imq})\mathcal{L}(\mathcal{I}_2 - e^{iq}\mathcal{M})^{-1}\begin{bmatrix}1\\0\end{bmatrix}\;,
\end{align}
where we have used Eq. \eqref{eq:coherence_periodic_simple} with $\epsilon=(-1)^p$.
%$\mathcal{M}^{m} \begin{bmatrix} 1\\0 \end{bmatrix}=(-1)^p\begin{bmatrix} 1\\0 \end{bmatrix}$
%
%\begin{equation*}
%(1-(-1)^p e^{imq})\times\mathcal{L}(q)\cdot(\mathcal{I}_2 - e^{iq}\mathcal{M})^{-1}\begin{bmatrix}1\\0\end{bmatrix}\;.
%\end{equation*}
%
We can simplify the matrix product further with the remark that the comatrix operation is linear for $2\times2$ matrices, and that $\det\mathcal{M}=1$, so that
%\begin{equation}
%(\mathcal{I}_2 - e^{iq}\mathcal{M})^{-1} = \frac{(\mathcal{I}_2 - e^{iq}\mathcal{M}^{-1})}{\det(\mathcal{I}_2 - e^{iq}\mathcal{M})}\;,
%\end{equation}
%given that $\det \mathcal{M} = 1$. Moreover, $\mathcal{M}^{-1} = \mathcal{R}^{-1}\mathcal{K}^{-1}$ and from $\mathcal{K}^{-1} = \mathcal{I}_2 - \mathcal{N}$ we finally get:
\begin{equation*}
\mathcal{L}(\mathcal{I}_2 - e^{iq}\mathcal{M})^{-1}\begin{bmatrix}1\\0\end{bmatrix}
=\frac{\mathcal{L}(\mathcal{I}_2 - e^{iq}\mathcal{M}^{-1})}{\det(\mathcal{I}_2 - e^{iq}\mathcal{M})}\begin{bmatrix}1\\0\end{bmatrix}
=\frac{\mathcal{L}(\mathcal{I}_2 - e^{iq}\mathcal{R}^{-1})}{\det(\mathcal{I}_2 - e^{iq}\mathcal{M})} \begin{bmatrix}1\\0\end{bmatrix}\;.
\end{equation*}
From the knowledge of the trace and determinant of the matrix $\mathcal{M}$ we compute
\begin{equation*}
\det(\mathcal{I}_2 - e^{iq}\mathcal{M}) = -2e^{iq}(\cos \psi - \cos q)\;.
\end{equation*}
Furthermore,
\begin{equation*}
\mathcal{L}(\mathcal{I}_2 - e^{iq}\mathcal{R}^{-1})\begin{bmatrix}1\\0\end{bmatrix} = -2ie^{iq}(\cos \alpha - \cos q) \frac{ql}{q^2-{\alpha}^2}\;.
\end{equation*}
Putting all the pieces together yields
\begin{equation}
\label{eq:Fourier_simple}
\int_0^L v_{j,p}(x)e^{i\gamma g \delta x} \,\mathrm{d}x = e^{imq/2}\frac{iql(e^{-imq/2}-(-1)^p e^{imq/2})\frac{\cos q -\cos\alpha_{j,p}}{\cos q -\cos p\pi/m}}{q^2-{\alpha_{j,p}}^2}\;.
\end{equation}
Note that the ratio is either real ($p$ even) or imaginary ($p$ odd) which is consistent with the symmetry or anti-symmetry of the mode (see Sec. \ref{section:symmetry}).

%%%%%%%%%%%%%%%%%%%%%%%%%%%%%%%%%%%%%%%%%%%%%%

\subsection{Complete expression of the signal}

Let us summarize our results. In the array of $m$ identical cells one has $D_i=D$ and $l_i=l$, $i=1,\ldots,m$. We thus introduce the dimensionless time $t=D\Delta/l^2$, where $\Delta$ is the diffusion time (see Fig. \ref{fig:PGSE}), and $q=\gamma g \delta l$. The combination of the previous results yields the formula:
%
%The eigenvalues $\lambda_n$ of the diffusion operator are equal to $\frac{D{\alpha_n}^2}{l^2}$. The $\alpha_n$ are solutions of the equation: $\sin\alpha \frac{\sin m\psi}{\sin \psi}=0$, where $\cos\psi=\cos\alpha-\frac{\tilde{r}}{2}\alpha \sin\alpha$. Thus $\alpha=j\pi$ or $\cos\alpha-\frac{\tilde{r}}{2}\alpha\sin\alpha=\cos p\pi/m$, for $p \in \left\lbrace1,\ldots,m-1\right\rbrace$. The latter gives an infinite array of solutions $\alpha_{j,p}$.
%
%\begin{align}
%S=&\frac{2(1-\cos mq)}{(mq)^2} + \sum_{j=1}^{\infty} \frac{4q^2(1-(-1)^{jm}\cos mq)}{m^2\left(q^2-(j\pi)^2\right)^2}e^{-(j\pi)^2t}\nonumber\\&+\sum_{j=0}^{\infty}\sum_{p=1}^{m-1} \frac{4q^2(1-(-1)^{p}\cos mq)\left(\frac{\cos q-\cos\alpha_{j,p}}{\cos q-\cos p\pi/m}\right)^2}{m^2\left(q^2-{\alpha_{j,p}}^2\right)^2\left[\left(1+\frac{\tilde{r}}{2}\right)\sin\alpha_{j,p}+\frac{\tilde{r}}{2}\alpha_{j,p}\cos\alpha_{j,p} \right]\frac{\sin\alpha_{j,p}}{\sin^2(p\pi/m)}}e^{-{\alpha_{j,p}}^2t}\;,
%\label{eq:S_simple_periodic}
%\end{align}
%
\begin{align}
S=&\frac{2(1-\cos mq)}{(mq)^2} + \sum_{j=1}^{\infty} \frac{4q^2(1-(-1)^{jm}\cos mq)}{m^2\left(q^2-(j\pi)^2\right)^2}e^{-(j\pi)^2t}\nonumber\\&+\sum_{j=0}^{\infty}\sum_{p=1}^{m-1} \frac{2lq^2}{m} \frac{1-(-1)^{p}\cos mq}{(\cos q-\cos p\pi/m)^2}\left(\frac{\cos q-\cos\alpha_{j,p}}{q^2-\alpha_{j,p}^2}\right)^2\beta_{j,p}^2 e^{-{\alpha_{j,p}}^2t}\;,
\label{eq:S_simple_periodic}
\end{align}
where $\beta_{j,p}^2$ is given by Eq. \eqref{eq:beta_j_p}.
%with
%\begin{equation*}
%\beta_{j,p}^2=\frac{2}{ml}\frac{\sin^2 p\pi/m}{\sin\alpha_{j,p}\left(\sin\alpha_{j,p}\left(1+\frac{\tilde{r}}{2}\right)+\frac{\tilde{r}}{2}\alpha_{j,p}\cos\alpha_{j,p}\right)}\;.
%\end{equation*}

%%%%%%%%%%%%%%%%%%%%%%%%%%%%%%%%%%%%%%%%%%%%%%%%%

If $m=1$, there is no double sum on the second line of Eq. \eqref{eq:S_simple_periodic},  and one retrieves the well-known result by Tanner \cite{Tanner68}:
\begin{equation}
S_1(q,t)=\frac{2(1-\cos q)}{q^2} + \sum_{j=1}^{\infty} \frac{4q^2(1-(-1)^{j}\cos q)}{\left(q^2-(j\pi)^2\right)^2}e^{-(j\pi)^2t}\;.
\label{eq:signal_m_1}
\end{equation}

The opposite limit $m\to \infty$ was the motivation of the subsequent article by Tanner \cite{Tanner78} and was derived analytically in \cite{Sukstanskii04}.
When $m\to \infty$, each term of the sum in Eq. \eqref{eq:S_simple_periodic} vanishes except the ones for which $\cos p\pi/m$ is close to $\cos q$. Let us write
\begin{equation*}
q=2k\pi+p_0\pi/m+\epsilon/m\;, \quad p_0\in \left\lbrace 0,\ldots,m-1\right\rbrace\;, \quad 0\leq \epsilon <\pi\;.
\end{equation*}
%with $k$, $n_0$ and $\epsilon$ depends on $m$). 
Then we have: 
\begin{equation*}
\frac{1-(-1)^p \cos(mq)}{m^2 (\cos q-\cos{p\pi/m})^2}\approx\frac{1-(-1)^{p_0-p} \cos \epsilon}{\pi^2\sin^2(q)(p_0-p+\epsilon/\pi)^2}\;.
\end{equation*}
To get the signal in the $m\to \infty$ limit, we thus have to compute the following sum:
\begin{equation*}
\sum_{p=-\infty}^{\infty} \frac{1}{\pi^2}\frac{1-(-1)^p\cos \epsilon}{(p+\epsilon/\pi)^2}=1.
\end{equation*}
%We conclude then that from the $m^2$ terms of the sum \eqref{eq:S_simple_periodic}, only the ones such that $\cos p\pi/m \approx \cos q$ do not vanish when $m\to \infty$. It means that the new equation of the spectrum of the diffusion operator is:
%
The new equation on $\alpha$ is
\begin{equation}
\cos\psi = \cos \alpha - \frac{\tilde{r}}{2} \alpha \sin \alpha = \cos q\;,
\label{eq:spectre_m_infini}
\end{equation}
and the expression of the signal becomes
\begin{equation}
S_\infty(q,t,\tilde{\kappa})
%=\sum_{n=1}^{\infty} \frac{4q^2 \left(\frac{cos q - \cos \alpha_n}{q^2-{\alpha_n}^2}\right)^2}{\left(\left(1+\frac{\tilde{r}}{2}\right)\sin\alpha_n+\frac{\tilde{r}}{2}\alpha_n\cos\alpha_n \right)\sin\alpha_n}e^{-{\alpha_n}^2t}
=\frac{2q^2}{\tilde{\kappa}}\sum_{n=1}^{\infty}\frac{{\alpha_n}^2\sin\alpha_n  e^{-{\alpha_n}^2 t}}{({\alpha_n}^2-q^2)^2\left((2\tilde{\kappa}+1)\sin\alpha_n+\alpha_n\cos\alpha_n\right)}\;.
\label{eq:signal_m_infini}
\end{equation}
This is exactly the formula derived in \cite{Sukstanskii04} by the computation of the Laplace transform of $\int G(x_0\to x)e^{i\gamma g \delta (x-x_0)} \,\mathrm{d}x_0$ on an infinite periodic geometry.
Note that although the geometry is infinite and thus the spectrum of the diffusion operator is continuous, the signal is expressed in terms of a discrete set of eigenvalues because of Eq. \eqref{eq:spectre_m_infini}: the Fourier transform selects only the modes that globally oscillate at the wavenumber $q$ (recall that $\alpha$ only describes the intra-block oscillations, whereas the global behavior of the mode is dictated by $\psi$, according to Eq. \eqref{eq:variation_inter_bloc}). This is consistent with the discreteness of the spectrum of the Airy operator $D\frac{\mathrm{d}^2}{\mathrm{d}x^2}+i\gamma g x$ on any (bounded or unbounded) interval segmented by semi-permeable barriers \cite{Grebenkov14-2,Grebenkov17}.
As a consequence, one has to compute $\alpha_n$, $n=1,2,\ldots$ for each value of $q$, in contrast to the finite geometry where the spectrum depends only on the geometry and needs to be calculated only once. This is an important numerical advantage of the finite geometry over the infinite one because the computation of the spectrum is one of the most time-consuming step (as explained in Sec.\ref{section:study_spectrum} and \ref{section:numerical_implementation}).

\subsection{Discussion: dependence of the signal on the permeability}

In this section we study the diffusion operator eigenvalues and the signal in various regimes in order to show the dependence of the signal on the dimensionless permeability of the inner barriers, $\tilde{\kappa}$, which characterizes the microstructure.
In biological tissues, one has typically: $D\sim 1\dunit$, $l=1-100\lunit$, $\kappa\sim 10^{-3}-1 \kunit$, and the experimental range of diffusion time is about $\Delta\sim 10-10^3 \tunit$.
Thus we have the following ranges of variation for our dimensionless parameters: $\tilde{\kappa}\sim 10^{-3}-10^2$ and $t\sim 10^{-3}-10^3$.

In the limit $\tilde{\kappa}\to\infty$, one obviously recovers the signal associated to the whole interval of length $ml$ with no barriers, whereas in the opposite limit $\tilde{\kappa}\to 0$ one gets the signal \eqref{eq:signal_m_1} associated to one interval of length $l$ (we detail the mathematical proof in Sec. \ref{section:Limite_permeabilite}).
In other words
\begin{equation*}
S(m,q,t,\tilde{\kappa})\xrightarrow[\tilde{\kappa}\to\infty]{} S_1(mq,t/m^2)\; \qquad \text{and } \qquad
S(m,q,t,\tilde{\kappa})\xrightarrow[\tilde{\kappa}\to 0]{} S_1(q,t)\;.
\end{equation*}
We are interested in the transition from one limit to the other, that is the dependence of the signal on the permeability. Expansions of $\alpha_{j,p}$ at low and high permeability are derived in Sec. \ref{section:DLs}.
%(\cref{eq:DL_petit_kappa_j0,eq:DL_petit_kappa,eq:DL_grand_kappa_j0,eq:DL_grand_kappa}).
%
They show that the transition from $\tilde{\kappa}=0$ to $\tilde{\kappa}=\infty$ does not occur at one fixed value of $\tilde{\kappa}$ but depends on the branch of eigenvalues that we consider. Typically for the branch $j$ the transition occurs at $\tilde{\kappa}\sim j\pi/2$ if $j>0$. As we have already seen, the $j=0$ branch is particular and exhibits a $\tilde{\kappa}^{1/2}$ dependence at low $\tilde{\kappa}$ (see Eqs. \eqref{eq:DL_petit_alpha} and \eqref{eq:DL_petit_kappa_j0}).
In order to refine our analysis we distinguish long-time and short-time regimes.

\subsubsection{Long-time regime}
\label{section:temps_long}
\begin{figure*}[tb]
\begin{center}
\includegraphics[width=0.9\linewidth]{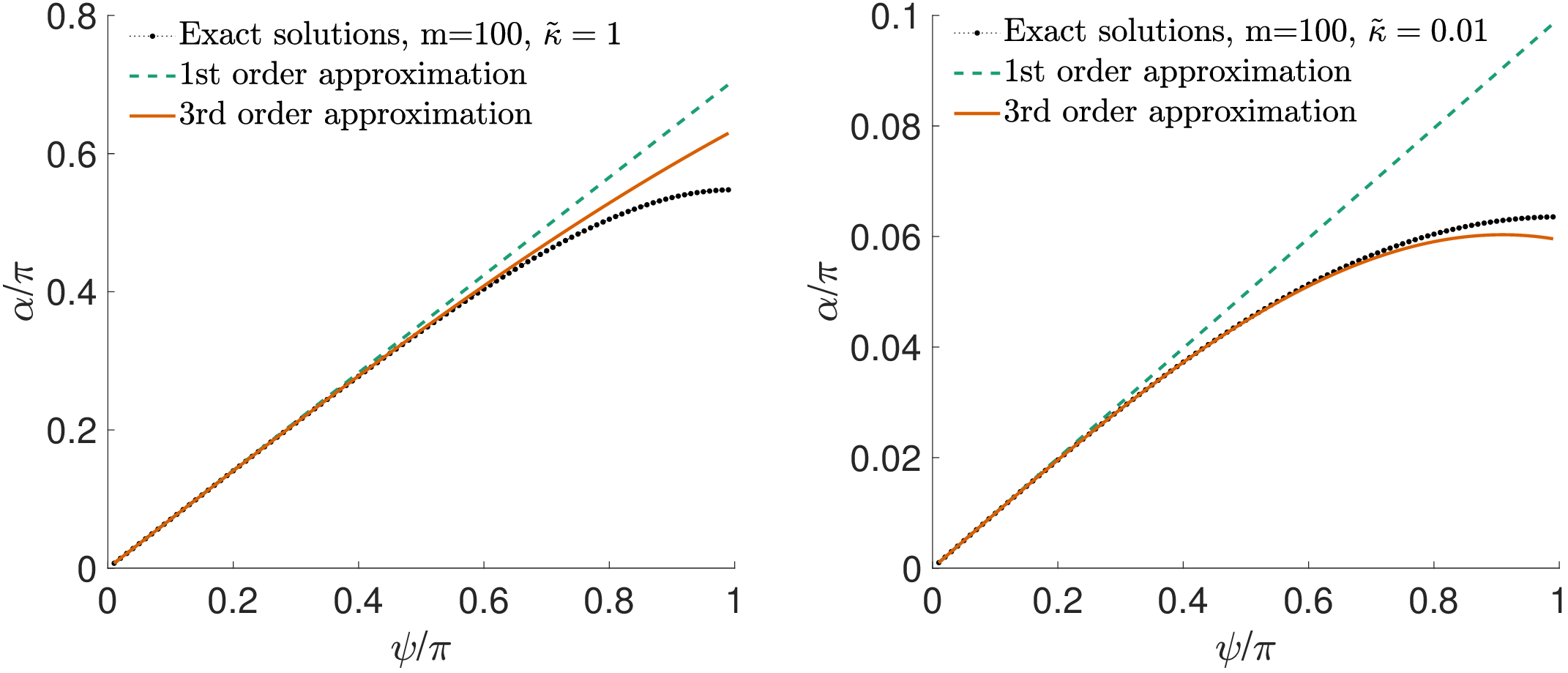}
\end{center}
\caption{The $j=0$ branch of solutions for $m=100$ compartments and its approximation by Eq. \eqref{eq:DL_petit_alpha}. (left) $\tilde{\kappa}=1$; (right) $\tilde{\kappa}=0.01$. One can see that the first order approximation formula is more accurate when $\tilde{\kappa}$ is higher which is consistent with Eq. \eqref{eq:DL_petit_alpha}.
}
\label{fig:DL_petit_alpha}
\end{figure*}

In the limit $t\to\infty$, all the modes with non-zero eigenvalues vanish and we are left with
\begin{equation}
S=\frac{2(1-\cos mq)}{(mq)^2}\;,
\label{eq:limite_temps_infini}
\end{equation}
which is a well-known formula \cite{Tanner68}. 
Note that relaxation at the outer boundaries would lead to zero signal in the long-time limit because $\lambda=0$ would not be an eigenvalue of the diffusion operator anymore. As expected at long times the details of the geometry are averaged out and the signal depends only on the length of the whole interval, $L=ml$.
The next terms are given by the first solutions of the $j=0$ branch. Let us study Eq. \eqref{eq:spectre_periodic} at small $\alpha$, $\psi$. Expanding the sine and cosine functions, one gets
\begin{equation}
\alpha=\psi \sqrt{\frac{\tilde{\kappa}}{\tilde{\kappa}+1}}\left(1-\frac{\psi^2}{24(\tilde{\kappa}+1)^2}\right)+O(\psi^5)\;.
\label{eq:DL_petit_alpha}
\end{equation}
Note that the third order correction is below $1\%$ if $\psi/\pi < 0.15 (\tilde{\kappa}+1)$ and approximately below $10\%$ if $\psi/\pi <0.5 (\tilde{\kappa}+1)$. In particular the accuracy of the first-order approximation is always better than $10\%$ for the first non-zero solution $\psi=\pi/m$ ($m>1)$.
This is illustrated in Fig. \ref{fig:DL_petit_alpha} for two values of $\tilde{\kappa}$: $1$ and $0.01$. As expected, the approximation is more accurate for larger $\tilde{\kappa}$.
Using this expansion we get the long-time asymptotic behavior
\begin{equation}
S\approx\frac{2(1-\cos mq)}{(mq)^2}+ A_1(q) \exp\left(-\frac{\pi^2\tilde{\kappa}t}{m^2(\tilde{\kappa}+1)}\right)\;,
\label{eq:developpement_temps_long}
\end{equation}
where $A_1(q)$ can be read on Eq. \eqref{eq:S_simple_periodic}: 
\begin{equation*}
%A_1(q)=\frac{4q^2(1+\cos mq)\left(\frac{\cos q-\cos\alpha_{1,0}}{\cos q-\cos \pi/m}\right)^2}{m^2\left(q^2-{\alpha_{1,0}}^2\right)^2\left[\left(1+\frac{\tilde{r}}{2}\right)\sin\alpha_{1,0}+\frac{\tilde{r}}{2}\alpha_{1,0}\cos\alpha_{1,0} \right]\frac{\sin\alpha_{1,0}}{\sin^2(\pi/m)}}\;.
A_1(q)
%%=\left|\int_0^L u_{0,1}(x)e^{iqx}\,\mathrm{d}x\right|^2
=\frac{2lq^2}{m} \frac{1+\cos mq}{(\cos q-\cos \pi/m)^2}\left(\frac{\cos q-\cos\alpha_{0,1}}{q^2-\alpha_{0,1}^2}\right)^2\beta_{0,1}^2\;.
%%\leq 1\;.
%%
\end{equation*}
%Note that because $\alpha_{0,1}$ is small, we have approximately:
Because $\alpha_{0,1}$ is small, we have approximately
\begin{equation*}
A_1(q)\approx \frac{4(1+\cos{mq})(1-\cos{q})^2}{q^2 m^2(\cos{q} - \cos{\pi/m})^2}\;,
\end{equation*}
which does not depend on $\alpha_{0,1}$ anymore but only on $\psi_{0,1}=\pi/m$. In other words, $A_1(q)$ weakly depends on $\tilde{\kappa}$. This approximation is especially accurate at high $m$ (we checked numerically that the error is less than $3\%$ for $m>10$, for example). This is a consequence of the remark that the global behavior  of the mode, hence its norm and Fourier transform, is dictated by $\psi$ (see Eq. \eqref{eq:variation_inter_bloc}).

From the expansion \eqref{eq:developpement_temps_long} we conclude that the parameter which controls the validity of the long-time limit is not $t$ but rather $\tilde{\kappa}t/((\tilde{\kappa}+1)m^2)$. The $m$-dependence is obvious: $m^2$ is in fact the (dimensionless) time required to diffuse through all the compartments if there are no barriers. One can then see that the effect of the barriers is to increase this diffusion time by a factor $({\tilde{\kappa}+1})/{\tilde{\kappa}}$. In other words, the time-dependence of the signal yields an apparent diffusion coefficient
\begin{equation}
D_{app}=D\frac{\tilde{\kappa}}{\tilde{\kappa}+1}=\frac{1}{1/D+1/(\kappa l)}\;.
\end{equation}
This formula is a well-known correction that can be derived by simple geometrical arguments \cite{Crick70}. When the permeability is high, the diffusion coefficient is slightly diminished. In the opposite limit $\tilde{\kappa} \ll 1$ one gets an apparent diffusion coefficient: $D_{app}=D\tilde{\kappa}=\kappa l$, which does not depend on the \textquote{true} diffusion coefficient anymore. In this regime, the kinetics of diffusion are governed by the crossing of the barriers and not by the (much faster) intra-compartment diffusion.

More generally, we have:
\begin{equation*}
S\approx\sum_{p=0}^{m-1} A_p(q) \exp\left(-\frac{p^2\pi^2\tilde{\kappa}t}{m^2(\tilde{\kappa}+1)}\right)\;,\quad
A_p(q)=\left|\int_0^L u_{0,p}(x)e^{iqx}\,\mathrm{d}x\right|^2\;,
\end{equation*}
where $A_p(q)$ weakly depends on $\tilde{\kappa}$. Thus in the long-time regime, the signal depends on $t$ and $\tilde{\kappa}$ via the combination $\tilde{\kappa}t/(\tilde{\kappa}+1)$.

If $1\ll t \ll 1/\tilde{\kappa}$, then $S\approx \sum_{p=0}^{m-1} A_p(q)$ and from Sec. \ref{section:Limite_permeabilite} we get:
\begin{equation}
S\approx \frac{2(1-\cos{q})}{q^2}\; \qquad (1\ll t \ll 1/\tilde{\kappa})\;.
\label{eq:signal_temps_long_faible_permeabilite}
\end{equation}
%The condition $t\gg 1$ means that the averaging due to diffusion has taken place in each compartment, on the other hand $t\ll 1/\tilde{\kappa}$ means that very few particles have leaked outside their compartments.
The condition $t\gg 1$ means that the diffusion has averaged the magnetization inside each compartment, whereas $t\ll 1/\tilde{\kappa}$ means that very few particles have crossed the inner barriers.
As a consequence we recover the signal in the long-time limit for one compartment of length $l$ and not of length $L=ml$ (as in Eq. \eqref{eq:limite_temps_infini}), even though $t\gg 1$.

\begin{figure*}[tb]
\begin{center}
\includegraphics[width=0.9\linewidth]{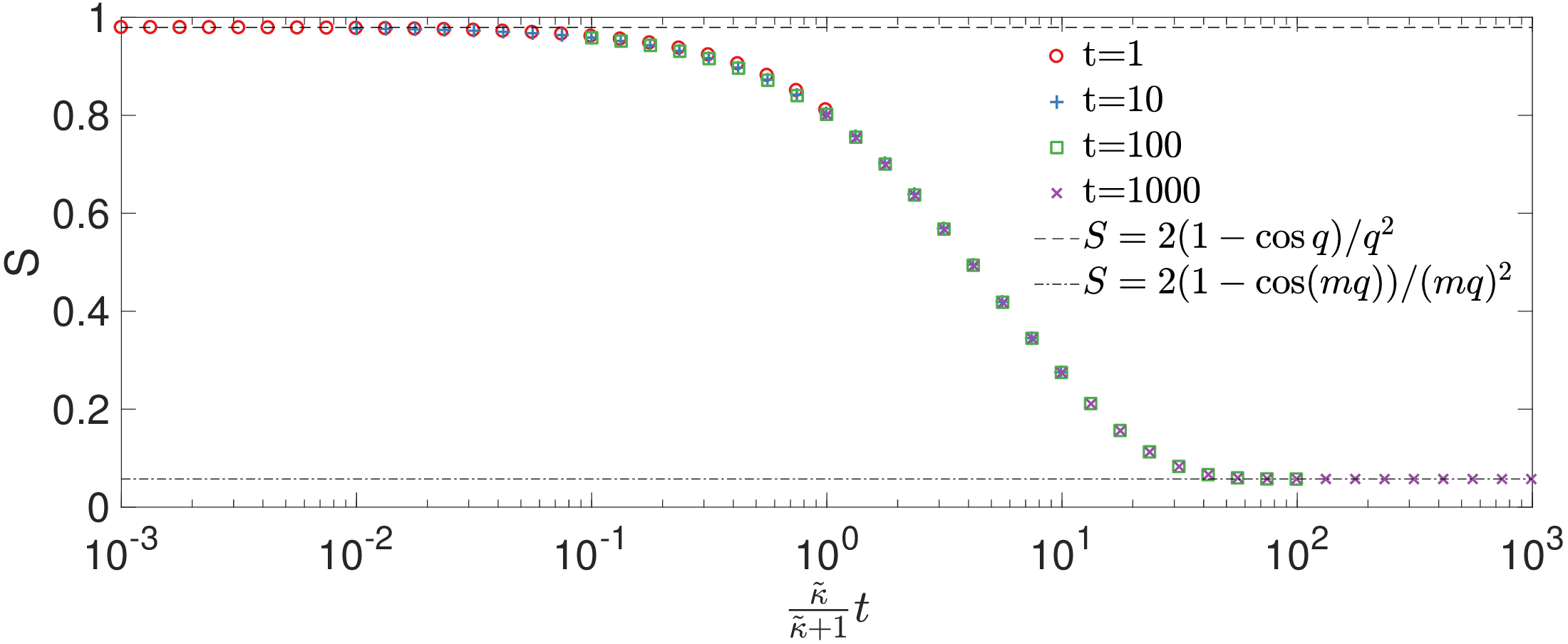}
\end{center}
\caption{Signal as a function of $\tilde{\kappa}t/(\tilde{\kappa}+1)$ at long diffusion times ($t>1$) for $m=10$ compartments and fixed $q=0.5$. One can see that the curves fall onto one master curve. The low- and high-permeability limits (Eqs. \eqref{eq:signal_temps_long_faible_permeabilite} and \eqref{eq:limite_temps_infini}, respectively) are plotted by dashed and dash-dotted line, respectively.
}
\label{fig:signal_m10_t_kappa_master_tlong}
\end{figure*}

Figure \ref{fig:signal_m10_t_kappa_master_tlong} illustrates the long-time regime ($t>1$) for an interval segmented into $m=10$ compartments. The signal is plotted as a function of $\tilde{\kappa}t/(\tilde{\kappa}+1)$ at fixed $q=0.5$ and different times.
The choice of $q$ is a compromise between the two limits given by Eqs. \eqref{eq:signal_temps_long_faible_permeabilite} and \eqref{eq:limite_temps_infini} (dashed and dash-dotted line, respectively).
In fact, $q$ should be small enough so that the signal in the limit $\tilde{\kappa}\to 0$ is close to $1$, and large enough so that the signal in the limit $\tilde{\kappa}\to\infty$ should be close to $0$, in order to maximize the variation of the signal with $\tilde{\kappa}$. 
One can see that all the symbols fall onto one master curve.
In particular, the transition from low- to high-permeability occurs at a fixed value of $\tilde{\kappa}t/(\tilde{\kappa}+1)$, which is around $1/q^2$.

\subsubsection{Short-time regime}

The short-time limit is the opposite case: all the branches of $j\lesssim 1/\sqrt{t}$ have to be taken into account in the formula \eqref{eq:S_simple_periodic} of the signal. However, we know that $j\pi <\alpha_{j,p} < (j+1)\pi$, so that increasing $\tilde{\kappa}$ from $0$ to $\infty$ produces a net increase of the $\alpha_{j,p}$ which is less than $\pi$. As a consequence, the relative decrease of $\exp(-\alpha_{j,p}^2t)$ is at most $\pi^2(2j+1)t \lesssim 2\pi^2\sqrt{t} \ll 1$. Thus, as expected, the signal weakly depends on the permeability.
As $\tilde{\kappa}$ increases the branches of solutions transform successively from the $\tilde{\kappa}=0$ limit to the $\tilde{\kappa}=\infty$ limit. Beyond $\tilde{\kappa}\sim 1/\sqrt{t}$, the increase of $\tilde{\kappa}$ produces little change on the most contributing branches, hence on the signal.
One can interpret this behavior in the following way: the dependence of the signal on the permeability is proportional to the fraction of particles which have reached a barrier. Indeed at short time, this fraction is given by $\sqrt{t}$. Among those particles, the ones that have crossed the barrier represent a fraction $\tilde{\kappa}t/\sqrt{t}=\tilde{\kappa}\sqrt{t}$. Hence $\tilde{\kappa}\sim 1/\sqrt{t}$ is the value of the permeability from which almost every particle that has reached a barrier has crossed it.

%%%%%%%%%%%%%%%%%%%%%%%%%%%%%%%%%%%%%%%%%%%%%%%%%%

\section{First exit time distribution}
\label{section:application_temps_sortie}
\subsection{Regular geometry}
\label{section:temps_sortie_regular}

Let us study the first exit time distribution \eqref{eq:formule_temps_sortie_f} for a geometry similar to the example of Sec. \ref{section:example_periodic} and \ref{section:application_dmri}: it consists of an array of $m$ identical cells of length $L/m$, where $L$ is independent of $m$, with \emph{perfectly relaxing} conditions at the outer boundaries ($K_\pm=\infty$).
%We use the same notations as before \eqref{eq:notations_simple}.
The computations are detailed in Sec. \ref{section:calculs_relax}. Since $u_n(0)=0$, one cannot use the normalization $v(0)=1$ from Sec. \ref{section:calcul_general}, so we write $u=\beta w$ with another normalization, $w'(0)=\sqrt{\lambda/D}$, which corresponds to $\begin{bmatrix} a^l_1\\b^l_1\end{bmatrix}=\begin{bmatrix}0\\1\end{bmatrix}$.
%
%The starting point is located at the center of the interval: $x=L/2$. 
%
Because the geometry is symmetric the eigenmodes of the diffusion operator $u_n$, $n=1,2,\ldots$ are alternately symmetric or anti-symmetric (see Sec. \ref{section:symmetry}); the latter give a zero contribution in the sum in Eqs. \eqref{eq:formule_temps_sortie_P} and \eqref{eq:formule_temps_sortie_f}. As for the symmetric eigenmodes, one obtains:
\begin{align}
\int_0^L& w(x)\,\mathrm{d}x=\frac{2l}{\alpha}\;,
\label{eq:Fourier_modes_relaxation}\\
\beta^{-2}&=\frac{-ml}{2}\frac{\sin\alpha\left(1+\frac{\tilde{r}}{2}\right)+\frac{\tilde{r}}{2}\alpha\cos\alpha}{\sin^2\psi}\left(\sin\alpha\cos m\psi+\frac{\tilde{r}\alpha (m-1)}{m}\cos((m-1)\psi)\right)\nonumber\\ &+\frac{ml}{2}\left(\frac{\sin\alpha}{\alpha}-\cos\alpha \right)\frac{\sin m\psi}{m\sin\psi}\;,
\label{eq:norme_modes_relaxation}
\end{align}
where $\alpha$ is a solution of the equation
\begin{equation}
\sin\alpha\frac{\sin{m\psi}}{\sin\psi} + \tilde{r}\alpha\frac{\sin((m-1)\psi)}{\sin\psi}=0\;.
\label{eq:coherence_condition_relaxation}
\end{equation}

We recall that 
\begin{equation}
\lambda=D\alpha^2/l^2=D\alpha^2 m^2/L^2\;,
\label{eq:lambda_m}
\end{equation}
and we introduce the dimensionless time:
\begin{equation}
t=D\tau/L^2\;.
\label{eq:temps_adimensionne_sortie}
\end{equation}
Note that the solutions $\alpha$ depend only on $m$ and $\tilde{\kappa}$, hence the tail distribution is a function of $t$, $m$, $\tilde{\kappa}$, and the starting point $x$:
\begin{align*}
&\mathbb{P}(T_x>\tau)=P_x(t,m,\tilde{\kappa})\;,\\
&\rho_{T_x}(\tau)=-\frac{\partial P_x}{\partial \tau}=-\frac{L^2}{D}\frac{\partial P_x}{\partial t}=\frac{L^2}{D} \rho_x(t,m,\tilde{\kappa})\;,
\end{align*}
$\rho_x(t,m,\tilde{\kappa})$ being the probability density function of the dimensionless random variable $D T_x/L^2$.

We consider now the limit $m\to \infty$.
We recall that $\tilde{\kappa}={\kappa l}/{D}={\kappa L}/{(mD)}$,
hence $\tilde{\kappa}$ depends on $m$ if $\kappa$, $D$, $L$ are fixed. However in what follows we consider $\tilde{\kappa}$ and $m$ as independent parameters.
From Eq. \eqref{eq:lambda_m} we get that only the smallest solutions $\alpha$ contribute to the sum in Eqs. \eqref{eq:formule_temps_sortie_P} and \eqref{eq:formule_temps_sortie_f}, hence we use Eq. \eqref{eq:DL_petit_alpha_relax} which immediately implies that in the $m\to \infty$ limit all the curves fall on a unique master curve of the variable ${\tilde{\kappa}t}/{(\tilde{\kappa}+1)}$:
\begin{align}
&P_x(t,m,\tilde{\kappa})\approx P^*_x\left(\frac{\tilde{\kappa}t}{\tilde{\kappa}+1}\right)\;, \label{eq:temps_sortie_master1}\\
&\rho_x(t,m,\tilde{\kappa})\approx \frac{\tilde{\kappa}}{\tilde{\kappa}+1}\rho^*_x\left(\frac{\tilde{\kappa}t}{\tilde{\kappa}+1}\right)\;.\label{eq:temps_sortie_master2}
\end{align}
This master curve ($P^*_x$, $\rho^*_x$) is precisely the one corresponding to an interval without any barriers ($\tilde{\kappa}\to\infty$). The interpretation is that a very large number of barriers can be modeled as an effective medium with the diffusion coefficient $D_{app}={D\tilde{\kappa}}/{(\tilde{\kappa}+1)}$. In particular, one obtains the formula for the mean first exit time:
\begin{equation}
\mathbb{E}[T_x]=\frac{x(L-x)}{2D_{app}}=\frac{x(L-x)}{2D}\frac{\tilde{\kappa}+1}{\tilde{\kappa}}\;.
\label{eq:temps_sortie_moyen}
\end{equation}

Note that from the second equality in Eq. \eqref{eq:DL_petit_alpha_relax} we get that one should replace $\tilde{\kappa}$ by $\tilde{\kappa}\left(1+\frac{2}{m}\right)$ in order to obtain the scaling laws \eqref{eq:temps_sortie_master1} and \eqref{eq:temps_sortie_master2}, and thus Eq. \eqref{eq:temps_sortie_moyen}, to the first order in $1/m$.

\subsection{Irregular geometry}

Now we turn to an irregular geometry: the lengths of the intervals and the permeabilities of the inner barriers are randomly distributed. We still impose that the whole interval has a constant length $L$.
If the number of compartments $m$ is sufficiently large, we expect that the effective medium description still holds, with an effective value of $\tilde{\kappa}$.
The formula for $\tilde{\kappa}$ should involve all the lengths $l_i$ and permeabilities $\kappa_{i,i+1}$. Moreover in the case of a regular geometry, $l_i=l$ and $\kappa_{i,i+1}=\kappa$, and one should retrieve $\tilde{\kappa}=\kappa l/D$.
If $l_i$ and $\kappa_{i,i+1}$ are independent, we find numerically that the formula
\begin{equation}
\tilde{\kappa}=\frac{\langle l \rangle }{\langle r \rangle D}\;,
\label{eq:kappa_tilde_aleatoire}
\end{equation}
where $\langle\cdot\rangle$ denotes arithmetic mean, works well for large values of $m$ (typically, $m\gtrsim 100$).
As a consequence, an irregular geometry does not differ from a regular geometry provided that the number of compartments is sufficiently large, when one replaces $l$ by $\langle l \rangle $ and $r$ by $\langle r \rangle $.
%
%Because $\langle l \rangle =L/m$, where $L$ is the fixed length of the whole interval, the probability density does not depend on the precise distribution of $l$ in the regime of large number of compartments.
%
%The dependence on the permeability distribution is more intricate. In particular, \cref{eq:kappa_tilde_aleatoire} holds only if $\langle r \rangle $ is finite.
%If that is the case, the law of large numbers ensures the self-averaging of the distribution. In turn, the variance of $r$ is related to the relative deviations of the distribution from the limit $m=\infty$. On the other hand, if $\langle r \rangle $ is infinite (for example if $\kappa$ is uniformly distributed on an interval $[0,2\kappa_0]$, with $\kappa_0$ a constant), there is no self-averaging and \cref{eq:kappa_tilde_aleatoire} does not hold.

\begin{figure*}[tbp]
\begin{center}
\includegraphics[width=0.9\linewidth]{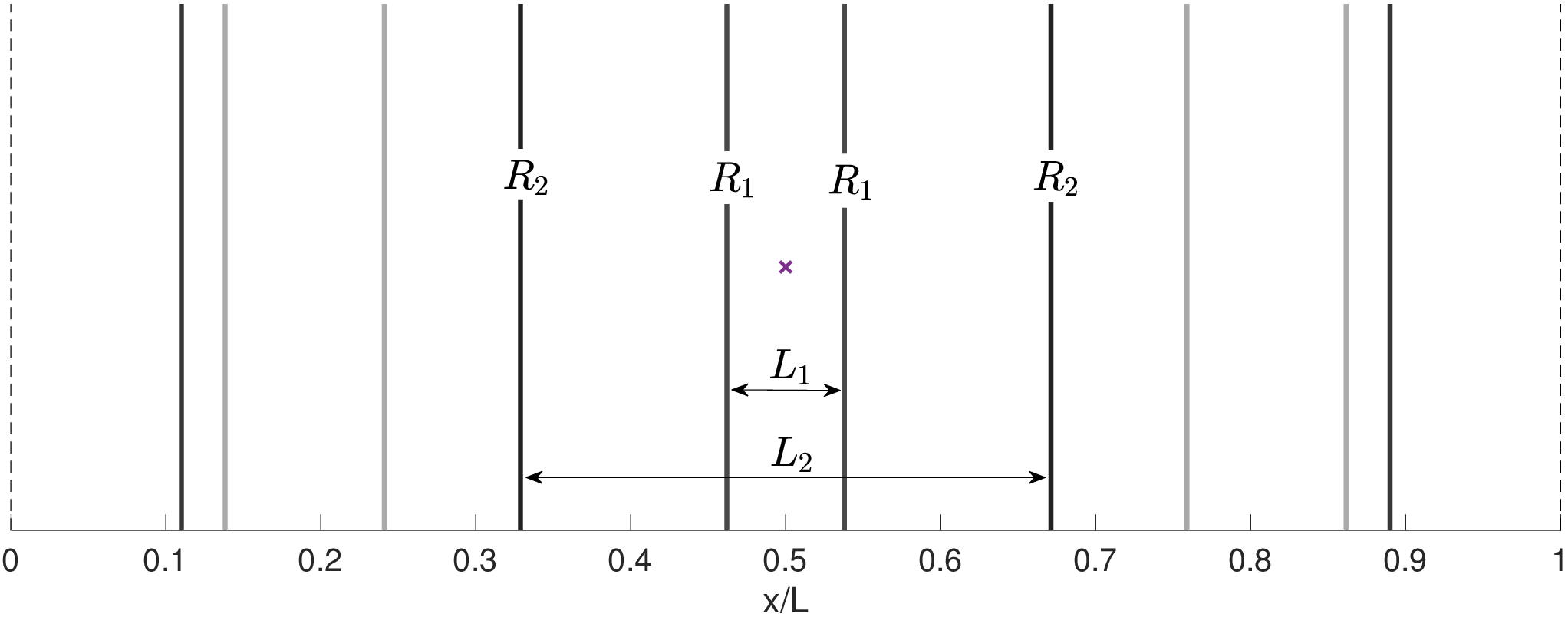}
\end{center}
\caption{An example of a random symmetric structure (with $m=11$ compartments). The solid vertical lines picture the barriers (the darker the line, the higher the resistance of the barrier). One can see this structure as nested subintervals of lengths $L_1<L_2<\ldots$ enclosed by barriers of resistances $R_1,R_2,\ldots$. The cross indicates the starting position of the particles, $x=L/2$.
}
\label{fig:Structure_test}
\end{figure*}

However, this formula fails at small values of $m$. The following reasoning suggests indeed that the formula of $\tilde{\kappa}$ should involve a correlation between the position of the barriers and their resistances. Let us assume for simplicity that the lengths of the compartments are randomly generated in such a way that the geometry is symmetric with respect to the middle of the interval (and that $m$ is odd). One can then see the structure as $(m-1)/2$ nested subintervals $I_1\subset I_2 \subset \dots \subset [0,L]$ of sizes $L_1 < L_2 <\dots <L$ and enclosed by barriers of resistances $R_1, R_2, \ldots, R_{(m-1)/2}$ (see Fig. \ref{fig:Structure_test}). 
We let a large number of particles diffuse from $x=L/2$. First they diffuse inside the first subinterval $I_1$, so that they \textquote{feel} $\tilde{r}_1=DR_1/L_1$. Let us assume that the barriers are quasi-impermeable, that is $\tilde{r}_1\gg 1$. According to Eq. \eqref{eq:temps_sortie_moyen}, after a time $T_1\sim L_1^2\tilde{r}_1(8/D)\sim L_1R_1$ they have crossed the first barriers. The particle density is then quite homogeneous inside the second subinterval $I_2$ and so the particles feel $\tilde{r}_2=DR_2/L_2$. After a time $T_2\sim L_2R_2$ they cross the second barriers, they homogenize inside the third subinterval, and so on. The mean exit time is thus proportional to $\sum_{i=1}^{(m-1)/2} R_i L_i$. According to Eq. \eqref{eq:temps_sortie_moyen} and to the condition that we recover $\tilde{r}=rD/l$ for a regular geometry in the $m=\infty$ limit, one can guess:
\begin{equation}
\tilde{\kappa}=\tilde{r}^{-1}=\frac{L^2}{4D}\left(\sum_{i=1}^{(m-1)/2}R_iL_i\right)^{-1}=\frac{L^2}{4D}\left(\sum_{i=1}^{m-1} r_{i,i+1}\left|x_{i,i+1}-L/2\right|\right)^{-1}\;.
\label{eq:kappa_tilde_aleatoire_complique}
\end{equation}
Interestingly, the correction $\tilde{\kappa}\to\tilde{\kappa}\left(1+\frac{2}{m}\right)$ is contained in this formula in case of a regular geometry (see Sec. \ref{section:temps_sortie_regular}).
This formula was obtained for a symmetric geometry and it has to be refined for asymmetric geometries. In particular, it is not clear how it should be changed if the starting point $x$ is not at the middle of the interval anymore. 
The same reasoning suggests a formula such as:
\begin{equation}
\tilde{\kappa}=\frac{x^2}{4D}\left(\sum_{i=1}^{i_0-1} r_{i,i+1}(x-x_{i,i+1})\right)^{-1} + \frac{(L-x)^2}{4D}\left(\sum_{i=i_0}^{m-1} r_{i,i+1}(x_{i,i+1}-x)\right)^{-1}\;,
\end{equation}
if $x\in \Omega_{i_0}$. However the numerical agreement is not as good as with a symmetric geometry and $x=L/2$. Therefore we focus on Eq. \eqref{eq:kappa_tilde_aleatoire_complique} in the following. Note that Eq. \eqref{eq:kappa_tilde_aleatoire_complique} gives different weights to the barriers depending on their position with respect to the middle of the interval, which is rather intuitive. Indeed one expects a barrier located exactly at the middle of the interval to have no effect at all (given the symmetry of the geometry) whereas barriers located near the exit points should have the greatest effect.

\begin{figure*}[tb]
\begin{center}
\includegraphics[width=0.9\linewidth]{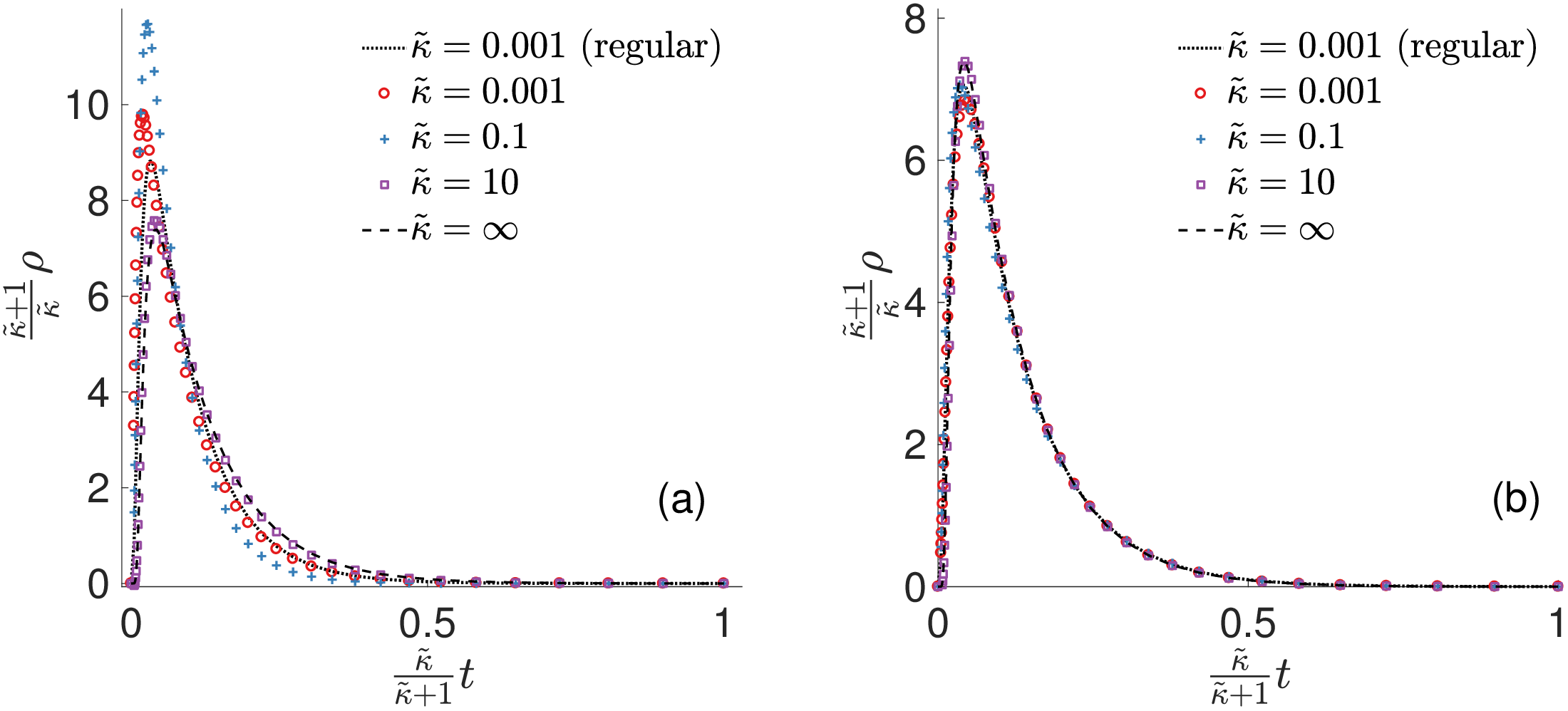}
\end{center}
\caption{The probability density of the first exit time from an interval segmented into $m=11$ compartments by random barriers of variable mean resistance (such as in Fig. \ref{fig:Structure_test}). We apply the scale change: $t\to\tilde{\kappa}t/(\tilde{\kappa}+1)$, where $\tilde{\kappa}$ is computed either with Eq. \eqref{eq:kappa_tilde_aleatoire} or Eq. \eqref{eq:kappa_tilde_aleatoire_complique}. The dotted and dashed lines correspond to a regular geometry with quasi-impermeable and permeable barriers, respectively.  
(a) $\tilde{\kappa}$ is computed with Eq. \eqref{eq:kappa_tilde_aleatoire}. One can see that the curves corresponding to the regular geometry do not coincide very well, while the curves corresponding to the random structures exhibit large deviations between each other. 
(b) $\tilde{\kappa}$ is computed with Eq. \eqref{eq:kappa_tilde_aleatoire_complique}. Visually, all the curves fall onto one master curve.
%(c) $m=100$ compartments, $\tilde{\kappa}$ is computed with \cref{eq:kappa_tilde_aleatoire}. Now the curves corresponding to the regular geometry coincide, but there are still some little deviations of the curves corresponding to the random structures. 
%(d) $m=10$ compartments, $\tilde{\kappa}$ is computed with \cref{eq:kappa_tilde_aleatoire_complique}. All the curves are visually superimposed.
%
}
\label{fig:Temps_sortie_master_comparaison}
\end{figure*}

If the permeabilities of the barriers and the lengths of the compartments are independent random variables and are distributed in a way that $\langle r \rangle$ is finite, then Eqs. \eqref{eq:kappa_tilde_aleatoire} and \eqref{eq:kappa_tilde_aleatoire_complique} are identical in the limit $m\to \infty$. Furthermore, according to the central limit theorem we expect their deviation to be of order $m^{-1/2}$. Figure \ref{fig:Temps_sortie_master_comparaison} shows a comparison of the two formulas. We have plotted the first exit time distribution for random structures such as the one shown in Fig. \ref{fig:Structure_test}, with $m=11$ compartments. The lengths of the compartments and the barrier resistances follow an exponential distribution. 
We choose various mean values of the barrier resistances and we compute $\tilde{\kappa}$ according to Eq. \eqref{eq:kappa_tilde_aleatoire} or Eq. \eqref{eq:kappa_tilde_aleatoire_complique}. Then we apply the scaling $t\to \tilde{\kappa}t/(\tilde{\kappa}+1)$. One can see that with Eq. \eqref{eq:kappa_tilde_aleatoire_complique} all the curves fall onto one master curve, whereas Eq. \eqref{eq:kappa_tilde_aleatoire} leads to significant deviations.
Even though Eq. \eqref{eq:kappa_tilde_aleatoire} is less accurate than Eq. \eqref{eq:kappa_tilde_aleatoire_complique}, the latter involves the correlation between the position of the barriers and their permeabilities, which may be unknown in actual experiments. In this case one should use Eq. \eqref{eq:kappa_tilde_aleatoire}, which is more \textquote{universal}.

Let us conclude this section by the investigation of the particular case $\tilde{\kappa}\ll 1$. As discussed previously, in this regime the intra-compartment diffusion is much faster than the inter-compartment exchange, hence our diffusion model becomes equivalent to a random walk process on a discrete one-dimensional lattice of size $m$. The hopping rate from site $i$ to site $i+1$ and from site $i$ to site $i-1$ are respectively given by:
\begin{equation}
W_{i\to i+1}=\frac{\kappa_{i,i+1}}{l_i}\;,\quad \text{ and } \quad
W_{i\to i-1}=\frac{\kappa_{i-1,i}}{l_i}\;.
\label{eq:taux_aleatoires}
\end{equation}
Such models of discrete random walks with random hopping rates have been considered by many authors \cite{Alexander81,Sinai83,Bernasconi82,Azbel82,Derrida83}, and in particular from the perspective of first exit times \cite{Noskowicz88,LeDoussal89,Murthy89,Kehr90,Raykin93,LeDoussal99}. In particular, Murthy and Kehr discuss in \cite{Murthy89} various cases for the distribution of the hopping rates $W_{i\to i+1}$. 
They consider discrete random walks starting from the left endpoint (site $0$, reflecting condition) of the lattice and analyze the first exit time through the right endpoint (site $N$, absorbing condition).
%
%They work with discrete random walks and they consider the first exit time from the left endpoint (site $0$, reflecting condition) of the lattice to the right endpoint (site $N$, absorbing condition).
%
By reflecting the whole lattice with respect to the left endpoint, it is equivalent to a symmetric geometry with a starting point at the middle of the interval (and $m=2N+2$).
In two particular cases they obtain exact formulas for the mean first exit time:

\begin{itemize}
\item \textquote{Symmetric case}, with $W_{i\to i+1}=W_{i+1\to i}$, which in our case corresponds to $l_i=l_{i+1}=l$. The mean exit time is then given by
\begin{align*}
\mathbb{E}[T]&\overset{MK 89}{=}\sum_{i=1}^{N}\frac{i}{W_{m/2+i\to m/2+i+1}}=\frac{1}{2}\sum_{i=1}^{m-1} |i-m/2|l r_{i,i+1}\\
&=\frac{1}{2}\sum_{i=1}^{m-1}|x_{i,i+1}-L/2|r_{i,i+1}=\frac{L^2}{8D\tilde{\kappa}}\;.
\end{align*}
The first equality is from \cite{Murthy89} (with suitable changes of notations). Using Eq. \eqref{eq:taux_aleatoires}, we obtain at the end the same formula as Eq. \eqref{eq:temps_sortie_moyen} (recall that $\tilde{\kappa}\ll 1$ and $x=L/2$), where $\tilde{\kappa}$ is given by Eq. \eqref{eq:kappa_tilde_aleatoire_complique}.

\item \textquote{Random sojourn probabilities}, with $W_{i\to i+1}=W_{i\to i-1}$, which translates into $r_{i,i+1}=r_{i-1,i}=r$. The mean exit time is given by
\begin{align*}
\mathbb{E}[T]&\overset{MK 89}{=}\sum_{i=1}^{N}\frac{i}{W_{m-i\to m-i+1}}=\sum_{i=1}^N \sum_{k=1}^{i} l_{m-i}r=\sum_{k=1}^{N}\left(\sum_{i=1}^{k}l_{m/2+i}\right) r\nonumber\\
& = \frac{1}{2}\sum_{k=1}^{m-1}|x_{k,k+1}-L/2|r=\frac{L^2}{8D\tilde{\kappa}}\;.
\end{align*}
Again, the first equality is from \cite{Murthy89}. By rearranging the sum, it transforms exactly into Eq. \eqref{eq:temps_sortie_moyen}.
\end{itemize}
We conclude that our formula Eq. \eqref{eq:kappa_tilde_aleatoire_complique} introduces an effective permeability $\tilde{\kappa}$ which is consistent with the predictions of the random hopping rate models and accurately describes the first exit time distribution even for moderate number of barriers.

\section{Mathematical proofs}
\label{section:non-degeneracy}

In this section we prove the non-degeneracy of the eigenvalues of the diffusion operator under the assumption that all inner membranes are semi-permeable $\kappa_{i,i+1} > 0, i=1,\ldots,m-1$. In fact this statement involves two facts: (i) the eigenvalues $\lambda_n$ of the diffusion operator are distinct; (ii) the zeros of $F$ are simple, that is $F'(\lambda_n)\neq 0, n=1,2,\ldots$ (in this section, prime denotes derivative with respect to $\lambda$). Furthermore we shall obtain as a corollary that there are infinitely many eigenvalues $\lambda_n$, that they grow monotonically with the inner and outer barrier permeabilities $\kappa_{i,i+1}$ and $K_\pm$, \textcolor{black}{as well as a Courant nodal theorem for the eigenmodes.}

The assumption of non-zero permeability is crucial. Indeed it is clear that any inner impermeable barrier would split the structure into two non-communicating parts. The eigenmodes for the whole structure would then  be given by the eigenmodes for one part and the other separately. If the two parts are identical, each eigenvalue is twice degenerate. We make no other assumption about the geometry and we consider general relaxing outer boundary conditions.

\subsection{Uniqueness of the eigenmodes}
\label{section:unicity}
Let us assume that there exist two eigenmodes $u$ and $\tilde{u}$ satisfying Eqs. \eqref{eq:equation_generale_modes}-\eqref{eq:equation_normalisation},
with the same eigenvalue $\lambda$. We shall prove that $u$ is proportional to $\tilde{u}$.
% by induction on the index of the compartment $i$.
%
Because $u$ and $\tilde{u}$ both satisfy Eq. \eqref{eq:equation_bords_1}, one has $\frac{u'(0)}{u(0)}=\frac{\tilde{u}'(0)}{\tilde{u}(0)}$ hence there exists a constant $A$ such that
\begin{equation*}
u(0)-A\tilde{u}(0)=0\; \quad \text{ and } \quad u'(0)-A\tilde{u}'(0)=0\;.
\end{equation*}
Let us denote $u-A\tilde{u}$ by $w$. This function satisfies Eqs. \eqref{eq:equation_generale_modes}-\eqref{eq:equation_normalisation}
because all these equations are linear. What remains to show is that $w$ is equal to $0$ over the whole interval $[0,L]$. We prove it by induction on the index of the compartment $i$. The main mathematical argument is Cauchy-Lipschitz uniqueness theorem for second order linear differential equations (U): \textquote{if $f$ satisfies a second order linear differential equation over an interval $\Omega$ and $f(c)=f'(c)=0$, with $c\in \Omega$, then $f(x)=0$ for every $x\in \Omega$}.

\begin{itemize}
\item We apply (U) to $w\restrict{\Omega_1}$: $w\restrict{\Omega_1}(0)=w\restrict{\Omega_1}'(0)=0$ and $D_1 w''\restrict{\Omega_1}+\lambda w\restrict{\Omega_1}=0$, hence $w\restrict{\Omega_1}=0$.
\item Let us assume that $w\restrict{\Omega_i}=0$, with $0<i<m-1$. Then, because $\kappa_{i,i+1}\neq 0$, the inner boundary conditions in Eqs. \eqref{eq:equation_membranes_1} and \eqref{eq:equation_membranes_2} imply that $w\restrict{\Omega_{i+1}}(x_{i,i+1})=w'\restrict{\Omega_{i+1}}(x_{i,i+1})=0$. Because $w\restrict{\Omega_{i+1}}$ obeys the equation $D_{i+1}w''\restrict{\Omega_{i+1}}+\lambda w\restrict{\Omega_{i+1}}=0$, one can apply again (U), which implies $w\restrict{\Omega_{i+1}}=0$.
\end{itemize}

\subsection{Simplicity of the zeros of $F$}
\label{section:simplicity}
Now we prove that $F'(\lambda_n) \neq 0$ for any eigenvalue $\lambda_n$. In order to simplify the notations we consider the case where $K_\pm$ are finite. However the proof follows the same steps in the case of infinite $K_\pm$. Throughout the proof we implicitly discard the case $\lambda=0$. Let us recall that if we consider the function $v(\lambda,x)$ which satisfies Eqs. \eqref{eq:equation_generale_modes}-\eqref{eq:equation_bords_1}
as well as the condition $v(0)=1$ (we have proven above that this function is unique), then 
\begin{equation}
F(\lambda)=\frac{K_+}{D_m}v(\lambda,L)+\frac{\partial v}{\partial x}(\lambda,L)\;.
\label{eq:formule_F_bis}
\end{equation}
Instead of writing $v$ as a sum of sine and cosine functions (see Eq. \eqref{eq:formule_modes_gauche}), we introduce an amplitude and phase representation:
\begin{equation}
{v\restrict{\Omega_i}}(x)
%= a^l_i\cos(\sqrt{\lambda/D_i}(x-x_{i-1,i}))+b^l_i\sin(\sqrt{\lambda/D_i}(x-x_{i-1,i}))
=A_i(\lambda) \cos(\sqrt{\lambda/D_i}x+\phi_i(\lambda))=A_i(\lambda)\cos(\Phi_i(\lambda,x))\;,
\end{equation}
with $A_i \geq0$.
It is clear from Eq. \eqref{eq:formule_modes_gauche} that $A_i$ and $\phi_i$ do not depend on $x$. Moreover we have proven in the above paragraph that $A_i(\lambda)$ is non-zero for all $i$ and $\lambda$.
We now translate the boundary conditions \eqref{eq:equation_membranes_1}-\eqref{eq:equation_bords_2}
in terms of $\Phi_i$.
Equation \eqref{eq:equation_bords_1} yields: $K_-A_i\cos\phi_1 + \sqrt{\lambda D_1}\sin\phi_1=0$, hence
\begin{equation}
\tan\phi_1=-\frac{K_-}{\sqrt{\lambda D_1}}\;\qquad (-\pi/2 \leq \phi_1 \leq 0)\;.
\label{eq:condition_initiale_ampli_phase}
\end{equation}
Equtaions \eqref{eq:equation_membranes_1} and \eqref{eq:equation_membranes_2} can be restated as
\begin{equation*}
-A_i\sqrt{\lambda D_i}\sin(\Phi_i)=-A_{i+1}\sqrt{\lambda D_{i+1}}\sin(\Phi_{i+1})
=\kappa_{i,i+1}(A_{i+1}\cos(\Phi_{i+1})-A_{i}\cos(\Phi_i)) \;
\end{equation*}
at $x=x_{i,i+1}$, hence by eliminating $A_i$ and $A_{i+1}$, we get
\begin{equation}
\frac{\cot{\Phi_{i}(\lambda,x_{i,i+1})}}{\sqrt{D_i}}-\frac{\cot{\Phi_{i+1}(\lambda,x_{i,i+1})}}{\sqrt{D_{i+1}}}=r_{i,i+1}\sqrt{\lambda}\;,
\label{eq:condition_barrieres_ampli_phase}
\end{equation}
with $0\leq \Phi_{i+1}(\lambda,x_{i,i+1})-\Phi_{i}(\lambda,x_{i,i+1})<\pi$.
Finally, one can rewrite Eq. \eqref{eq:formule_F_bis} as
\begin{align}
F(\lambda)&=A_m(\lambda)\left(\frac{K_+}{D_m}\cos{\Phi_m(\lambda,L)} -\sqrt{\lambda/D_m}\sin{\Phi_m(\lambda,L)}\right)\nonumber\\
&=A_{m+1}(\lambda)\cos(\Phi_m(\lambda,L)+\phi_{m+1}(\lambda))\;,
\label{eq:F_ampli_phase}
\end{align}
with:
\begin{equation}
A_{m+1}(\lambda)=A_m(\lambda)\sqrt{\left(\frac{K_+}{D_m}\right)^2+\frac{\lambda}{D_m}}\;,\qquad \qquad \cot{\phi_{m+1}(\lambda)}= \frac{K_+}{\sqrt{\lambda D_m}}\;
\label{eq:ampli_phase_bord_droit}
\end{equation}
and $0\leq\phi_{m+1}\leq\pi/2$.
We have $A_{m+1}(\lambda)\neq 0$ for any $\lambda$ and $-\pi/2<\Phi_m(0,L)+\phi_{m+1}(0)\leq\pi/2$, hence Eq. \eqref{eq:coherence_condition3} is equivalent to $\Phi_m(\lambda_n,L)+\phi_{m+1}(\lambda_n)=(2n-1)\pi/2$.
The derivative of $F$ at $\lambda=\lambda_n$ is then given by
%\begin{align}
%F'(\lambda_n)&=-A_m(\lambda_n)\left(\Phi'_m(\lambda_n,L)\left(\frac{K_+}{D_m}\sin{\Phi_m(\lambda_n,L)} +\sqrt{\frac{\lambda_n}{D_m}}\cos{\Phi_m(\lambda_n,L)}\right)+\frac{1}{2\sqrt{\lambda D_m}} \sin{\Phi_m(\lambda_n,L)}\right)\nonumber\\
%&=-A_m(\lambda_n)\left(\sqrt{\frac{\lambda_n}{D_m}}\frac{\Phi'_m(\lambda_n,L)}{\cos{\Phi_m(\lambda_n,L)}}+\frac{K_+}{2\lambda_n D_m}\sin{\Phi_m(\lambda_n,L)}\right)\;,
%\end{align}
\begin{equation}
F'(\lambda_n)=(-1)^{n}A_{m+1}(\lambda_n)\left(\Phi_m'(\lambda_n,L)+\phi_{m+1}'(\lambda_n)\right)\;.
\end{equation}
It is clear from Eq. \eqref{eq:ampli_phase_bord_droit} that $\phi_{m+1}'(\lambda)\geq 0$ for any $\lambda$. In order to prove that $F'(\lambda_n)\neq 0$, it is then sufficient to show that $\Phi_m'(\lambda,L)>0$.
We prove by induction on the index of the compartment $i$ that $\Phi_i'(\lambda,x)$ is positive for any $\lambda$ and any $x\in \Omega_i$:
\begin{itemize}
\item From Eq. \eqref{eq:condition_initiale_ampli_phase} we get that $\phi_1$ is an increasing function of $\lambda$. As $\Phi_1(\lambda,x)=\sqrt{\lambda/D_1}x+\phi_1(\lambda)$, we immediately get that $\Phi_1'(\lambda,x)>0$ for any $x\in \Omega_1$.
\item Let us assume that $\Phi_i(\lambda,x_{i,i+1})$ is an increasing function of $\lambda$. 
According to Eq. \eqref{eq:condition_barrieres_ampli_phase}, let us introduce the function: 
\begin{equation}
f(\lambda,y)=\cot^{-1}\left(\sqrt{\frac{D_{i+1}}{D_i}}\cot{y}-r_{i,i+1}\sqrt{\lambda D_{i+1}}\right)\;.
\label{eq:fonction_f}
\end{equation}
Because $\cot$ is a decreasing function, $f$ is an increasing function of $y$ and a non-decreasing function of $\lambda$, which implies that $\Phi_{i+1}(\lambda,x_{i,i+1})=f(\lambda,\Phi_i(\lambda,x_{i,i+1}))$ is an increasing function of $\lambda$. 
It is then clear that $\Phi_{i+1}(\lambda,x)=\Phi_{i+1}(\lambda,x_{i,i+1})+\sqrt{\lambda/D_{i+1}} (x-x_{i,i+1})$ is an increasing function of $\lambda$ for any $x \in \Omega_{i+1}$.
\end{itemize}
This proves the simplicity of the zeros of $F$. Moreover, we also obtain that $\Phi_m(\lambda,L)$ grows indefinitely with $\lambda$. According to Eq. \eqref{eq:F_ampli_phase}, this implies that there are infinitely many values of $\lambda$ such that $F(\lambda)=0$. In other words, there are infinitely many eigenvalues $\lambda_n$.

\subsection{Monotonicity of the eigenvalues with respect to the permeabilities}
\label{section:monotonicity}
The previous computations enable us to show that the eigenvalues grow monotonically with the inner and outer permeabilities $\kappa_{i,i+1}$ and $K_\pm$. In fact, because $\Phi_m(\lambda,L)+\phi_{m+1}(\lambda)$ is an increasing function of $\lambda$, we just have to prove that $\Phi_m(\lambda,L)+\phi_{m+1}(\lambda)$ is a non-increasing function of $\kappa_{i,i+1}$ and $K_\pm$, which follows immediately from Eqs. \eqref{eq:condition_initiale_ampli_phase}, \eqref{eq:fonction_f} and \eqref{eq:ampli_phase_bord_droit}.

{\color{black}
\subsection{Courant nodal theorem}
\label{section:courant_nodal_theorem}
Let us define the nodal domains of an eigenmode $u_n$ as connected components on which $u_n$ does not change sign. We prove here that $u_n$ has exactly $n$ nodal domains, which means that it changes sign $n-1$ times (recall that we numbered the modes $n=1,2,\ldots$). Note that these sign changes can occur at discontinuity points of $u_n$. The proof relies on the amplitude and phase representation detailed above. Let us then write
\begin{equation}
u_n(x)=A(\lambda_n,x)\cos(\Phi(\lambda_n,x))\;,
\end{equation}
where $A$ and $\Phi$ are piecewise continuous functions of $x$ defined by $A\restrict{\Omega_i}=A_i$ and $\Phi\restrict{\Omega_i}=\Phi_i$. The changes of sign of the eigenmode occur when the phase $\Phi$ crosses an odd multiple of $\pi/2$. Indeed, $A(\lambda_n,x)$ has a constant sign, and from Eq. \eqref{eq:condition_barrieres_ampli_phase} we get that the jumps of $\Phi$ at the barriers are always less than $\pi$ (which means that $\Phi$ cannot cross two odd multiples of $\pi/2$ at the same time). 

Moreover, we know the phase at the left endpoint: $\Phi(\lambda_n,0)=\phi_1(\lambda_n) \in [-\pi/2;0]$ and the  phase at the right endpoint: $\Phi(\lambda_n,L)=(2n-1)\pi/2 - \phi_{m+1}(\lambda_n) \in [(n-1)\pi;(n-1)\pi+\pi/2]$.
We conclude that the interval $(\Phi(\lambda_n,0);\Phi(\lambda_n,L))$ contains exactly $n-1$ odd multiple of $\pi/2$, thus the eigenmode has $n$ nodal domains.
}
%%%%%%%%%%%%%%%%%%%%%%%%%%%%%%%%%%%%%%%%%%%%%%%%%%%%%%

%%%%%%%%%%%%%%%%%%%%%%%%%%%%%%%%%%%%%%%%%%%%%%%%%%%%%%%%%%%%%%%%%%%%%%%%
\section{Computations for an array of identical cells with symmetric relaxation conditions at the outer boundaries}
\label{section:autres_structures_relax}
In this section we extend the computation presented in Sec. \ref{section:modes_simple_periodic} by allowing relaxation or leakage at the endpoints of the interval. In other words, we relax the reflecting boundary conditions $K_\pm=0$ at the outer membranes. In particular we will also study the limit $K_\pm\to \infty$ which is the perfectly relaxing case that we use in Sec. \ref{section:application_temps_sortie}.
The cells are the same: $l_i=l, D_i=D, \kappa_{i,i+1}=\kappa$, and the relaxation coefficients are identical: $K_+=K_-=K$. In addition to the notations \eqref{eq:notations_simple}, we introduce: $\tilde{K}=Kl/D$.
%
%We consider in this appendix an array of identical cells with symmetric relaxing conditions at the outer boundaries ($K_-=K_+=K$).

\subsection{Eigenmodes}

Because the geometry is symmetric we know that $\epsilon=\pm1$.
%We use again the notations \eqref{eq:notations_simple}, additionally we introduce:
%\begin{equation}
%\tilde{K}=\frac{Kl}{D}
%\label{eq:notations_relaxation}
%\end{equation}
In this case we need to solve the general equation \eqref{eq:coherence_periodic}
\begin{equation}
\mathcal{K}^{-1}
\mathcal{M}^{m}
\begin{bmatrix}
{\alpha}\\\tilde{K}
\end{bmatrix}
=\epsilon
\begin{bmatrix}
{\alpha} \\ -\tilde{K}
\end{bmatrix}\;.
\label{eq:coherence_periodic_relaxation}
\end{equation}
With the help of Eq. \eqref{eq:M_puissance_simple} we can compute the matrix $\mathcal{K}^{-1}\mathcal{M}^{m}$:
\begin{equation}
\mathcal{K}^{-1}\mathcal{M}^{m}=
\begin{bmatrix}
\cos\alpha \frac{\sin m\psi}{\sin \psi}-\frac{\sin (m-1)\psi}{\sin \psi}&
\sin\alpha\frac{\sin m\psi}{\sin \psi}+\tilde{r}\alpha\frac{\sin (m-1)\psi}{\sin \psi}\\
-\sin\alpha\frac{\sin m\psi}{\sin \psi}&
\cos\alpha \frac{\sin m\psi}{\sin \psi}-\frac{\sin (m-1)\psi}{\sin \psi}
\end{bmatrix}\;.
\label{transition_periodic}
\end{equation}
Thus Eq. \eqref{eq:coherence_periodic_relaxation} yields the system
\begin{equation}
\begin{cases}
\left(\cos\alpha+\tilde{K}\frac{\sin\alpha}{\alpha}\right)\frac{\sin m\psi}{\sin\psi}-\left(1-\tilde{r}\tilde{K}\right)\frac{\sin(m-1)\psi}{\sin\psi}=\pm1\\
\left(\cos\alpha-\frac{1}{\tilde{K}}\alpha\sin\alpha\right)\frac{\sin m\psi}{\sin\psi}-\frac{\sin(m-1)\psi}{\sin\psi}=\mp1
\end{cases}\;,
\label{eq:coherence_relaxation_1}
\end{equation}
which is equivalent to the equation
\begin{equation}
\left(\cos\alpha+\frac{1}{2}\left(\frac{\tilde{K}}{\alpha}-\frac{\alpha}{\tilde{K}}\right)\sin\alpha\right)\frac{\sin m\psi}{\sin \psi}-\left(1-\frac{\tilde{r}\tilde{K}}{2}\right)\frac{\sin (m-1)\psi}{\sin \psi}=0\;.
\label{eq:coherence_relaxation_2}
\end{equation}
Combined with Eq. \eqref{eq:equation_alpha_psi} it forms a system whose solutions $\alpha_n$ determine the eigenvalues $\lambda_n$.
%It will be useful to identify the solutions $\alpha$ by the sign $\pm$ on the right-hand side of the equation \eqref{eq:coherence_periodic_relaxation}. We will thus refer to them as $\alpha^+_n$ and $\alpha^-_n$ in the following. 
%The presence of the relaxation modifies the solutions $\alpha$ compared to the impermeable case. In general, the $\alpha$ increase with $\tilde{K}$.
Compared to the $K=0$ case from Sec. \ref{section:modes_simple_periodic}, the solutions $\alpha_n$ are modified and in general increase with $\tilde{K}$.

In the particular case $\tilde{K}=2\tilde{\kappa}$, Eq. \eqref{eq:coherence_relaxation_2} simplifies into
\begin{equation}
\frac{\sin m\psi}{\sin \psi}=0\quad \text{ or }\quad \cos\alpha+\frac{1}{2}\left(\frac{\tilde{K}}{\alpha}-\frac{\alpha}{\tilde{K}}\right)\sin\alpha=0\;.
\label{eq:W=2k}
\end{equation}
The first equation gives the $\alpha_{j,p}$ ($p=1,\ldots,m-1$) from the earlier considered $K=0$ case. The second equation gives the solutions of $\cos\psi=\pm1$ that are not multiple of $\pi$ (that we denote as $\alpha_{j,m}$ if $j$ is even and $\alpha_{j,0}$ if $j$ is odd, to be consistent with our previous notations).
The condition $\tilde{K}=2\tilde{\kappa}$ can be interpreted as \textquote{one inner barrier is equivalent to two stacked outer barriers} or equivalently \textquote{the crossing of one inner barrier transforms $\begin{bmatrix}-1\\ \frac{\tilde{K}}{\alpha}\end{bmatrix}$ into $\begin{bmatrix}1\\ \frac{\tilde{K}}{\alpha}\end{bmatrix}$}. In this way the reason why the $\alpha_{j,p}$ are solutions becomes clear: the matrix $\mathcal{K}\left(\mathcal{K}^{-1}\mathcal{M}^m\right)=\mathcal{M}^m$ should send $\begin{bmatrix}1\\ \frac{\tilde{K}}{\alpha}\end{bmatrix}$ onto plus or minus itself. The $\alpha_{j,p}$ (with $1<p<m$) are solutions of $\mathcal{M}^m=\pm\mathcal{I}_2$ and the $\alpha_{j,0}$ and $\alpha_{j,m}$ are such that $\begin{bmatrix}1\\ \frac{\tilde{K}}{\alpha}\end{bmatrix}$ is an eigenvector of $\mathcal{M}$.

\begin{figure*}[tbp]
\begin{center}
\includegraphics[width=0.9\linewidth]{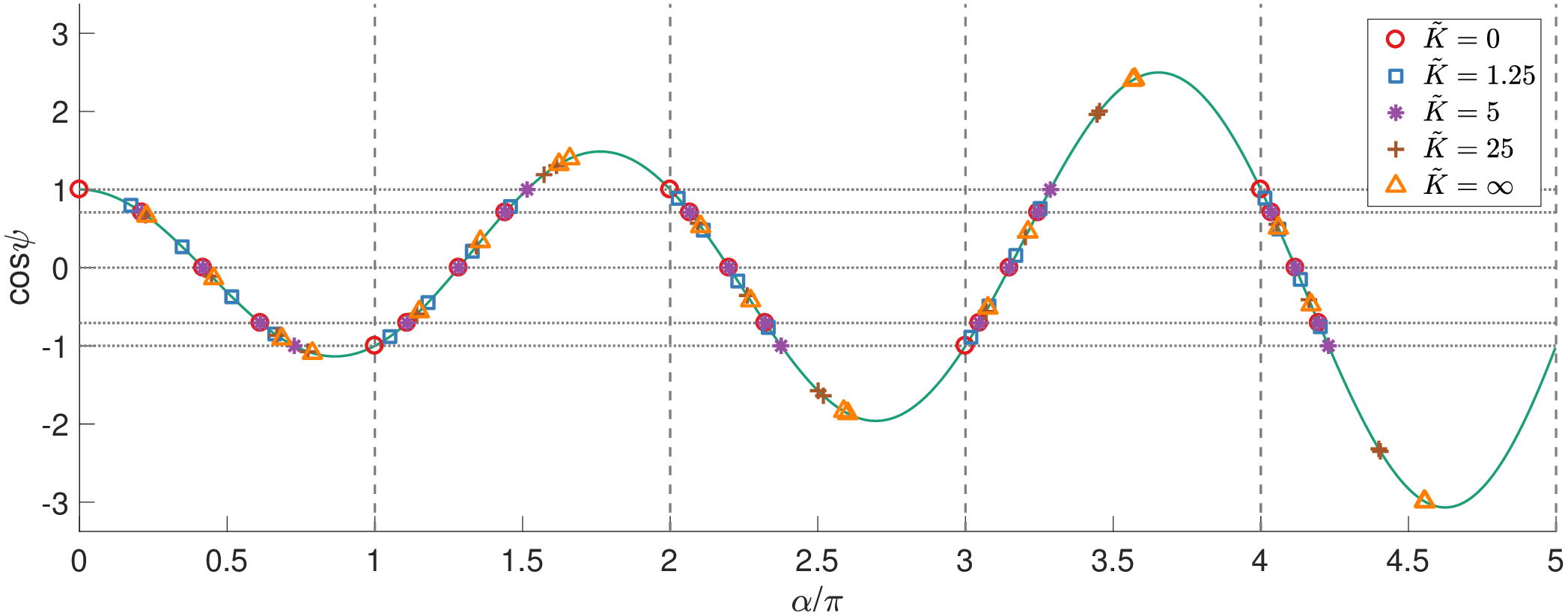}
\end{center}
\caption{Spectrum of the finite periodic geometry with $m=4$ compartments and $\tilde{r}=0.4$, for $\tilde{K}=0$ (circles), $\tilde{K}=\tilde{\kappa}/2=1.25$ (squares), $\tilde{K}=2\tilde{\kappa}=5$ (asterisks), $\tilde{K}=10\tilde{\kappa}=25$ (pluses) and $\tilde{K}=\infty$ (triangles). The values of $\alpha$ increase with $\tilde{K}$. Notice how the spectra for $\tilde{K}=0$ and $\tilde{K}=2\tilde{\kappa}$ coincide except at the beginning and the end of the branches.
}
\label{fig:spectre_W}
\end{figure*}

As a consequence, the spectrum for the case $\tilde{K}=2\tilde{\kappa}$ differs little from the spectrum for the impermeable outer boundary condition. The only difference lies in the beginning and the end of the branches (see Fig. \ref{fig:spectre_W}). This is nevertheless not a small difference because the eigenvalue $\lambda=0$ (which is absent of the spectrum if $\tilde{K}>0$) plays an important role in the long-time limit of the diffusion propagator as we have discussed in Sec.  \ref{section:temps_long}.

%Beyond this special value of $\tilde{K}$, the $\alpha$ continue to increase so that the values at the end of the branches correspond to $\psi=ix$ or $\psi=\pi+ix$, with $x \in \mathbb{R}$.

Beyond this special value of $\tilde{K}$, the solutions $\alpha_n$ continue to increase so that some values of $\psi_n$ become complex (because $|\cos\psi| > 1$, which is apparent in Fig. \ref{fig:spectre_W}). More precisely they have the general form $\psi=ix$ or $\psi=\pi+ix$, with $x \in \mathbb{R}$. These values correspond to eigenmodes strongly localized inside the outer compartments. Indeed, Eq. \eqref{eq:variation_inter_bloc} implies that the coefficients $a$ and $b$ vary like linear combinations of $\cosh$ and $\sinh$ functions of the compartment index $i$. The physical interpretation is simple: when $\tilde{K}\gg\tilde{\kappa}$ we are indeed in a regime where the leakage through the outer membranes is much faster than the exchange through the inner barriers. As a consequence the outer compartments evolve separately from the inner compartments, which corresponds mathematically to the existence of localized eigenmodes. On the other hand, when $\tilde{K} \ll \tilde{\kappa}$, the outer leakage is much slower than the inner exchange, thus all compartments are coupled.
We treat the limit $\tilde{K}\to\infty$ below in Sec. \ref{section:calculs_relax}.

\subsection{Computation of the norm}

The general formula \eqref{eq:normalisation1} reads
%\begin{equation}
%\beta^{-2}=\frac{l}{2}\left|\left(\begin{bmatrix}\frac{\tilde{K}}{\alpha} & 1\end{bmatrix}\frac{\mathrm{d}\mathcal{T}}{\mathrm{d}\alpha}\begin{bmatrix}1\\\frac{\tilde{K}}{\alpha}\end{bmatrix}\right)\mp 2\frac{\tilde{K}}{\alpha^2}\right|_{\alpha=\alpha_n^\pm}\;.
%\end{equation}
\begin{equation}
\beta^{-2}=\frac{l}{2}\left|\frac{\mathrm{d}}{\mathrm{d}\alpha}\left(\begin{bmatrix}\frac{\tilde{K}}{\alpha} & 1\end{bmatrix}\mathcal{T}(\alpha)\begin{bmatrix}1\\\frac{\tilde{K}}{\alpha}\end{bmatrix}\right)\right|_{\alpha=\alpha_n}\;.
\end{equation}
After lengthy computations, one gets
\begin{align}
\beta^{-2}&=\frac{\epsilon l}{2}\frac{\sin\alpha\left(1+\frac{\tilde{r}}{2}\right)+\frac{\tilde{r}}{2}\alpha\cos\alpha}{\sin^2\psi}\left(\left(\frac{\tilde{K}}{\alpha}\right)^2 \sin\alpha +2\frac{\tilde{K}}{\alpha}\cos\alpha-\sin\alpha\right)\left(\frac{(m-1)\sin\psi}{\sin (m-1)\psi}-\cos m\psi\right)\nonumber\\
&+\frac{\epsilon l}{2}\frac{\sin m\psi}{\sin \psi}\left(\frac{\sin\alpha}{\alpha}\left(1+2\tilde{K}+\left(\frac{\tilde{K}}{\alpha}\right)^2\right)+\left(1-\left(\frac{\tilde{K}}{\alpha}\right)^2\right)\cos\alpha\right)\;.
\label{eq:calcul_relax_general_norme}
\end{align}
Note that when $\tilde{K}=2\tilde{\kappa}$ we have to compute separately the cases $\psi=0$ and $\psi=\pi$. We get
\begin{align*}
&\beta^{-2} = 
\frac{ml}{2}\left(-\cos\alpha-2\frac{\tilde{K}}{\alpha}\sin\alpha+\left(\frac{\tilde{K}}{\alpha}\right)^2\left(\cos\alpha+\frac{m-1}{m\tilde{\kappa}}\right)+2\frac{\tilde{K}}{m\alpha^2}\right)\;\qquad \text{if $\psi=0$,}\\
&\beta^{-2} = 
\frac{ml}{2}\left(\cos\alpha+2\frac{\tilde{K}}{\alpha}\sin\alpha-\left(\frac{\tilde{K}}{\alpha}\right)^2\left(\cos\alpha-\frac{m-1}{m\tilde{\kappa}}\right)+2\frac{\tilde{K}}{m\alpha^2}\right)\;\qquad \text{if $\psi=\pi$.}
\end{align*}

\subsection{Fourier transform}

Except for the conditions at the outer boundaries, the geometry is the same as in Sec. \ref{section:Fourier_simple}. Hence the computation follows the same steps. Using the condition \eqref{eq:coherence_periodic_relaxation}, we are led to compute the product
\begin{equation*}
\mathcal{L} \left(\mathcal{I}_2-e^{iq}\mathcal{R}^{-1}\mathcal{K}^{-1}\right)\left(\mathcal{I}_2-\epsilon e^{imq}\mathcal{K}\mathcal{S}\right)\begin{bmatrix}\alpha\\\tilde{K}\end{bmatrix}\;.
\end{equation*}
Skipping the technical computations, one gets depending on $\epsilon =\pm1$
%\begin{equation}
%%\sum_i e^{ikq}\mathcal{L}(q) \begin{bmatrix}a^l_{i}\\b^l_{i} \end{bmatrix} =
%\int_0^L v(x) e^{i\gamma g \delta x} \, \mathrm{d}x= 
%\frac{e^{imq/2} 2l}{(q^2-\alpha)^2(\cos q - \cos \psi)}\times
%\begin{cases}
%\cos(mq/2)\tilde{K}\left[(\cos\alpha-\cos q)+\frac{\tilde{r}}{2}(q\sin q-\alpha \sin \alpha)\right] & \text{if $\epsilon = +1$}
%\\+\sin(mq/2) q(\cos q-\cos \alpha)(1-\frac{\tilde{r}\tilde{K}}{2})\\
%\\
%-i\sin(mq/2){K}\left[(\cos\alpha - \cos q)+\frac{\tilde{r}}{2}(q \sin q - \alpha \sin \alpha)\right]& \text{if $\epsilon = -1$}\\
%+i\cos(mq/2)  q(\cos q - \cos\alpha)(1-\frac{\tilde{r}\tilde{K}}{2})
%\end{cases}
%\end{equation}

\begin{align}
%\sum_i e^{ikq}\mathcal{L}(q) \begin{bmatrix}a^l_{i}\\b^l_{i} \end{bmatrix} =
\int_0^L v(x) e^{i\gamma g \delta x} \, \mathrm{d}x= 
\frac{e^{imq/2} 2l}{(q^2-\alpha)^2(\cos q - \cos \psi)}(A\cos(mq/2) 
+B\sin(mq/2))\;\qquad \text{if $\epsilon = +1$,}\label{eq:calcul_relax_general_fourier0}\\
\int_0^L v(x) e^{i\gamma g \delta x} \, \mathrm{d}x= 
\frac{-ie^{imq/2} 2l}{(q^2-\alpha)^2(\cos q - \cos \psi)}
(A\sin(mq/2)
-B\cos(mq/2))\;\qquad \text{if $\epsilon = -1$,}
\label{eq:calcul_relax_general_fourier1}
\end{align}
where
\begin{equation}
A=\tilde{K}\left((\cos\alpha-\cos q)+\frac{\tilde{r}}{2}(q\sin q-\alpha \sin \alpha)\right)\;,\quad B=q(\cos q-\cos \alpha)\left(1-\frac{\tilde{r}\tilde{K}}{2}\right)\;.
\label{eq:calcul_relax_general_fourier2}
\end{equation}

%\paragraph{Intra- and extra-cellular spaces: } Here the computation is a little different because there are two different $\mathcal{L}$ matrices. We recall that the transition matrix is:
%\begin{equation}
%\mathcal{M}=
%\underbrace{\begin{bmatrix}
%1&\frac{\sqrt{\lambda_n Di}}{{\kappa}}\\
%0&\sqrt{D_i/D_{e}}
%\end{bmatrix}
%\mathcal{R}(-\alpha_{n,i})}_{\mathcal{M}_i}
%\underbrace{\begin{bmatrix}
%1&\frac{\sqrt{\lambda_n D_e}}{{\kappa}}\\
%0&\sqrt{D_e/D_{i}}
%\end{bmatrix}
%\mathcal{R}(-\alpha_{n,e})}_{\mathcal{M}_e} \;.
%\end{equation}
%We thus have:
%\begin{align}
%\sum_i e^{ikql}\mathcal{L}_i(q)\cdot \begin{bmatrix}\tilde{c}_{n,k}\\\tilde{d}_{n,k} \end{bmatrix} &= \sum_{i=0}^{m-1} e^{ikq(l_e+l_i)}\left( \mathcal{L}_e(q) + e^{iql_e}\mathcal{L}_i(q) \cdot \mathcal{M}_e\right)\cdot\mathcal{M}^k \begin{bmatrix}1\\0\end{bmatrix}\\
%&= \mathcal{L}(q) \cdot (\mathcal{I}_2 - e^{iql}\mathcal{M})^{-1}(\mathcal{I}_2 - e^{imql}\mathcal{M}^{m})\begin{bmatrix}1\\0\end{bmatrix}\;,
%\end{align}
%with $\mathcal{L}(q) =\mathcal{L}_e(q) + e^{iql_e}\mathcal{L}_i(q) \cdot \mathcal{M}_e$ and $l=l_i+l_e$.
%We can already see something interesting: the computation will be somewhat identical to the simple periodic case, except that the row vector $\mathcal{L}$ is modified by the inner structure of the repeated block.

\subsection{Complete expression of the dMRI signal}

According to Eq. \eqref{eq:NPA_modes_formula}, the signal is expressed as a sum over all eigenmodes $u_n$. We recall that the eigenmodes are alternately symmetric (odd $n$) and anti-symmetric (even $n$). Combining the above results \eqref{eq:calcul_relax_general_norme}-\eqref{eq:calcul_relax_general_fourier2}, one gets
%\begin{equation}
%S=\sum_{j,\pm} \frac{\left[\splitfrac{\left(\substack{\cos(mq/2)\\ \sin(mq/2)}\right)\tilde{K}\left[(\cos{\alpha_j}-\cos q)\pm\tilde{r}(q\sin q-{\alpha_j} \sin {\alpha_j})\right]}{+\left(\substack{\sin(mq/2)\\ \cos(mq/2)}\right) q(\cos q-\cos {\alpha_j})(1-\frac{\tilde{K}}{\tilde{\kappa}})}\right]^2\frac{8 e^{-{{\alpha_j}}^2t}}{(q^2-({\alpha_j})^2)^2(\cos q - \cos {\psi_j})^2}}{\left[\splitfrac{\frac{\sin{\alpha_j}\left(1+\frac{\tilde{r}}{2}\right)+\frac{\tilde{r}}{2}{\alpha_j}\cos{\alpha_j}}{\sin^2{\psi_j}}\left[\sin{\alpha_j} -2\frac{\tilde{K}}{{\alpha_j}}\cos {\alpha_j}-\frac{\tilde{K}^2}{{{\alpha_j}}^2} \sin{\alpha_j}\right]\left[\frac{(m-1)\sin{\psi_j}}{\sin (m-1){\psi_j}}-\cos m{\psi_j}\right]}{-\frac{\sin m{\psi_j}}{\sin {\psi_j}}\left[\frac{\sin{\alpha_j}}{{\alpha_j}}\left(1+2\tilde{K}+\frac{\tilde{K}^2}{{{\alpha_j}}^2}\right)+\left(1-\frac{\tilde{K}^2}{{{\alpha_j}}^2}\right)\cos{\alpha_j}\right]} \right]}
%\end{equation}
%
\begin{equation}
S=\sum_{n=1}^{\infty} \frac{\left(A_n^2+B_n^2+(-1)^{n-1}(A_n^2-B_n^2)\cos{mq}+(-1)^{n-1}2A_nB_n\sin{mq}\right)4\beta_n^2 e^{-\alpha_n^2t}}{(q^2-{\alpha_n}^2)^2 (\cos{q}-\cos{\psi_n})^2}\;,
\end{equation}
where $\beta_n$ is given by Eq. \eqref{eq:calcul_relax_general_norme}, $A_n$ and $B_n$ by Eq. \eqref{eq:calcul_relax_general_fourier2}, $\psi_n$ by Eq. \eqref{eq:equation_alpha_psi} and $\alpha_n$ are solutions of Eq. \eqref{eq:coherence_relaxation_2}. For $m=1$, we recover the signal derived by Coy and Callaghan \cite{Coy94}.

%%%%%%%%%%%%%%%%%%%%%%%%%%%%%%%%%%%%%%%%%%%%%%%%%%%%%%%%%%%%%%

\subsection{Perfectly relaxing outer boundaries}
\label{section:calculs_relax}

Note that the limit $\tilde{K}\to\infty$ is singular because of the chosen normalization \eqref{eq:u_beta_v}. This is particularly clear in Eq. \eqref{eq:conditions_limites_matrice} where $b^l_1 \to \infty$. In fact, $\tilde{K}=\infty$ represents Dirichlet conditions at the outer boundaries: $u(0)=u(L)=0$. To avoid the singularity we use another normalization:
\begin{equation}
u=\beta w\;, \quad w'(0)=\sqrt{\lambda/D}\;,
\end{equation}
which corresponds to the coefficients (for $w$)
\begin{equation*}
\begin{bmatrix}a^l_1\\b^l_1\end{bmatrix}=\begin{bmatrix}\frac{\sqrt{\lambda D_1}}{K_-}\\1\end{bmatrix}\;.
\end{equation*}
%%%%%%%%%%%%%%%%%%%%%%%%%%%%%%%%%%%%%%%%%%%%%%%%%%%%%%%%%%%%%%%%%%%%

\subsubsection{Study of the spectrum}

When $\tilde{K}\to \infty$, Eq. \eqref{eq:coherence_relaxation_2} simplifies into
\begin{equation}
\sin\alpha \frac{\sin{m\psi}}{\sin{\psi}}+\tilde{r}\alpha\frac{\sin{(m-1)\psi}}{\sin{\psi}}=0\;.
\label{eq:coherence_relaxation_relax}
\end{equation}
We now study the solutions of this equation in three different regimes: high-permeability, low-permeability, and very large number of compartments. We rely on the discussion developed in Sec. \ref{section:study_spectrum}, which leads us to the following conclusions.

\paragraph{High-permeability regime}

In the high-permeability regime ($\tilde{r}\ll1$), the solutions are located near the limits $\alpha_0={n\pi}/{m}$, which correspond also to $\psi_0=n\pi/m$ ($n=1,2,\ldots$). More precisely one can compute the first-order expansion:
\begin{equation*}
\begin{cases}
\displaystyle
\alpha_n\approx \frac{n\pi}{m}\left(1-\frac{\tilde{r}(m-2)}{2m}\right)&\text{if $n$ is not a multiple of $m$,}\\
\displaystyle
\alpha_n\approx \frac{n\pi}{m}\left(1-\frac{\tilde{r}(m-1)}{m}\right)&\text{otherwise.}
\end{cases}
\end{equation*}
As already noted this case presents no difficulty from the numerical point of view.

\paragraph{Low-permeability regime}

In the low-permeability regime ($\tilde{\kappa}\ll 1$), the solutions are divided into two categories. 
%First the solutions corresponding to all the \textquote{inner} compartments: $1 < k <m$. These solutions form groups located at a distance of order $\tilde{\kappa}/n\pi$ from $\alpha_0=n\pi/l$ ($n$ being an integer); moreover each group typically spreads over a distance of order $\tilde{\kappa}/n\pi$. 

$\bullet$ First, the solutions corresponding to the \textquote{inner} compartments: $1 < k <m$. These solutions form groups located around $\alpha_0=j\pi$ ($j$ being an integer). In fact they correspond to $\psi \in \mathbb{R}$, at which $\sin(m\psi)$ and $\sin((m-1)\psi)$ are of the same order. This implies that Eq. \eqref{eq:coherence_condition_relaxation} becomes in the low-permeability limit
\begin{equation*}
\frac{\sin((m-1)\psi)}{\sin\psi}=0\;,
\end{equation*}
which is (almost) the equation of the spectrum of $m-1$ identical cells with impermeable outer boundaries \eqref{eq:coherence_periodic_explicite}. One gets simply the solutions $\psi_0={p\pi}/{(m-1)}$, $p=1,\ldots,m-2$, thus the solutions in the first category are approximately determined by
\begin{equation*}
\cos\alpha - \frac{\tilde{r}}{2}\alpha\sin\alpha=\cos(p\pi/(m-1)), \quad p=1,\ldots,m-2\;.
\end{equation*}
We study this equation in details in Sec. \ref{section:DLs}. In particular, applying Eq. \eqref{eq:DL_petit_kappa_j0} one gets for the $m-2$ first solutions:
\begin{equation}
\alpha_n \approx 2\sqrt{\tilde{\kappa}}\sin(\frac{n\pi}{2(m-1)})\;, \quad n=1,\ldots,m-2 \;.
\label{eq:DL_petit_kappa_j0_relax}
\end{equation}

$\bullet$ Second, the solutions corresponding to the outer compartments $k=1,m$. These solutions form pairs $\alpha_\pm$ such that
\begin{align*}
\left(n+\frac{1}{2}\right)\pi-\alpha_+ &\approx \left(n+\frac{1}{2}\right)\pi - \alpha_- \sim \frac{\tilde{\kappa}}{(n+1/2)\pi}\;,\\
\alpha_+-\alpha_- &\sim \left(\frac{\tilde{\kappa}}{(n+1/2)\pi}\right)^{m-1}\;, 
\end{align*}
with $n=1,2,\ldots$.
Therefore in the low-permeability limit ($\tilde{\kappa}\to 0$) these pairs are very difficult to detect, especially when one is dealing with a large number of compartments $m$. As explained in Sec. \ref{section:numerical_implementation}, even if one finds the roots, the subsequent computation of the eigenmodes and their norm may be inaccurate.
%Moreover the fact that these solutions are difficult to detect is related to the very fast variations of the left-hand side of \eqref{eq:coherence_condition_relaxation}) with $\alpha$. This implies that in this regime it may be difficult to accurately compute the modes (because of the fast variation of the $(a_i, b_i)$ with $\alpha$) as well as their norm. 
%The integral of the mode (and in general its Fourier transform) is little sensitive to small changes in $\alpha$ and thus may be computed very accurately.
%However we shall see in the following that these solutions have little influence on the first exit time distribution.
However in this regime these solutions are much larger than the smallest one from the first category which go to zero according to Eq. \eqref{eq:DL_petit_kappa_j0_relax}. Hence they have little influence on the first exit time distribution \eqref{eq:formule_temps_sortie_f} because of the very fast exponential decay compared to the first terms of the sum.

\paragraph{Limit $m\to \infty$}

From the above discussion we get that the $m-2$ first solutions of Eq. \eqref{eq:coherence_condition_relaxation}, $\alpha_1,\ldots,\alpha_{m-2}$,  satisfy
\begin{equation*}
n\pi/m < \psi_n < n\pi/(m-1)\;, \quad n=1,\ldots,m-2\;.
\end{equation*}
Thus one may write $\psi_n=\frac{n\pi}{m-x}$, with $0 < x <1$. Let us rewrite Eq. \eqref{eq:coherence_condition_relaxation} as
\begin{align*}
\sin\alpha_n \sin(m\psi_n) + \tilde{r}\alpha_n \sin((m-1)\psi_n)=(-1)^n\left(\sin\alpha_n\sin(\frac{xn\pi}{m-x})-\tilde{r}\alpha_n\sin(\frac{(1-x)n\pi}{m-x})\right)=0\;.
\end{align*}
Now we study the limit $m\to\infty$ with fixed $n$. Then $\psi_n, \alpha_n \ll 1$ and the above equation transforms into
\begin{equation*}
\frac{(-1)^n\alpha_n n\pi}{m-x}(x-\tilde{r}(1-x))=0\;,
\end{equation*}
from which we get $x=\tilde{r}/(1+\tilde{r})=1/(1+\tilde{\kappa})$.
Let us use the expansion \eqref{eq:DL_petit_alpha}:
\begin{equation}
\alpha_n\approx \sqrt{\frac{\tilde{\kappa}}{\tilde{\kappa}+1}} \frac{n\pi}{m-\frac{1}{\tilde{\kappa}+1}}
\approx\sqrt{\frac{\tilde{\kappa}\left(1+\frac{2}{m}\right)}{\tilde{\kappa}\left(1+\frac{2}{m}\right)+1}} \frac{n\pi}{m}\;,
\quad n=1,\ldots,m-2\;.
\label{eq:DL_petit_alpha_relax}
\end{equation}

%\paragraph{Ratio $\lambda_3/\lambda_1$}
%Now we do not make any assumption about $m$ (except $m>2$). Recall that the symmetric modes $u_n$, $n=1,3,5,\ldots$ are the only ones that contribute to the sum \labelcref{eq:formule_temps_sortie}. We study in particular the ratio $\lambda_3/\lambda_1$ of the first two contributing eigenvalues.
%
%In the high-permeability limit, the solutions $\alpha$ are equally spaced: $\alpha_3=3\alpha_1$, which implies $\lambda_3 = 9 \lambda_1$. When the permeability decreases:
%\begin{itemize}
%\item if $m>4$ the ratio between $\alpha_3$ and $\alpha_1$ decreases; when $\tilde{\kappa}\to 0$ one gets from \cref{eq:DL_petit_kappa_j0_relax}:
%\begin{equation*}
%\frac{\alpha_3}{\alpha_1}=\frac{\sin(\frac{3\pi}{2(m-1)})}{\sin(\frac{\pi}{2(m-1)})}> 2\;,
%\end{equation*}
%hence $\lambda_3 > 4\lambda_1$. Thus one has always: $4<\lambda_3/\lambda_1<9$.
%%
%\item if $m=3$ or $m=4$, the ratio between $\alpha_3$ and $\alpha_1$ decreases then increases and goes to infinity when $\tilde{\kappa}\to 0$ ($\alpha_1 \to 0$ and $\alpha_3 \to \pi/2$). The minimal value is reached around $\tilde{\kappa}=1$ and we check numerically that it is greater than $2$.
%Hence once again $\lambda_3 > 4\lambda_1$. However in this particular case the ratio $\lambda_3/\lambda_1$ is unbounded.
%\end{itemize}

%%%%%%%%%%%%%%%%%%%%%%%%%%%%%%%%%%%%%%%%%%%%%%%%%%%%%%%%%%%

\subsubsection{Computation of the norm}

The formula \eqref{eq:normalisation1} for the norm  becomes
\begin{equation*}
\beta^{-2}=\int_0^L w^2 = \frac{-\sqrt{D_1}}{2\eta}\left.\frac{\mathrm{d}}{\mathrm{d}\sqrt{s}}\left(\begin{bmatrix}\frac{K_+}{K_-}&\frac{\sqrt{D_m s}}{K_-}\end{bmatrix}\mathcal{T}(s)\begin{bmatrix}\frac{\sqrt{D_1 s}}{K_-} \\ 1 \end{bmatrix}\right)\right|_{s=\lambda}\;.
\end{equation*}
In the particular geometry we are dealing with and in the case $\tilde{K}=\infty$, this gives
\begin{align}
\beta^{-2}&=\frac{-\epsilon l}{2}\left|\begin{bmatrix}1&0\end{bmatrix}\frac{\mathrm{d}\mathcal{T}}{\mathrm{d}\alpha}\begin{bmatrix}0\\1\end{bmatrix}\right|\\
&=\frac{-\epsilon ml}{2}\frac{\sin\alpha\left(1+\frac{\tilde{r}}{2}\right)+\frac{\tilde{r}}{2}\alpha\cos\alpha}{\sin^2\psi}\left[\sin\alpha\cos m\psi+\frac{\tilde{r}\alpha (m-1)}{m}\cos((m-1)\psi)\right]\nonumber\\
&+\frac{\epsilon ml}{2}\left(\frac{\sin\alpha}{\alpha}-\cos\alpha \right)\frac{\sin m\psi}{m\sin\psi}\;.
\end{align}

\subsection{Computation of the Fourier transform}

In the same way, the computation of the Fourier transform of $w$ simplifies into
\begin{equation}
\frac{e^{imq/2} 2l\alpha}{(q^2-\alpha)^2(\cos q - \cos \psi)}\times
\begin{cases}
A\cos(mq/2) + B \sin(mq/2)& \text{if $\epsilon = +1$}\\
-i(A\sin(mq/2)-B\cos(mq/2))& \text{if $\epsilon = -1$}
\end{cases}\;,
\end{equation}
with
\begin{equation}
A=\left[(\cos\alpha-\cos q)+\frac{\tilde{r}}{2}(q\sin q-\alpha \sin \alpha)\right]\;,\quad B=\frac{\tilde{r}}{2} q(\cos \alpha-\cos q)\;.
\end{equation}

%%%%%%%%%%%%%%%%%%%%%%%%%%%%%%%%%%%%%%%%%%%%%%%%%%%%%%%%%%%%%%%%%%%%%%%%

%%%%%%%%%%%%%%%%%%%%%%%%%%%%%%%%%%%%%%%%%%

\section{Bi-periodic geometry}
\label{section:bi-periodique}
In this section, we briefly apply our method to the computation of the spectrum of the diffusion operator on a finite periodic geometry where the elementary block is made of two different compartments (repeated $M$ times). Such a system may model laminated steel coils in industrial processes \cite{Yuen94,Hickson09-1} or intra- and extra-cellular spaces in biology \cite{Kuchel99,Novikov98,Fieremans10}. This is also a good example of the numerical simplifications that our method enables.
The lengths of the compartments are denoted by $l_e$ and $l_i$, their diffusion coefficients by $D_e$ and $D_i$ and the barrier between the two compartments has a permeability $\kappa$ (or equivalently a resistance $r=1/\kappa$). For simplicity we assume reflecting boundary conditions at the outer boundaries. 
Let us introduce the notations
\begin{equation}
\tau_i=l_i^2/D_i\; \qquad \text{and} \qquad \tau_e=l_e^2/D_e\;.
\end{equation}
In that case, the equation \eqref{eq:coherence_condition} on the spectrum is $\mathcal{M}^M\begin{bmatrix}1\\0\end{bmatrix}=\epsilon\begin{bmatrix}1\\0\end{bmatrix}$, with
\begin{align}
\mathcal{M}=
\begin{bmatrix}1&r\sqrt{\lambda D_i}\\0&\sqrt{D_i/D_e}\end{bmatrix}&
\begin{bmatrix}\cos(\sqrt{\lambda\tau_i})&\sin(\sqrt{\lambda\tau_i})\\-\sin(\sqrt{\lambda\tau_i})&\cos(\sqrt{\lambda\tau_i})\end{bmatrix}
\begin{bmatrix}1&r\sqrt{\lambda D_e}\\0&\sqrt{D_e/D_i}\end{bmatrix}
\begin{bmatrix}\cos(\sqrt{\lambda\tau_e})&\sin(\sqrt{\lambda\tau_e})\\-\sin(\sqrt{\lambda\tau_e})&\cos(\sqrt{\lambda\tau_e})\end{bmatrix}\;.
\label{eq:matrice_transition_biperiodique}
\end{align}
Because the geometry is \emph{not} symmetric, $\epsilon$ is not necessary equal to $\pm 1$. Moreover we have $\epsilon \eta =\sqrt{D_e/D_i}$. Following the same reasoning as in Sec. \ref{section:modes_simple_periodic}, we obtain that the solutions of Eq. \eqref{eq:coherence_condition} can be decomposed into two types:
\begin{itemize}
\item the ones such that $\begin{bmatrix}1\\0\end{bmatrix}$ is an eigenvector of the transition matrix of one block, $\mathcal{M}$, from Eq. \eqref{eq:matrice_transition_biperiodique}. This gives the condition:
\begin{equation}
r\sqrt{\lambda D_iD_e} = \sqrt{D_i}\cot(\sqrt{\lambda\tau_e})\sin(\sqrt{\lambda\tau_i})+\sqrt{D_e}\sin(\sqrt{\lambda\tau_e})\cot(\sqrt{\lambda\tau_i})\;.
\label{eq:spectre_biperiodique_1}
\end{equation}
Moreover, one has
\begin{equation}
\epsilon=\left(\!\cos(\sqrt{\lambda\tau_e})\cos(\sqrt{\lambda\tau_i})\!-\!\sqrt{\frac{D_i}{D_e}}\sin(\sqrt{\lambda\tau_e})\sin(\sqrt{\lambda\tau_i})\!-\! r\sqrt{\lambda D_i}\cos(\sqrt{\lambda\tau_e})\sin(\sqrt{\lambda\tau_i})\!\!\right)^{-M}\;;
\end{equation}

\item the ones such that $\Tr(\mathcal{M})=2\cos{p\pi/M}$, with $p=1,\ldots,M-1$, which corresponds to $\mathcal{M}^M=(-1)^p \mathcal{I}_2$ and thus to $\epsilon=(-1)^p$. This gives the equation
\begin{align}2\cos{p\pi/M}&=2\cos(\sqrt{\lambda\tau_e})\cos(\sqrt{\lambda\tau_i})-\left(\sqrt{\frac{D_i}{D_e}}+\sqrt{\frac{D_e}{D_i}}\right)\sin\left(\sqrt{\lambda\tau_e}\right)\sin\left(\sqrt{\lambda\tau_i}\right)
\nonumber\\&-2r\sqrt{\lambda}\left(\sqrt{D_e}\sin\left(\sqrt{\lambda\tau_e}\right)\cos\left(\sqrt{\lambda\tau_i}\right)\!+\!\sqrt{D_i}\cos\left(\sqrt{\lambda\tau_e}\right)\sin\left(\sqrt{\lambda\tau_i}\right)\!\right)
\nonumber\\&+r^2\lambda\sqrt{D_iD_e}\sin\left(\sqrt{\lambda\tau_e}\right)\sin\left(\sqrt{\lambda\tau_i}\right) \;, \qquad p=1,\ldots,M-1\;.
\label{eq:spectre_biperiodique_2}
\end{align}
\end{itemize}
It is interesting to compare the above equations with the analysis conducted in Sec. \ref{section:study_spectrum}. Indeed, one can see that in the limit of quasi-impermeable barriers ($r\to\infty$), Eq. \eqref{eq:spectre_biperiodique_1} yields approximately
\begin{equation}
\sqrt{\lambda/D_e}\approx \frac{n\pi}{l_e} + \frac{1}{n\pi rD_e} \qquad \text{and} \qquad \sqrt{\lambda/D_i}\approx \frac{n\pi}{l_i} + \frac{1}{n\pi rD_i}\;, \quad n=1,2,\ldots,
\end{equation}
which is exactly Eq. \eqref{eq:formule_approchee_racine_isolee} with $\zeta=1$, that is for the outer compartments. In the same way, Eq. \eqref{eq:spectre_biperiodique_2} yields approximately
\begin{align}
\sqrt{\lambda/D_e}&\approx \frac{n\pi}{l_e} + \frac{2}{n\pi rD_e}+\frac{l_e\sqrt{D_e/D_i} X_p}{(n\pi rD_e)^2}\;,\nonumber\\
\sqrt{\lambda/D_i}&\approx \frac{n\pi}{l_i} + \frac{2}{n\pi rD_i}+\frac{l_i\sqrt{D_i/D_e} Y_p}{(n\pi rD_i)^2}\;,
\end{align}
where $n=1,2,\ldots$, and $X_p,Y_p$ are dimensionless coefficients which depend on the value of $\cos p\pi/M$, with $p=1,\ldots,M-1$. One recognizes the first order correction from Eq. \eqref{eq:formule_approchee_racine_isolee} for inner compartments. The second order correction is also discussed in Eq. \eqref{section:study_spectrum} and arises from the next-nearest neighbor coupling between the compartments of the same type. Therefore, in the low-permeability limit, the spectrum is made of groups of $M$ closely packed eigenvalues located around $\lambda=D_e(n\pi/l_e)^2$ or $\lambda=D_i(n\pi/l_i)^2$: one eigenvalue is given by Eq. \eqref{eq:spectre_biperiodique_1} then the following $M-1$ eigenvalues are given by Eq. \eqref{eq:spectre_biperiodique_2}.
These groups correspond to eigenmodes localized inside all compartments of type \textquote{$e$} or \textquote{$i$}, respectively. More precisely, the first eigenvalue of each group corresponds to an eigenmode localized inside an outer compartment and the $M-1$ following eigenvalues correspond to eigenmodes localized inside all inner compartments.

Equations \eqref{eq:spectre_biperiodique_1} and \eqref{eq:spectre_biperiodique_2} \textquote{disentangle} these groups of eigenvalues, that allows one to compute very fast the spectrum of the diffusion operator for any number of repetitions $M$ and any barrier permeability. This is a major simplification of the numerical problem of the determination of the spectrum (see Sec. \ref{section:study_spectrum} and \ref{section:numerical_implementation}).
The same remark applies to any finite periodic geometry, provided that the repeated elementary block is not too long.

%%%%%%%%%%%%%%%%%%%%%%%%%%%%%%%%%%%%%%%%%%%%%%%%%%

%%%%%%%%%%%%%%%%%%%%%%%%%%%%%%%%%%%%%%%%%%%%%%%%%%%%%%%%%%%%%%%%%%%%%%%%

\section{Two-scale geometry}
\label{section:autres_structures_imbrique}
\subsection{Eigenmodes}

\begin{figure*}[tbp]
\begin{center}
\includegraphics[width=0.9\linewidth]{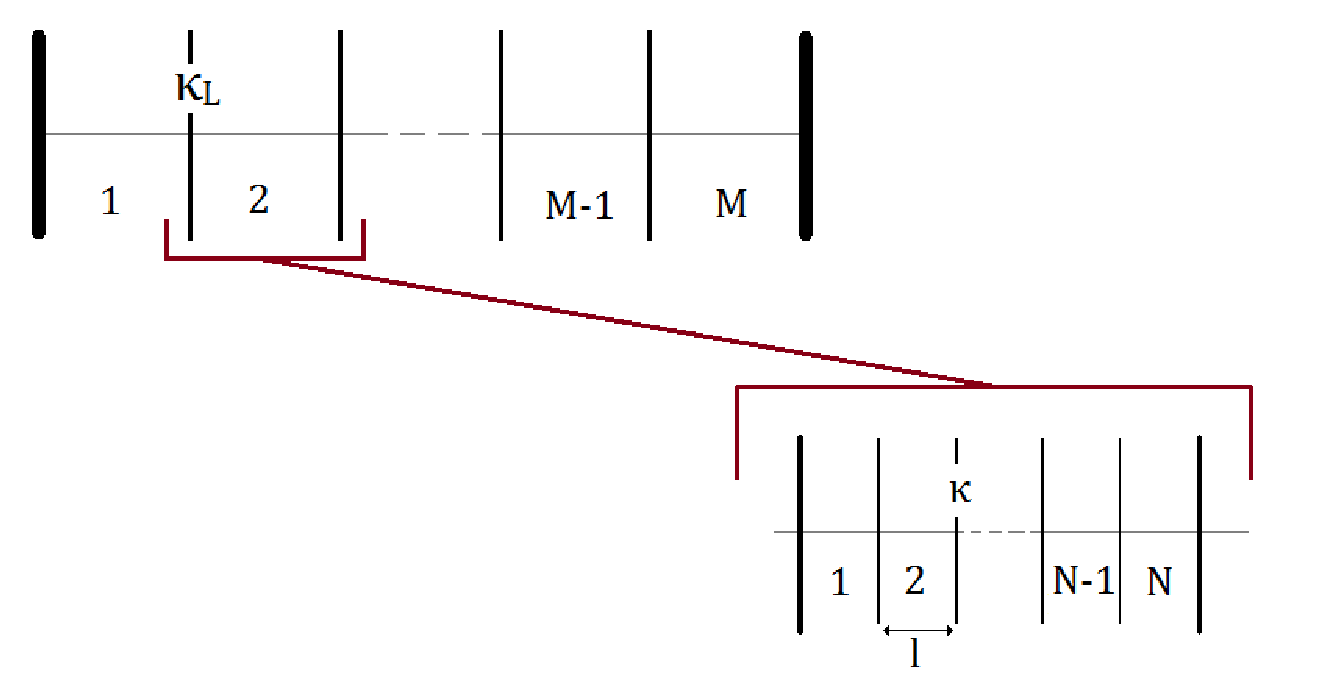}
\end{center}
\caption{Illustration of the two-scale geometry, which is a repetition of $M$ blocks of $N$ cells. All the cells have the same length $l$ and diffusion coefficient $D$ and are separated by barriers of permeability $\kappa$. The blocks are separated by barriers of permeability $\kappa_L$.
}
\label{fig:couches2}
\end{figure*}

We consider again the repetition of an elementary block but without restricting ourselves to a small block. Indeed the structure is the repetition of $M$ arrays of $N$ identical cells, each array being separated from others by a ``large barrier'' (see Fig. \ref{fig:couches2}). For simplicity we assume reflecting boundary conditions at the endpoints. The cells are of length $l$, the barriers are of permeability $\kappa$, the diffusion coefficient is $D$, and the ``larger barriers'' are of permeability $\kappa_L$. In addition to the notations \eqref{eq:notations_simple}, we introduce:
\begin{equation}
\tilde{r}_L=1/\tilde{\kappa}_L = D/(\kappa_Ll) \quad \text{ and }\quad \tilde{\rho}=\tilde{r}_L-\tilde{r}\;.
\end{equation}
Strictly speaking, $\tilde{\rho}$ may be negative, however we have in mind the opposite case where the \textquote{larger barriers} are less permeable than the inner barriers.

We have two different matrices to consider:
\begin{itemize}
\item the matrix associated to the microstructure is $\mathcal{M}_1 = \begin{bmatrix}
1&\tilde{r}\alpha\\
0&1
\end{bmatrix}
\begin{bmatrix}\cos\alpha & \sin\alpha \\-\sin\alpha & \cos\alpha \end{bmatrix}$.

\item the matrix associated to the macrostructure is $\mathcal{M}_2=\begin{bmatrix}
1&\tilde{\rho}\alpha\\
0&1
\end{bmatrix}{\mathcal{M}_1}^N$.
\end{itemize}
Thanks to the formula \eqref{eq:M_puissance_simple}, we can compute the matrix $\mathcal{M}_2$:
\begin{equation}
\mathcal{M}_2=\frac{1}{\sin\psi}
\begin{bmatrix}
\left(\splitfrac{\sin(N+1)\psi}{-\left(\cos\alpha+\tilde{\rho}\alpha\sin\alpha\right)\sin N\psi}\right) & \left(\splitfrac{(\sin\alpha + \tilde{R}\alpha\cos\alpha)\sin N\psi}{- \tilde{\rho}\alpha\sin(N-1)\psi}\right)
\\
-\sin\alpha \sin N\psi & \cos\alpha \sin N\psi - \sin(N-1)\psi
\end{bmatrix}\;.
\end{equation}
Since the geometry is symmetric, Eq. \eqref{eq:coherence_condition} of the spectrum is 
\begin{equation}
{\mathcal{M}_2}^M
\begin{bmatrix}
1\\0
\end{bmatrix}
=\epsilon
\begin{bmatrix}
1\\0
\end{bmatrix}\;,
\label{eq:coherence_imbrique}
\end{equation}
with $\epsilon=\pm1$, and by analogy with the finite periodic geometry from Sec. \ref{section:modes_simple_periodic} we have two cases:
\begin{itemize}
\item $\sin\alpha\frac{\sin N\psi}{\sin\psi}=0$: the vector $\begin{bmatrix}1\\0\end{bmatrix}$ is an eigenvector of the matrix $\mathcal{M}_2$. This condition gives exactly the solutions $\alpha_{j,p}$, $j=0,1,\ldots$ and $p=0,\ldots,N$ (Sec. \ref{section:modes_simple_periodic}). One has $\epsilon=(-1)^{pM}$.
\item  The trace of the matrix ${\mathcal{M}_2}$ is $2\cos P\pi/M$, for $P\in\left\{1,\ldots,M-1\right\}$: ${\mathcal{M}_2}^M$ is plus or minus the identity matrix $\mathcal{I}_2$, which gives the condition:
\begin{equation}
\cos N\psi - \frac{\tilde{r}}{2}\alpha\sin\alpha \frac{\sin N\psi}{\sin\psi} = \cos P\pi/M\;, \quad P=1,\ldots,M-1\;.
\end{equation} In this case $\epsilon = (-1)^P$.
%One can note the similarity between this equation and equation \eqref{equation_relaxation_2}.
%
Again, we use a special notation for the solutions: $\alpha_{j,p,P}$, where the index $j$ means $j\pi \leq \alpha_{j,p,P} < (j+1)\pi$ and the index $p$ means $p\pi/N \leq \psi_{j,p,P} < (p+1)\pi/N$. The $P=0$ (resp., $P=M$) case corresponds then to the solutions for the finite periodic case $\alpha_{j,p}$ if $p$ is even (resp. if $p$ is odd).
\end{itemize}
%\begin{figure*}[tbp]
%\begin{center}
%\includegraphics[width=120mm]{Spectre_imbrique3.eps}
%\end{center}
%\caption{Solutions $\alpha_{j,p,P}$ for the two-scale geometry plotted versus $\psi_{j,p,P}$ as circles of different colors depending on the value of $P$, from $P=0$ (dark) to $P=M$ (bright). Each elementary block is made of $N=4$ cells, and the block is repeated $M=6$ times. The barrier resistances are $\tilde{r}=0.4$ and $\tilde{\rho}=1$. The solutions form little groups (indexed by $p$) inside larger branches (indexed by $j$) that are identical to the solutions for the array of $N$ identical cells ($\psi_{p,j}=p\pi/N$). Each group of $\alpha_{j,p,P}$ starts with the corresponding $\alpha_{j,p}$. Depending on the parity of $p$, this value corresponds to $P=0$ or $P=M$, and thus the $\alpha_{j,p,P}$ inside the group increase with increasing or decreasing $P$.
%\textit{[Bon.. je reconnais que cette figure n'est pas tres lisible. Je ne sais pas trop. Peut-etre faut-il mettre une figure mais une differente ? En l'etat actuel je pense qu'on peut la supprimer]}
%}
%\label{fig:spectre_imbrique}
%\end{figure*}

The interpretation of the indices $j,p,P$ follows the same line of reasoning as with the simple periodic geometry: they give the intra-compartment, inter-compartment (or intra-block) and inter-block variation of the mode, respectively.
%The index $j$ gives the intra-compartment variation of the mode; the index $p$ the inter-compartment (that is, the intra-block) variation and the index $P$ gives the inter-block variation over the whole interval.

\subsection{Computation of the norm:}
We use again Eq. \eqref{eq:M_puissance}:
\begin{align}
\begin{bmatrix}0&1\end{bmatrix}\mathcal{T}\begin{bmatrix}1\\0\end{bmatrix}&=\begin{bmatrix}0&1\end{bmatrix}(\mathcal{K}_2\mathcal{M}^N)^M\begin{bmatrix}1\\0\end{bmatrix}=\frac{\sin M\phi}{\sin\phi}\begin{bmatrix}0&1\end{bmatrix}\mathcal{K}_2\mathcal{M}^N\begin{bmatrix}1\\0\end{bmatrix}\nonumber\\
&=\frac{\sin M\phi}{\sin\phi}\frac{\sin N\psi}{\sin \psi}\begin{bmatrix}0&1\end{bmatrix}\mathcal\mathcal{M}\begin{bmatrix}1\\0\end{bmatrix}=-\frac{\sin M\phi}{\sin\phi}\frac{\sin N\psi}{\sin \psi}\sin\alpha\;,
\end{align}
where we have introduced $\phi$ defined by
\begin{equation}
\cos\phi=\frac{1}{2}\Tr(\mathcal{K}_2\mathcal{M}^m)=\cos N\psi - \frac{\tilde{r}}{2}\alpha\sin\alpha \frac{\sin N\psi}{\sin\psi}\;.
\label{eq:equation_phi}
\end{equation}
Now we have three cases:
\begin{enumerate}
\item $\sin\alpha=0$, which corresponds to $\alpha_{j,0}$ and $\alpha_{j,N}$. One gets
\begin{equation*}
\beta^2=\frac{2}{MNl}\;.
\end{equation*}
\item $\frac{\sin N\psi}{\sin \psi}=0$, which corresponds to $\alpha_{j,p}, p=1,\ldots,N-1$.
In this case we get
\begin{equation*}
\beta_{j,p}^2=\frac{2}{ml}\frac{\sin^2 p\pi/N}{\sin\alpha_{j,p}\left(\sin\alpha_{j,p}\left(1+\frac{\tilde{r}}{2}\right)+\frac{\tilde{r}}{2}\alpha_{j,p}\cos\alpha_{j,p}\right)}\;.
\end{equation*}
\item $\frac{\sin M\phi}{\sin \phi}=0$, which corresponds to the general case. We use the chain rule again to compute the derivative with respect to $\alpha$:
\begin{align*}
&\frac{\mathrm{d}}{\mathrm{d}\alpha}\left(\frac{\sin M\phi}{\sin \phi}\right)=\frac{\mathrm{d}\cos\phi}{\mathrm{d}\alpha}\frac{\mathrm{d}\phi}{\mathrm{d}\cos\phi}\frac{\mathrm{d}}{\mathrm{d}\phi}\left(\frac{\sin M\phi}{\sin \phi}\right)\;,\nonumber\\
&\frac{\mathrm{d}\cos\phi}{\mathrm{d}\alpha}=-N\frac{1-\cos N\psi \cos P\pi/M}{\sin N\psi \sin \psi}\left[\left(1+\frac{\tilde{r}}{2}\right)\sin\alpha+\frac{\tilde{r}}{2}\alpha\cos\alpha\right]\nonumber\\
&+\frac{\cos N\psi-\cos P\pi/M}{\sin^2\psi}\left[\frac{\sin^2\alpha}{\alpha}+\frac{\tilde{r}}{2}(\alpha+\sin\alpha\cos\alpha)\right] \;, \nonumber\\
&\frac{\mathrm{d}\phi}{\mathrm{d}\cos\phi}\frac{\mathrm{d}}{\mathrm{d}\phi}\left(\frac{\sin M\phi}{\sin \phi}\right)=\left(\frac{-1}{\sin{P\pi/M}}\right)\left(\frac{(-1)^P}{\sin{P\pi/M}}\right)\;.
\end{align*}
\end{enumerate}
Hence we get the normalization constant:
\begin{equation}
\beta_{j,p,P}^2=\left.\frac{\frac{2\sin^2\left(P\pi/M\right) \sin\psi}{ml\sin\alpha \sin{N\psi}}}{\left[\splitfrac{\frac{1-\cos N\psi \cos P\pi/M}{\sin N\psi \sin \psi}\left(\left(1+\frac{\tilde{r}}{2}\right)\sin\alpha+\frac{\tilde{r}}{2}\alpha\cos\alpha\right)}{+\frac{\cos N\psi-\cos P\pi/M}{N\sin^2\psi}\left(\frac{\sin^2\alpha}{\alpha}+\frac{\tilde{r}}{2}(\alpha+\sin\alpha\cos\alpha)\right)}\right]}\right|_{\alpha=\alpha_{j,p,P}}\;.
\label{eq:normalisation_imbrique}
\end{equation}
\subsection{Fourier transform}

In the same way as for the finite periodic geometry, we have only one $\mathcal{L}$ to consider, so we need to compute
\begin{align}
&\sum_i e^{ikq}\mathcal{L}_i \begin{bmatrix}a^l_{i}\\b^l_{i} \end{bmatrix} = \mathcal{L} \sum_{i=0}^{M-1}\sum_{i=0}^{N-1} e^{iq(KN + k)} {\mathcal{M}_1}^k{\mathcal{M}_2}^K \begin{bmatrix}1\\0\end{bmatrix}\nonumber\\
&= \mathcal{L}  (\mathcal{I}_2 - e^{iq}\mathcal{M}_1)^{-1}(\mathcal{I}_2 - e^{iNq}{\mathcal{M}_1}^{N}) (\mathcal{I}_2 - e^{iqN}\mathcal{M}_2)^{-1}(\mathcal{I}_2 - e^{iNMq}{\mathcal{M}_2}^{M})\begin{bmatrix}1\\0\end{bmatrix}\;.
\end{align}
Using Eq. \eqref{eq:coherence_imbrique} on the spectrum and the linearity of the comatrix operation, we get to simplify a lot the above expression:
\begin{equation}
\sum_i e^{ikq}\mathcal{L}_i \begin{bmatrix}a^l_{i}\\b^l_{i} \end{bmatrix} =(1-(-1)^Pe^{iNMq}) \frac{\det(\mathcal{I}_2 - e^{iNq}{\mathcal{M}_1}^N)}{\det(\mathcal{I}_2 - e^{iNq}\mathcal{M}_2)} \mathcal{L}  (\mathcal{I}_2 - e^{iq}\mathcal{M}_1)^{-1}\begin{bmatrix}1\\0\end{bmatrix}\;.
\end{equation}
And finally
\begin{equation}
\int_0^L v	(x)e^{i\gamma g \delta x} \,\mathrm{d}x = \frac{iql \left(1-(-1)^Pe^{iNMq}\right)\frac{\cos N\psi_{j,p,P} - \cos Nq}{\cos P\pi/M - \cos Nq}\frac{\cos\alpha_{j,p,P} - \cos q}{\cos \psi_{j,p,P} - \cos q}}{q^2 - {\alpha_{j,p,P}}^2} \;.
\end{equation}

\subsection{Complete expression of the dMRI signal}
We gather the above expressions to obtain the signal as a function of $q=\gamma g \delta l$ and $t=D\Delta/l^2$:

%\begin{align}
%S=&\frac{2(1-\cos mq)}{(mq)^2} + \sum_{j=1}^{\infty} \frac{4q^2(1-(-1)^{j\cdot m}\cos mq)}{m^2\left(q^2-(j\pi)^2\right)^2}e^{-D(j\pi)^2t/l^2}\nonumber\\
%&+\sum_{j=1}^{\infty}\sum_{p=1}^{N-1} \frac{4q^2\frac{(1-(-1)^{pM}\cos mq)}{m^2\left(\cos{q} - \cos{p\pi/N}\right)^2}\left(\frac{\cos q-\cos\alpha_{j,p}}{q^2-{\alpha_{j,p}}^2}\right)^2}{\left[\left(1+\frac{\tilde{r}}{2}\right)\sin\alpha_{j,p}+\frac{\tilde{r}}{2}\alpha_{j,p}\cos\alpha_{j,p} \right]\frac{\sin\alpha_{j,p}}{\sin^2(p\pi/N)}}e^{-D{\alpha_{j,p}}^2t/l^2}\nonumber\\
%&+\sum_{j=1}^{\infty}\sum_{p=0}^{N}\sum_{P=1}^{M-1}\left.\frac{4q^2\frac{(1-(-1)^{P}\cos mq)}{M^2\left({\cos Nq-\cos P\pi/M}\right)^2 }\left(\frac{\cos q-\cos\alpha}{q^2-{\alpha}^2}\right)^2
%\left(\frac{\cos Nq-\cos N\psi}{N(\cos q-\cos \psi)}\right)^2 \frac{\sin\psi \sin^2 P\pi/M}{\sin\alpha\sin N\psi}e^{-{\alpha_{j,p,P}}^2t}}{
%\frac{1-\cos N\psi \cos P\pi/M}{\sin N\psi \sin \psi}\left[\left(1+\frac{\tilde{r}}{2}\right)\sin\alpha+\frac{\tilde{r}}{2}\alpha\cos\alpha\right]+\frac{\cos N\psi-\cos P\pi/M}{N\sin^2\psi}\left[\frac{\sin^2\alpha}{\alpha}+\frac{\tilde{r}}{2}(\alpha+\sin\alpha\cos\alpha)\right]}\right|_{\alpha=\alpha_{j,p,P}} 
%\label{eq:S_twoscale}
%\end{align}

\begin{align}
S&=\frac{2(1-\cos mq)}{(mq)^2} + \sum_{j=1}^{\infty} \frac{4q^2(1-(-1)^{j m}\cos mq)}{m^2\left(q^2-(j\pi)^2\right)^2}e^{-(j\pi)^2t}&\; \nonumber\\
&+\sum_{j=0}^{\infty}\sum_{p=1}^{N-1} \frac{2lq^2}{m} \frac{1-(-1)^{pM}\cos mq}{(\cos q-\cos p\pi/N)^2}\left(\frac{\cos q-\cos\alpha_{j,p}}{q^2-\alpha_{j,p}^2}\right)^2\beta_{j,p}^2 e^{-{\alpha_{j,p}}^2t}\nonumber\\
&+\sum_{j=1}^{\infty}\sum_{p=0}^{N}\sum_{P=1}^{M-1}\frac{2mlq^2(1-(-1)^{P}\cos mq)}{M^2\left({\cos Nq-\cos P\pi/M}\right)^2 }\left(\frac{\cos q-\cos\alpha_{j,p,P}}{q^2-{\alpha_{j,p,P}}^2}\right)^2
\left(\frac{\cos Nq-\cos N\psi_{j,p,P}}{N(\cos q  -\cos \psi_{j,p,P})}\right)^2
\nonumber\\
& \qquad\qquad\qquad\qquad\qquad\qquad\qquad\qquad
\times \beta^2_{j,p,P} e^{-{\alpha_{j,p,P}}^2t}\;,
\label{eq:S_twoscale}
\end{align}
where $\beta_{j,p}^2$ and $\beta_{j,p,P}^2$ are given by Eqs. \eqref{eq:normalisation_simple_periodic} and \eqref{eq:normalisation_imbrique}, respectively.

\section{Limit of the dMRI signal for the periodic geometry as $\tilde{\kappa}\to 0$ and $\tilde{\kappa}\to \infty$}
\label{section:Limite_permeabilite}

\subsection{High-permeability limit: $\tilde{\kappa}\to\infty$}

In this limit, one has:
\begin{equation}
\begin{cases}
\alpha_{j,p}=j\pi+p\pi/m & \text{if $j$ is even,}\\
\alpha_{j,p}=j\pi+(m-p)\pi/m & \text{if $j$ is odd.}
\end{cases}
\end{equation}
In particular, $\cos\alpha_{j,p}=\cos\psi_{j,p}$, so the expression of the signal simplifies into
\begin{align*}
S&=\frac{2(1-\cos mq)}{(mq)^2} + \sum_{j=1}^{\infty} \frac{4q^2(1-(-1)^{jm}\cos mq)}{m^2\left(q^2-(j\pi)^2\right)^2}e^{-(j\pi)^2t}\\
&+\sum_{j=0}^{\infty}\sum_{p=1}^{m-1} \frac{2lq^2}{m} \frac{1-(-1)^{p}\cos mq}{(q^2-\alpha_{j,p}^2)^2}\beta_{j,p}^2 e^{-{\alpha_{j,p}}^2t}\;,
\end{align*}
with $\beta_{j,p}^2={2}/{(ml)}$. Hence:
\begin{equation}
S=\frac{2(1-\cos mq)}{(mq)^2} + \sum_{n=1}^{\infty} \frac{4q^2(1-(-1)^{n}\cos mq)}{\left((mq)^2-(n\pi)^2\right)^2}e^{-(n\pi)^2t/m^2}\;,
\end{equation}
which is the formula of the signal for one interval of length $L=ml$, as expected.

\subsection{Low-permeability limit: $\tilde{\kappa}\to 0$}

Although the result is intuitively expected, the computation is more complicated. The mathematical reason is that in the limit $\tilde{\kappa}\to 0$, $\alpha_{j,p}=j\pi$ so that the eigenmodes of the branch $j$ are degenerate. Using Eq. \eqref{eq:spectre_periodic}, one gets the expression of the signal:
\begin{align*}
S=&\frac{2(1-\cos mq)}{(mq)^2} + \sum_{j=1}^{\infty} \frac{4q^2(1-(-1)^{jm}\cos mq)}{m^2\left(q^2-(j\pi)^2\right)^2}e^{-(j\pi)^2t}\nonumber\\&+\sum_{j=0}^{\infty}\sum_{p=1}^{m-1} \frac{2lq^2}{m} \frac{1-(-1)^{p}\cos mq}{(\cos q-\cos p\pi/m)^2}\left(\frac{\cos q-(-1)^j}{q^2-(j\pi)^2}\right)^2\beta_{j,p}^2 e^{-{(j\pi)}^2t}\;,
\end{align*}
with
\begin{equation*}
\begin{cases}
\beta_{j,p}^2%=\frac{2\sin^2{p\pi/m}}{ml(1-(-1)^j\cos{p\pi/m})}
=\frac{2}{ml}(1+(-1)^j\cos{p\pi/m}) & \text{if $j>0$,}\\
\beta_{0,p}^2=\frac{1}{ml}(1+(-1)^j\cos{p\pi/m})\;.
\end{cases}
\end{equation*}
Gathering all the terms, we obtain
\begin{equation}
S= S_0(q) \frac{2(1-\cos{q})^2}{m^2q^2} +\frac{4q^2}{m^2}\sum_{j=1}^{\infty}S_j(q) \frac{(1-(-1)^j \cos{q})^2}{(q^2-(j\pi)^2)^2} e^{-(j\pi)^2t} \;,
\label{eq:limite_signal_k0}
\end{equation}
with
\begin{equation}
S_j(q)=\sum_{p=0}^m \frac{(1-(-1)^p\cos{mq})(1+(-1)^j\cos{p\pi/m})}{(\cos{q} -  \cos{p\pi/m})^2(1+\theta_p)}\;, \quad j=0,1,\ldots\;,
\end{equation}
where $\theta_p=1$ if $p=0$ or $m$, and $\theta_p=0$ otherwise. To compute $S_j(q)$, we introduce the following polynomial:
\begin{equation}
\mathcal{P}(X)=\prod_{p=0}^m \left(X-\cos{p\pi/m}\right)\;.
\end{equation}
The analysis of its roots and degree leads to the following formula:
\begin{equation}
\mathcal{P}(\cos{q})=\mathcal{N} \sin(mq)\sin{q}\;,
\label{eq:formule_P_cosq}
\end{equation}
where $\mathcal{N}$ is an unknown proportionality coefficient whose value is not needed in the following. This allows us to compute
\begin{align}
&\mathcal{P}'(\cos{q})=\left(\frac{-1}{\sin{q}}\right)\mathcal{N}\left(m\cos(mq)\sin{q} + \sin(mq)\cos{q}\right)\;,\\
&\mathcal{P}'(\cos{p\pi/m})=\mathcal{N}m(-1)^{p+1}(1+\theta_p)\;.
\end{align}
Now we use the standard partial fraction expansion formula, for any polynomial $\mathcal{Q}$ such that $\deg{\mathcal{Q}} \leq \deg{\mathcal{P}}$:
\begin{equation}
\frac{\mathcal{Q}(X)}{\mathcal{P}(X)}=C+\sum_{p=0}^{m} \frac{\mathcal{Q}(\cos{p\pi/m})}{\mathcal{P}'(\cos{p\pi/m})(X-\cos{p\pi/m})}\;,
\label{eq:elements_simples}
\end{equation}
where prime denotes the derivative with respect to $X$ and $C$ is a constant. With the polynomial $\mathcal{R}(\cos{q})=\cos{mq}$, we get according to Eq. \eqref{eq:elements_simples}
\begin{align*}
S_j(q)&=\mathcal{N}m\left[\left(\frac{\mathcal{R}(X)(1+(-1)^jX)}{\mathcal{P}(X)}\right)' - \mathcal{R}(X)\left(\frac{1+(-1)^jX}{\mathcal{P}(X)}\right)'\right]_{X=\cos{q}}\\
&=\mathcal{N}m\mathcal{R}'(\cos{q})\frac{1+(-1)^j\cos{q}}{\mathcal{P}(\cos{q})}\;.
\end{align*}
Computing the derivative of $\mathcal{R}$ and using Eq. \eqref{eq:formule_P_cosq}, one finally gets
\begin{equation}
S_j(q)=\frac{m^2}{1-(-1)^j\cos{q}}\;.
\end{equation}
Now we come back to Eq. \eqref{eq:limite_signal_k0}, which yields
\begin{equation}
S=\frac{2(1-\cos{q})}{q^2}+\sum_{j=1}^{\infty}\frac{4q^2(1-(-1)^j\cos{q})}{(q^2-(j\pi)^2)^2} e^{-(j\pi)^2t}\;,
\end{equation}
which is the expected formula of the signal for one interval of length $l$.

%%%%%%%%%%%%%%%%%%%%%%%%%%%%%%%%%%%%%%%%%%%%%%%%%%%%%%%%%%%%%%%%%%%%%%%%

%%%%%%%%%%%%%%%%%%%%%%%%%%%%%%%%%%%%%%%%%%%%%%%%%%%%%%%%%%%%%%%%%%%%%%%%%
\section{Expansions for $\alpha_{j,p}$ for the periodic geometry}
\label{section:DLs}

\paragraph{Low-permeability limit: $\tilde{\kappa}\to 0$} In this case we rewrite Eq. \eqref{eq:equation_alpha_psi} as $\alpha\sin\alpha=2\tilde{\kappa}(\cos\alpha-\cos\psi)$.
We start with the branch $j=0$. Let us write $\alpha=u \sqrt{2\tilde{\kappa}(1-\cos\psi)}$.
Then
\begin{align*}
\alpha\sin\alpha&=2\tilde{\kappa}(1-\cos\psi)u^2\left(1-\frac{1}{3}\tilde{\kappa}(1-\cos\psi)u^2\right)+O(\tilde{\kappa}^3)\;,\\
(\cos\alpha - \cos \psi)&=(1-\cos\psi)-\tilde{\kappa}(1-\cos\psi)u^2+O(\tilde{\kappa}^2)\;,
\end{align*}
from which we derive
\begin{equation}
\alpha_{0,p}=2{\tilde{\kappa}}^{1/2}\sin (p\pi/2m)-\tilde{\kappa}^{3/2}\left(\sin (p\pi/2m)-\frac{2}{3}\sin^2(p\pi/2m)\right)+O(\tilde{\kappa}^{5/2})\;.
\label{eq:DL_petit_kappa_j0}
\end{equation}
Now, if $\alpha=j\pi+\epsilon$, one has
\begin{equation*}
\alpha\sin\alpha=(-1)^j (j\pi\epsilon+\epsilon^2+O(\epsilon^3))\;, \qquad
(\cos\alpha-\cos\psi)=(-1)^j(1-(-1)^j\cos\psi+O(\epsilon^2))\;,
\end{equation*}
which gives
\begin{equation}
\alpha_{j,p}=\begin{cases}\displaystyle
j\pi+\frac{4\tilde{\kappa}}{j\pi}\sin^2(p\pi/2m)-\frac{(4\tilde{\kappa})^2}{(j\pi)^3}\sin^4(p\pi/2m)+O(\tilde{\kappa}^3)& \text{if $j$ is even,}\\ \displaystyle
j\pi+\frac{4\tilde{\kappa}}{j\pi}\sin^2((m-p)\pi/2m)-\frac{(4\tilde{\kappa})^2}{(j\pi)^3}\sin^4((m-p)\pi/2m)+O(\tilde{\kappa}^3)& \text{if $j$ is odd.}
\end{cases}
\label{eq:DL_petit_kappa}
\end{equation}
This is consistent with the idea that at very low permeability the compartments become independent so that $\alpha_{j,p}$ (with $p=1,\ldots,m-1$) are identical and equal to $j\pi$. One notices that the deviation from this limit decreases with $j$ which is consistent with previous observations (Fig. \ref{fig:spectre_ref02}).

\paragraph{High permeability limit: $\tilde{r}\to 0$} Again, we start with the $j=0$ branch. Let us write $\alpha=\psi-u$. Then we have the equations:
\begin{align*}
\cos\alpha&=\cos\psi\left(1-\frac{u^2}{2}+O(u^4)\right)+\sin\psi (u+O(u^3))\;,\\
\alpha\sin\alpha&=\psi\sin\psi+u\sin\psi+u\psi\cos\psi +O(u^3)\;,
\end{align*}
which yield
\begin{equation}
\alpha_{0,p}=\frac{p\pi}{m}\left(1-\frac{\tilde{r}}{2}+\frac{\tilde{r}^2}{4}\left[1+\frac{p\pi/m}{2\tan(p\pi/m)}\right]+O\left(\tilde{r}^3\right)\right)\;.
\label{eq:DL_grand_kappa_j0}
\end{equation}
For the other branches, the computations are similar:
\begin{equation}
\alpha_{j,p}=\begin{cases}\displaystyle
(j\pi+p\pi/m)\left(1-\frac{\tilde{r}}{2}+\frac{\tilde{r}^2}{4}\left[1+\frac{j\pi+p\pi/m}{2\tan(p\pi/m)}\right]\right)+O\left(\tilde{r}^3\right)&\text{$j$ even,}\\ \displaystyle
 (j\pi+(m-p)\pi/m)\left(1-\frac{\tilde{r}}{2}+\frac{\tilde{r}^2}{4}\left[1+\frac{j\pi+(m-p)\pi/m}{2\tan((m-p)\pi/m)}\right]\right)+O\left(\tilde{r}^3\right)&\text{$j$ odd.}\end{cases}
\label{eq:DL_grand_kappa}
\end{equation}
Again, the interpretation is quite clear. When the permeability is very high, $\tilde{r}\to 0$ and the $\alpha_{j,p}$ approach the solutions for one interval of length $ml$, for which $\alpha_n=n\pi/m$ ($n=0,1,\ldots$). Consistently with the above low-permeability regime, the deviation from the limit $\tilde{\kappa}=\infty$ increases with $j$.

\end{document}